\documentclass[article,11pt,nofootinbib,singlecolumn,superscriptaddress,preprintnumbers,floatfix]{revtex4-2} 
\usepackage[nodisplayskipstretch]{setspace}
\usepackage{cancel}
\usepackage{mathtools}
\usepackage{soul}
\usepackage{physics}
\usepackage{amssymb}
\usepackage{braket}
\usepackage{amsmath}
\usepackage{epsfig}
\usepackage{hyperref}
\usepackage{cleveref} 
\usepackage{breakurl}
\usepackage{bm}
\usepackage{url}
\usepackage{color}
\usepackage{slashed}
\usepackage{tikz}
\usepackage{mathrsfs}
\usepackage{enumerate}
\usepackage{fontawesome5}
\usepackage[mathcal]{euscript}
\usetikzlibrary{decorations.markings}
\usepackage[english]{babel}
\usepackage{xcolor}
\makeatletter
\newcommand{\cz}{%
  \mathord{\mathpalette\vaggelis@z{z}}%
}
\addto\captionsenglish{}

\newcommand\beq{\begin{equation}}
\newcommand\eeq{\end{equation}}
\def\bea{\begin{eqnarray}}
\def\eea{\end{eqnarray}}

\DeclareRobustCommand{\SkipTocEntry}[4]{}

\newcommand\beal{\begin{aligned}}
\newcommand\eeal{\end{aligned}}

\usepackage{xcolor}

\newcommand{\bul}[1]{{\normalfont\scriptsize\color{black!65}#1}}

\newcommand\pd{\partial}


\definecolor{darkorange}{rgb}{1,0.549,0}
\definecolor{darkgreen}{rgb}{0.3,0.6.2}

\begin{document}

\preprint{DESY-26-095\\\phantom{~}}
\title{Perturbing Gravitational Atoms: \\ Negative Love, Resonant Tides and Shifted Resonances }
\author{Mateja Bo\v{s}kovi\'c}
\affiliation{Deutsches Elektronen-Synchrotron DESY, Notkestr. 85, 22607 Hamburg, Germany}
\author{Nikola Savi\'c}
\affiliation{Université Paris-Saclay, CNRS, CEA, Institut de Physique Théorique, 91191 Gif-sur-Yvette, France}

\begin{abstract}

The superradiant instability of rotating black holes can generate a significant overdensity of bosonic matter around them, together forming a gravitational atom. This mechanism allows one to probe a large part of the parameter space of scalars, axions and vectors that lies beyond the reach of traditional detection strategies. Modelling the dynamics of these systems in binaries is, however, subtle, due to the competing nature of the different perturbations. In this work, we provide a consistent scheme to treat these perturbations, including both the minimal set of internal ones (relativistic corrections and self-gravity) and the external tidal field. Within the worldline effective field theory framework, we then calculate the Love numbers of gravitational atoms, for the first time also for the most interesting case of spinning clouds. Certain spinning states are found to have \textit{negative} (static) Love numbers, with a magnitude parametrically enhanced relative to the scaling for the non-spinning states -- in phenomenologically relevant scenarios by a factor $\mathcal{O}(10\text{--}10^3)$. Finally, we consider the binary evolution in the early inspiral, assessing the impact of the competing   perturbations on \textit{shifted resonances}, which for dense clouds can alter the nature of some of the   hyperfine and fine transitions (sinking instead of floating). This dynamical picture allows us to   identify at which stages the permanent multipoles are the strongest indicators of new light bosons, and   at which the induced ones take over, while keeping track of both types of finite-size effects even
during the resonance.  More broadly, our results demonstrate that gravitational atoms are a useful toy model for studying the theoretical aspects of tidal response in gravitational-wave physics.

\end{abstract}

\maketitle\global\count\footins=1000 %

\newpage

\tableofcontents

\newpage

\section{Introduction}

The accelerated development of gravitational-wave (GW) astronomy across the frequency bands~\cite{LIGOScientific:2016aoc,LIGOScientific:2025slb,NANOGrav:2023gor,EPTA:2023fyk,Reardon:2023gzh,Xu:2023wog,LISA:2022kgy,Maggiore:2019uih,Abac:2025saz} is opening a new channel into the Universe, allowing us to probe the formation, structure, and environments of compact objects by studying their coalescence~\cite{Maggiore:2018sht}. From the perspective of Worldline Effective Field Theory (WEFT), as long as the orbital separation $R$ is larger than the radius of the object $r_\mathrm{c}$, the effect of the complicated dynamics of the object's interior on the orbit is encoded in mass, spin and multipoles of the object~\cite{Goldberger:2004jt,Goldberger:2005cd,Porto:2005ac,Porto:2007qi,Porto:2016pyg,Goldberger:2022rqf}. In this regime multipoles capture the finite-size effects like tidal deformations of the compact object. Notably, the properties of black holes (BHs) are incredibly simple: the permanent (spin-induced) multipoles depend only on their mass $M$ and (dimensionless) spin $\tilde{a}=S/M^2$ and are bounded by the spin-extremality condition $\tilde{a} < 1$, while the multipoles induced by the stationary tidal field are zero for both non-spinning and spinning BHs, even when considering non-linear tidal distortions~\cite{Fang:2005qq,Damour:2009vw,Binnington:2009bb,Kol:2011vg,Chia:2020yla,Riva:2023rcm,Iteanu:2024dvx,Combaluzier-Szteinsznaider:2024sgb,Kehagias:2024rtz,Parra-Martinez:2025bcu,Rodriguez:2026iot}. On the other hand, the microphysics of astrophysical compact objects (white dwarfs and neutron stars) does not allow masses above $\simeq 2\text{--}3\,M_\odot$~\cite{Shapiro:1983du}. These two facts in combination imply strong evidence for physics beyond the Standard Model if an object heavier than $\simeq 3 M_\odot$ with non-trivial internal structure encoded in the finite-size parameters is discovered~\cite{Giudice:2016zpa,Cardoso:2017cfl,Chakraborty:2026qru}. \vskip 4pt 

One of the best candidates for such objects is a BH endowed with an overdensity of ultra-light bosons (a bosonic cloud), constituting together a \textit{gravitational atom} (GA)~\cite{Arvanitaki:2009fg,Arvanitaki:2010sy}. The existence of such particles is well motivated as axion(-like) particle(s) can, to name a few of the strongest motivations from high-energy physics and cosmology: solve the strong CP problem in quantum chromodynamics~\cite{Peccei:1977hh,Weinberg:1977ma,Wilczek:1977pj}; constitute a component or the totality of the dark matter budget in the Universe~\cite{Marsh:2015xka,Hui:2016ltb,Hui:2021tkt}; be one of the by-products of compactifications in string theory~\cite{Arvanitaki:2009fg,Demirtas:2018akl,Mehta:2021pwf,Baryakhtar:2026oun}. Scattering a low-frequency bosonic field off a rotating BH allows for its enhancement at the expense of the BH spin, via the process of superradiance (SR)~\cite{1971JETPL..14..180Z,1972JETP...35.1085Z,Endlich:2016jgc}. As the field is confined to the BH by its own mass, this leads to an avalanche process ending in the formation of a \textit{spinning} cloud with mass $M_\mathrm{c}/M \lesssim \mathcal{O}(0.1)$ and the (partial) de-spinning of the BH~\cite{Press:1972zz,Damour:1976kh,Zouros:1979iw,Arvanitaki:2009fg,East:2018glu,Brito:2015oca}. This simple picture becomes more complicated if the self-interactions~\cite{Gruzinov:2016hcq,Baryakhtar:2020gao,Witte:2024drg,Gavilan-Martin:2026zzw} (or couplings to other species~\cite{Rosa:2017ury,Boskovic:2018lkj,Fukuda:2019ewf}) are non-trivial, typically leading to less dense clouds compared to the baseline scenario, while still generating non-trivial overdensities in a large part of the motivated parameter space, compared to e.g.\ background dark matter density. In contrast, \textit{non-spinning}  clouds can be produced by dark matter environmental processes, albeit leading to less dense clouds compared to SR~\cite{Hui:2019aqm,Hui:2022sri,Budker:2023sex}.
\vskip 4pt

SR-induced large overdensities and BH de-spinning have led to many avenues for constraining ultra-light bosons and in particular to tentative constraints in the stellar-mass BH range, e.g.~\cite{Arvanitaki:2014wva,Brito:2014wla,Arvanitaki:2016qwi,Brito:2017zvb,Cardoso:2018tly,Palomba:2019vxe,Baryakhtar:2020gao, Zhu:2020tht, Ng:2020ruv,Tsukada:2020lgt,Khalaf:2024nwc,Witte:2024drg,Caputo:2025oap,Ning:2026ebu}. However, since the most precise probes of compact objects come from binary coalescences observed by present and future GW detectors (notably LISA~\cite{LISA:2022kgy} and the Einstein Telescope~\cite{Maggiore:2019uih,Abac:2025saz}), it is of paramount importance to understand the behaviour of boson clouds in a binary setting. Tidal coupling, starting from the quadrupole, leads to mixing of GA energy levels and can be resonantly enhanced~\cite{Baumann:2018vus}. Energy and angular momentum shifts of the atom must balance those of the binary, thus leading to an intricate non-linear dynamics, including such smoking-gun phenomena as floating and sinking orbits~\cite{Baumann:2019ztm,Boskovic:2024fga,Tomaselli:2024bdd,Boskovic:2025ixx} and ionization-driven dynamical friction~\cite{Baumann:2021fkf,Tomaselli:2023ysb}. Considering eccentric orbits and spin–orbit misalignment (obliquity), this dynamics can lead to fixed points in the orbital-element evolution, driving the growth of eccentricity~\cite{Boskovic:2024fga,Tomaselli:2024bdd,Boskovic:2025ixx} and off-equatorial spin–orbit misalignment~\cite{Boskovic:2025ixx}. Typically, the binary-induced mixing transfers bosons to a SR-decaying state and thus leads to a disruption of the cloud~\cite{Baumann:2018vus,Tong:2022bbl,Takahashi:2023flk,Boskovic:2024fga,Tomaselli:2025jfo,Boskovic:2025ixx}. A robust modelling of the dynamics is therefore of high phenomenological relevance, determining whether cloud disruption can occur inside the observing band of GW detectors~\cite{DellaMonica:2025zby,Kim:2025wwj,Boskovic:2025ixx}, or whether the trails of the cloud should be sought in the imprint of fixed-point structure in the binary orbital elements distribution~\cite{Boskovic:2024fga,Boskovic:2025ixx}.\vskip 4pt 

The WEFT provides a state-of-the-art framework to study the inspiral dynamics of binary coalescence, even if the constituent objects have complex structure, as in the case of GAs~\cite{Baumann:2018vus}. Recently it has been applied to study the orbital dynamics of GAs, by focusing on the permanent electric multipoles of the atom~\cite{Boskovic:2025ixx}. In this work, we take this task further by including tidally induced electric multipoles and classifying the importance of other terms (magnetic multipoles). On the other hand, in the description of the microphysics, a careful treatment of competing perturbations is necessary, including internal relativistic dynamics~\cite{Baumann:2018vus,Baumann:2019eav,Cannizzaro:2023jle}, self-gravity~\cite{Baryakhtar:2020gao,Takahashi:2021yhy,May:2024npn,Kim:2025wwj}, tidal coupling~\cite{Baumann:2018vus,Baumann:2019ztm,Brito:2023pyl,Boskovic:2025ixx} as well as self-interactions~\cite{Baryakhtar:2020gao,Witte:2024drg} more generally. Focusing on the minimal scenario where the self-interactions are negligible, we will provide a more systematic approach to (competing) perturbations in the UV description, borrowing concepts and techniques from atomic physics. This will allow us to refine, correct and generalise some of the previous statements in the literature on the impact of the microphysics on the onset and development of binary resonances. Besides resonances, due to the extended nature of GAs, the large finite-size coefficients that encode their tidal response (Love numbers) are clear indicators of the presence of a cloud around the BH~\cite{Baumann:2018vus,Baumann:2019ztm}. As we shall see, our careful treatment of the perturbations allows us both to confirm the results of~\cite{Arana:2024kaz} for Love numbers of non-rotating clouds and to calculate gravitational Love numbers for spinning GAs for the first time. \vskip 4pt 

Aside from the phenomenological motivations, the well-defined and controllable microphysics of boson clouds (or the self-gravitating limit of boson star configurations~\cite{Kaup:1968zz,Friedberg:1986tq,Liebling:2012fv,Giudice:2016zpa}) makes these systems valuable toy models in gravitational-wave physics. For example, they allow one to explore extreme values of a compact object's macroscopic parameters consistent with causality and energy conditions~\cite{Boskovic:2021nfs,Russo:2025ivk}, or to provide controlled setups for numerical-relativity tests of analytical inspiral calculations~\cite{Damour:2025oys}. Another recent interest has been the use of the ``positivity bounds'' program of EFTs~\cite{Adams:2006sv} to put prior constraints on the Love numbers from minimal consistency requirements on the UV completion~\cite{Correia:2025enx,Creminelli:2026xxx}. This is, however, nontrivial to achieve for at least two reasons. The first is the long-range nature of gravity, which affects analyticity properties and causes IR divergences of many observables, such as the time delay and scattering amplitudes (see~\cite{Chang:2025cxc,Bellazzini:2025bay} for recent progress). Secondly, EFTs realised on the worldline spontaneously break Lorentz invariance, challenging traditional amplitude-based approaches to deriving positivity bounds~\cite{Creminelli:2022onn,Creminelli:2023kze,Hui:2023pxc,Creminelli:2024lhd,Hui:2025aja,Creminelli:2025rxj}. Thus, having a well-defined toy model of the UV physics is welcome. It is with this aim that we also discuss in which scenarios GAs' Love numbers satisfy the ``positivity'' expectation. \vskip 4pt

This paper is organised as follows. In Sec.~\ref{sec:setup} we describe the WEFT and the microphysics at leading order, and in Sec.~\ref{sec:pert} we present the perturbative framework that we employ, where we also discuss the internal cloud perturbations (relativistic and self-gravity). In Sec.~\ref{sec:love} we introduce (gravitational) Love numbers at the WEFT level and calculate them for both non-spinning and spinning clouds. In Sec.~\ref{sec:pheno} we apply our results to the binary inspiral, where one of the objects is a GA, and discuss how our findings influence the phenomenology, including the orbital resonances. Finally, we conclude in Sec.~\ref{sec:conc} with a summary of our contributions. Some of the more technical details are deferred to the Appendices, referenced throughout the text: in App.~\ref{app:nr_limit} we give a consistent EFT treatment of the non-relativistic limit of \textit{real} scalars/axions in the (perturbed) Kerr background; in App.~\ref{app:bottom} we sketch the bottom-up construction of a broader class of WEFTs with occupancy dynamics and connect it to the oscillations of compact objects such as neutron stars; in App.~\ref{app:magnetic} we describe the permanent magnetic multipoles of GAs; in App.~\ref{app:spectrum} we provide beyond-hyperfine corrections to the (conservative) spectra of scalar clouds;  in App.~\ref{app:DL} we review the Dalgarno-Lewis summation technique used in the perturbative calculations; in App.~\ref{sec:dis} we review when the adiabatic condition in the treatment of states is necessary.

\vskip 4pt  \newpage

\noindent\textit{\textbf{Guide to the reader.}} One of the aims of this work is to provide a coherent and concise description of GAs (in binaries) at the level of both the worldline and the microphysics, in passing correcting, generalising and putting on a more formal footing previous results in the literature. This necessarily entails some degree of review, although we have tried to keep it to a minimum. The reader already familiar with the baseline results on the spectral shift due to internal relativistic corrections~\eqref{eq:hf_spectrum}~\cite{Baumann:2018vus} may wish to proceed, after setting the stage in Sec.~\ref{sec:setup}, directly to this work's central results: the Love numbers of (spinning) GAs (Sec.~\ref{sec:love}) and their phenomenological relevance (Sec.~\ref{sec:love_dyn}). Instead, the reader  familiar with the cloud-orbit phase space dynamics~\cite{Boskovic:2025ixx} and interested in the effect of self-gravity on resonant dynamics in the dense cloud regime can proceed to Sec.~\ref{sec:self_gravity} and Sec.~\ref{sec:pheno}.\vskip 2pt

\noindent\textit{\textbf{Notations and conventions.}} We work in geometric units $G = c  =  1$, where the scalar ``mass'' (frequency) has mass-dimension  $[\mu]=-1$. Greek indices run over spacetime coordinates and Latin indices over spatial ones. We use the composite notation $\mu_L \equiv \langle \mu_1 \ldots \mu_\ell \rangle$ for a symmetric trace-free (STF) combination of indices, denote Euclidean integrals by $\int_{\bm{x}} \equiv \int dr \, d\Omega_{\hat{r}} \, r^2$, and write the Euclidean Laplacian as $\nabla^2 \equiv \pd_r^2 + \frac{2}{r}\pd_r + \frac{1}{r^2}\left(\pd_\theta^2 + \frac{1}{\sin^{2}\theta} \, \pd_\phi^2\right)$. \vskip 2pt

Angular decompositions are performed in spherical harmonics $Y_{\ell m}(\theta_x,\phi_x)$, normalised as \\ $\int d\Omega \, Y^\ast_{\ell m} Y_{\ell' m'} =\delta_{\ell\ell'}\delta_{mm'}$, and we relate them to the STF basis $\mathcal{Y}^{i_L}_{\ell m}$ via
\begin{equation}
x^i \equiv n^{i}_x r_x \,, \qquad n^{i_L}_x = \mathsf{N}_{\ell} \sum_{m=-\ell}^{\ell} \mathcal{Y}^{i_L}_{\ell m} Y_{\ell m}(\theta,\phi) \,, \qquad \mathsf{N}_{\ell} \equiv \frac{4\pi\,\ell!}{(2\ell+1)!!} \,.
\end{equation}

\section{Setup} \label{sec:setup}

\subsection{Worldline EFT} \label{sec:weft}

The WEFT provides a description of a compact object coupled to a gravitational field varying on scales $R$ much larger than the size of the object $r_\mathrm{c}$. In the strict $r_\mathrm{c} \to 0$ limit, the object reduces to a point particle
\begin{eqnarray} \label{eq:action_pp}
S_{\rm pp} = - \int d \tau {\cal M}(\tau) + S_\mathrm{spin} \,,
\end{eqnarray}
where $\mathcal{M}$ is the GA mass, $\tau$ is the proper time, and $S_\mathrm{spin}$ describes the spin dynamics. In the case of a GA, the mass and the spin, as well as the multipole moments that we will shortly introduce, include contributions from both the BH and the bosonic cloud. In the rest of this work, we focus on the case where the body frame remains fixed (i.e.\ equatorial perturbations of the GA), so that the spin dynamics is not activated (see~\cite{Porto:2005ac,Porto:2007qi,Porto:2016pyg,Goldberger:2020fot} for the spin dynamics in the context of WEFT and~\cite{Boskovic:2025ixx} for the case of the GA). Corrections beyond the point-particle level, encoding finite-size effects (i.e.\ tidal effects), are described by $S_{\rm tidal}$, which contains higher-derivative operators attached to the worldline~\cite{Goldberger:2004jt,Goldberger:2005cd,Porto:2016pyg}
\begin{equation}
\label{tidalaction}
    S_{\rm tidal} = -\sum_{\ell \ge 2} \int d \tau \left( \mathcal{Q}_E^{\mu_L} E_{\mu_L} + \mathcal{Q}_B^{\mu_L} B_{\mu_L} \right)
\end{equation}
with $E_{\mu_L}, B_{\mu_L}$ defined in terms of covariant derivatives of the gravitational field at the worldline. In more detail, we introduce the even (electric) and odd (magnetic) components of the Weyl tensor:
\begin{equation}
    E_{\mu\nu} \equiv C_{\mu\rho\nu\sigma} u^{\rho} u^{\sigma}, \hspace{0.5cm} B_{\mu\nu} \equiv C^{*}_{\mu\gamma\nu\delta}u^{\gamma}u^{\delta} = \frac{1}{2} \epsilon_{\gamma\alpha\beta\langle\mu}C_{\nu\rangle\delta}^{\alpha\beta}u^{\gamma}u^{\delta} \,,
\end{equation}
using the four-velocity of the point particle $u^{\mu}$, the totally antisymmetric tensor $\epsilon_{\mu\nu\alpha\beta}$, and the covariant derivative orthogonal to the four-velocity $\nabla^{\perp}_{\mu} \equiv \left( g_{\mu\nu} + u_\mu u_{\nu} \right)\nabla^\nu$. Finally, the tidal field $E_{\mu_L}$ is defined as a symmetric trace-free combination of covariant derivatives:
\begin{equation}
    E_{\mu_L} \equiv \nabla^{\perp}_{\langle\mu_1}\ldots\nabla^{\perp}_{\mu_{\ell-2}}E_{\mu_{\ell-1}\mu_{\ell}\rangle}
\end{equation}
(with $B_{\mu_L}$ defined in a similar manner). Since both $E$ and $B$ are orthogonal to the four-velocity, the only non-zero components in the rest frame of the particle are purely spatial. We will thus write them in the rest frame using spatial indices (i.e.\ $E_{i_L}$). \vskip 4pt

The $\mathcal{Q}^{E/B}_{\mu_L}$ are the (electric/magnetic) multipoles of the compact object, encoding finite-size effects. In general, these are composite operators depending on the short-scale degrees of freedom confined within the object. In the case of neutron stars, they correspond to the macroscopic motion of the fluid. For a GA, there are degrees of freedom associated with the BH (capturing absorption due to the horizon, among other effects~\cite{Goldberger:2005cd,Porto:2007qi,Goldberger:2020fot}), varying on the scale of the horizon $\sim M$, and others associated with the configuration of the scalar field, having support on scales $\sim r_\mathrm{c}$. \vskip 4pt

Knowing the dynamics of these short-scale degrees of freedom coupled to gravity, one could in principle find the full evolution of the multipoles. If the characteristic frequency of the tidal field is much smaller than the one corresponding to the microscopic degrees of freedom, the multipoles are given by linear response theory~\cite{Rodriguez:2026iot}:
\begin{equation}
\label{eq:linresp}
    \mathcal{Q}^{E}_{\mu_L} = (\mathcal{Q}^{\rm perm})^{E}_{\mu_L}  -2\sum_{\ell'}\lambda_{\mu_L \nu_{L'}}E^{\nu_{L'}} + \mathcal{O}(E^2,\pd_t E)
\end{equation}
where $(\mathcal{Q}^{\rm perm})^{E}_{\mu_L}$ is the permanent multipole (i.e.\ the one present also for an object in isolation) and the second term captures the multipoles induced by the external tidal field. Notice that for an object that is spherically symmetric in isolation, $(\mathcal{Q}^{\rm perm})^{E}_{\mu_L}$ has to vanish. \vskip 4pt

In most phenomenologically interesting situations, the tidal field varies faster than the characteristic frequency of the cloud evolution~\footnote{This characteristic frequency can be as low as the hyperfine splitting in the energies of the states of the cloud (see Sec.~\ref{subsecmicro} for details).}. Thus, to have a description that is local in time, we introduce the relevant degrees of freedom associated with the ``fast'' dynamics explicitly in the worldline EFT. In the following, by integrating out the spatial profile of the scalar field, we will derive the action $S_{\rm micro}$, containing their kinetic term as well as their contribution to the permanent multipoles $(\mathcal{Q}^{\rm perm})^{E}_{\mu_L}$, and, in Sec.~\ref{sec:love}, to the induced multipoles determined by the operator $\lambda_{\mu_L \nu_{L'}}$. \vskip 4pt

Putting the above together, our WEFT action for the GA is
\begin{equation} \label{eq:action}
    S_{\rm WEFT} = S_{\rm EH} + S_{\rm pp} + S_{\rm tidal} + S_{\rm micro} \,, 
\end{equation}
where $S_{\rm EH}$ is the Einstein-Hilbert action of GR for the bulk. \vskip 4pt

\subsection{Microphysics \& matching}
\label{subsecmicro}

To set the stage, let us briefly review, at the order-of-magnitude level, the basic scaling relations of the scalar\footnote{Henceforth, we will focus on spin-$0$ particles, although the extension to the spin-$1$ case~\cite{Baryakhtar:2017ngi,East:2017mrj,Baumann:2019eav,Baumann:2019ztm} should be straightforward. Furthermore, the parity of the spin-$0$ field is not relevant for our discussion; thus, we switch between the scalar and axion labels throughout.} cloud that we shall use in the following for power counting~\cite{Arvanitaki:2009fg,Arvanitaki:2010sy}. The cloud consists of a set of virialised scalars with velocities $v^2_\mathrm{c} \sim M/r_\mathrm{c}$, where $M$ is the BH mass. Its size is set by the de Broglie wavelength of the scalar, giving $r_\mathrm{c} \sim M/\alpha^2$, where $\alpha \equiv G \mu M/(\hbar c) \sim v_\mathrm{c}$ is the \textit{structure constant}. On the other hand, the amount of scalar overdensity, which depends on the formation mechanism, controls the backreaction on the geometry, $q_\mathrm{c} \equiv M_\mathrm{c}/M$, where $M_\mathrm{c}$ is the integrated (Komar) mass of the scalar field. In this way, $\{\alpha,q_\mathrm{c}\}$ measure the relativistic corrections in the microphysics of the GA.\vskip 4pt 

The action of a real and non-interacting scalar of mass $\mu$ minimally coupled to gravity is
\begin{eqnarray}
    S_\Psi =  \int d^4x \left( -\frac{1}{2} g^{\alpha\beta}\pd_\alpha\Psi\pd_\beta \Psi -\frac{1}{2}\mu^2\Psi^2 \right) \,.
\end{eqnarray}
To study the non-relativistic limit, one typically substitutes the real scalar field $\Psi$ with a complex one $\psi$ as
\begin{equation} \label{eq:naivesub}
    \Psi \rightarrow \frac{1}{\sqrt{2\mu}}\left(\psi e^{-i\mu t} + \bar{\psi} e^{i\mu t}\right)
\end{equation}
where $\psi$ is assumed to vary more slowly than the extracted oscillatory factor. At leading order in derivatives, the substitution above gives the Schr\"odinger equation, provided one drops the terms containing rapidly oscillating factors $e^{\pm i \mu t}$, which average to zero in the leading approximation (e.g.~\cite{Baumann:2018vus}). This procedure, however, artificially increases the number of degrees of freedom of the theory and is not a controlled way to perform the non-relativistic reduction. Following~\cite{Salehian:2020bon}, we develop in App.~\ref{app:nr_limit} a systematic way to integrate out the high-frequency oscillations and obtain an EFT for the \textit{slow scalar} $\psi_\mathrm{s}$ with an approximate $U(1)$ symmetry, valid for energy scales below $\mu$, leading to
\begin{eqnarray}\label{eq:S_micro}
S_{\mathrm{micro}} &=& \int dt \int_{\bm{x}} \left[ \bar\psi_\mathrm{s} \left( i\partial_t - \mathcal{H}_\mathcal{B}- \mathcal{H}_\mathcal{F} - \mathcal{H}_\mathcal{H} \right) \psi_\mathrm{s} \right] + \mathcal{O}(\mu \alpha^6) \,,    \\
\mathcal{H}_\mathcal{B} &=& -\frac{\nabla^2}{2\mu} - \frac{\alpha}{r} \,, \quad \mathcal{H}_\mathcal{F} = -\frac{5\alpha^2}{2\mu r^2}-\frac{2i\alpha}{\mu r}\pd_t - \frac{\alpha}{2\mu^2 r}(\nabla^2 - 2\pd_r^2 -3 \frac{\pd_r}{r}) - \frac{1}{8\mu^3}\nabla^4 \,, \quad \mathcal{H}_\mathcal{H} =\frac{2 i a  M}{r^3}\pd_\phi \nonumber
\end{eqnarray}
in Boyer-Lindquist coordinates, with $\nabla^2$ being the Laplacian in three-dimensional Euclidean space. Here the $\mathcal{H}_i$ correspond to $i=\{\mathcal{B},\mathcal{F},\mathcal{H}\}$, i.e.\ the Bohr (non-relativistic limit), fine and hyperfine contributions (the last two being the leading order relativistic corrections, the hyperfine one due to the presence of the spin $\tilde{a}$ of the host BH). The spectrum that follows from~\eqref{eq:S_micro} is the same as in~\cite{Baumann:2018vus}, that is formally valid for complex-scalar clouds. The reason is the following: if the scalar theory is \textit{linear} in a background space-time, then, by the superposition principle, real and complex scalars (the latter having an exact $U(1)$ symmetry and thus particle conservation) share the same spectrum. This identification is, however, broken by the scalar-graviton interactions at $\mathcal{O}(\mu q_\mathrm{c} \alpha^4)$.\vskip 4pt 

Let us now expand $\psi_\mathrm{s}(\bm{x})$ in the basis of hydrogenic spatial wavefunctions
\begin{equation}
    \psi_\mathrm{s} = \sum_{N} C_{N}(t) \psi_{N}(\bm{x}) + \sum_{K} D_{K}(t) \psi_{K}(\bm{x})
\label{eq:ampdecomp} \,,
\end{equation}
where $\psi_{N}(\bm{x}) = \braket{\bm{x}| n \ell m} \equiv  \braket{\bm{x}|N} $ and $ \psi_{K}(\bm{x}) = \braket{\bm{x}|k; \ell m} \equiv \braket{\bm{x}|K}$ label the bound and the continuum states, respectively. As we focus on external field frequencies lower than the Bohr scale, the excitation of continuum states will not be efficient, and we set from now on $D_K(t) \to 0$~\cite{Baumann:2021fkf}. This allows us to integrate out the spatial profile of the cloud and to identify the microphysical degrees of freedom with the occupancy dynamics $C_N(t)$, where, at the Bohr level,
\begin{eqnarray} \label{eq:nreft_h0}
S_{\mathrm{micro}}  &=& \int dt  \sum_{N}\left\{ i \bar C_N \dot C_N  - \sum_{M}\bar C_M C_N \langle \mathcal{H} \rangle_{MN} \right\}  \,, \quad \langle \mathcal{H} \rangle_{MN} = \epsilon^\mathcal{B}_N \delta_{MN} \,, \quad       \epsilon^\mathcal{B}_N = -\mu\frac{\alpha^2}{2 n^2} \,,
\end{eqnarray}
implying $C_N \propto e^{-i \epsilon^\mathcal{B}_N t}$. This is analogous to the way the microphysics is described in tidal perturbations of neutron stars in binaries, where the internal fluid modes are modelled as a collection of oscillators~\cite{Flanagan:2007ix,Chakrabarti:2013xza,Martinez-Rodriguez:2026omk,HegadeKR:2026kku}. We develop this parallel further in App.~\ref{app:bottom}, where we outline general rules for constructing WEFTs that include the occupancy dynamics, with (and without) an approximate $U(1)$ symmetry. \vskip 4pt

In the cloud sector, we introduce action-angle variables, relying on the $U(1)$ symmetry, with a reference state $a$ (e.g.\ the one grown by the SR instability)
\begin{eqnarray} \label{eq:aa_var}
 \quad C_N \equiv \sqrt{N_\mathrm{c} \, I_N} e^{-i \phi_N }   \,, \quad \sqrt{I_a} \equiv  D = \sqrt{1-\sum_{J \neq a} I_J} \,, \quad \phi_a = 0 \,,
\end{eqnarray}
where $N_\mathrm{c}$ is the cloud occupancy. Ignoring the corrections to the Bohr spectrum $\epsilon^\mathcal{B}_N$, only the phases have a flow, $ \dot{\phi}_N = \Delta \epsilon_{aN}^\mathcal{B} $, where $\Delta \epsilon^\mathcal{B}_{aN} \equiv \epsilon^\mathcal{B}_N -  \epsilon^\mathcal{B}_a$. \vskip 4pt

\subsubsection{Permanent multipoles} \label{sec:spin_mult}

We can now proceed to calculate explicitly the permanent multipoles of the GA. Due to the large extent of the cloud, we treat it as a perturbation of the Kerr BH. The BH limit can be taken formally by setting $C_N =0$. The leading cloud contribution to the mass, the spin and the multipoles is $\mathcal{O}(q_\mathrm{c})$ for a free scalar field, while the backreaction produces contributions that are $\mathcal{O}(q_\mathrm{c}^2)$ and higher, even without intrinsic non-linearities of the scalar field. Thus, for $\mathcal{X} = \{\mathcal{M},(\mathcal{Q}^\mathrm{perm})^{E}_{\ell m}, (\mathcal{Q}^\mathrm{perm})^{B}_{\ell m},\dots \}$ we have $\mathcal{X} \approx X+X_\mathrm{c}$, where $X$ and $X_{\mathrm c}$ denote the BH and cloud contributions to $\mathcal{X}$, respectively (see~\cite{Herdeiro:2015gia} for the calculation of multipoles in the numerically constructed GA spacetime). 

Focusing first on the cloud sector, the leading-order contributions to the multipoles for weakly-gravitating systems are (e.g.~\cite{Ross:2012fc}):
\begin{eqnarray}
   (Q^\mathrm{perm}_\mathrm{c})_E^{i_L} &=& \frac{1}{\ell!} \int_{\bm{x}}  T_\mathrm{c}^{00} \, x^{i_L} + \mathcal{O} \left(\mu \alpha^2 \right) \,,  \label{eq:I_L} \\ 
      (Q^\mathrm{perm}_\mathrm{c})_B^{i_L} &=& - \frac{2 \ell}{(\ell+1)!} \int_{\bm{x}} \epsilon^{i_{\ell}ab}\, T_\mathrm{c}^{0a}\, x^{b\, i_{L-1}}  + \mathcal{O} \left(\mu  \alpha^3  \right)  \,,
  \label{eq:J_L}
\end{eqnarray}
while for the cloud energy-momentum tensor we use the non-relativistic expansion (see App.~\ref{app:nr_limit}):
\begin{eqnarray}
    T_\mathrm{c}^{00} &=& \mu\Bar\psi_{\rm s} \psi_{\rm s} + \mathcal{O}( \mu \alpha^2) \,, \label{eq:T00} \\
    T_\mathrm{c}^{0a} &=& -\frac{i}{2}(\Bar\psi_{\rm s} \pd^a \psi_{\rm s} - \psi_{\rm s} \pd^a \Bar\psi_{\rm s}) + \mathcal{O}(\mu \alpha^3) \,, \quad
    \label{eq:T0b}
\end{eqnarray} 
Projecting the electric multipoles onto the spherical-harmonic basis, and using the variables~\eqref{eq:aa_var},
\begin{eqnarray} \label{eq:E_mult_moment}
 (Q^\mathrm{perm}_\mathrm{c})^E_{\ell m} &\approx& \frac{1}{\ell!}\mu N_\mathrm{c} r_\mathrm{c}^\ell \sum_{MN} \sqrt{I_N I_M} e^{-i  \Delta \phi_{MN} } \mathcal{I}^{(MN|\ell m)}_r \mathcal{I}^{(MN|\ell m)}_{\Omega,E} \,, \\
 \quad \mathcal{I}^{(MN|\ell m)}_r &\equiv& \int^{\mathsf{R}\to \infty} d\mathsf{r} \mathsf{r}^2 \hat{\mathcal{R}}_M \hat{\mathcal{R}}_N \mathsf{r}^{\ell}\,,\quad \mathcal{I}^{(MN|\ell m)}_{\Omega,E} \equiv \int d\Omega_{\hat{r}} Y^\ast_M  Y_{\ell m}  Y_{N} \,,
\end{eqnarray}
where $\mathsf{r} \equiv r/r_\mathrm{c}$, $\mathsf{R} \equiv R/r_\mathrm{c}$ and
$\hat{\mathcal{R}}_N = r_\mathrm{c}^{3/2} \mathcal{R}_N$, with $\mathcal{R}_N$ the hydrogenic radial wavefunction (see~\cite{Boskovic:2025ixx} for a more detailed discussion in the two-state case). Note that the instantaneous mass of the cloud corresponds to $\ell=0$, i.e.\ $M_\mathrm{c} \approx \mu N_\mathrm{c}$, as the $N \neq M$ terms in the sum average out. The angular integral implies the selection rules:
\begin{eqnarray}
&& m = -\Delta m_{MN} \equiv - (m_N - m_M) \,, \nonumber \\
&&  \ell+\ell_M+\ell_N \in 2\mathbb{Z} \,, \nonumber \\
&& |\ell_M-\ell_N| \leq \ell \leq \ell_M+\ell_N \,. \label{eq:selection_electric}
\end{eqnarray}
Consider first the diagonal terms $M=N$, and thus $m=0$. From the selection rules it follows that, although the wavefunctions $\ket{n \ell m}$ are pure multipoles, any spinning cloud (density) $\ell_N > 0$ will generate a quadrupole moment, even when $\ell_N \neq 2$. The increasing spatial extent of the cloud as $\alpha \ll 1$ allows the cloud contribution to dominate the BH one, $(Q^\mathrm{perm})^E_{\ell m} \simeq M^{\ell + 1} \tilde{a}^\ell \delta_{\ell, 2\mathbb{Z}}$, for non-trivial overdensities $q_\mathrm{c}$, i.e.\ $(\mathcal{Q}^\mathrm{perm})^E_{\ell m} \approx (Q^\mathrm{perm}_\mathrm{c})^E_{\ell m}$~\cite{Baumann:2018vus,Baumann:2019ztm,Boskovic:2025ixx}. The cloud also supports permanent magnetic multipoles for $\ell_N \geq 2$ (which again dominate the BH contribution for non-trivial overdensities), as we discuss in App.~\ref{app:magnetic}. \vskip 4pt

\section{Cloud Perturbations}  \label{sec:pert}

We now focus on the static perturbations of the cloud. The unperturbed Bohr spectrum exhibits an accidental $SO(4)$ symmetry (e.g.~\cite{Weinberg_2015}), with the energies $\epsilon^\mathcal{B}_N$ being the same for a given $n_N$ regardless of the $(\ell,m)_N$ quantum numbers. If the perturbation $V$ induces a non-zero overlap $\braket{P|V|N}_\mathcal{B}$ only between different Bohr levels, $n_P \neq n_N$, ordinary (Rayleigh-Schr\"odinger) perturbation theory (PT) can be used (the subscript in $ \langle \dots \rangle_\mathcal{B}$ indicates that the product is computed using the unperturbed wavefunctions). In general, however, degenerate perturbation theory (DPT) is necessary. There, after decomposing the Hilbert space as $\mathcal{S} = \mathcal{S}_\mathrm{d} \bigoplus \mathcal{S}_\perp$, with $\mathcal{S}_\mathrm{d}$ the degenerate subspace and $\mathcal{S}_\perp$ the orthogonal complement ($\epsilon^\mathcal{B}_P \neq \epsilon^\mathcal{B}_N$), one transforms to a ``good basis'' $\ket{N} \to \ket{\chi_N}$ that diagonalises the total perturbation. If the perturbation lifts all the degeneracies at first order, the corrections to the states (to first order) and to the energies (to second order) are given\footnote{The second term describes a virtual transition $\mathcal{S}_\mathrm{d} \to \mathcal{S}_\perp \to \mathcal{S}_\mathrm{d}$, which is second order in $\xi$. However, the initial $\ket{\chi_N}$ and final $\ket{\chi_J}$ states within $\mathcal{S}_\mathrm{d}$ are split only by the first-order lift, giving the term an overall power counting of $\xi^1$. Note that this term is often erroneously omitted in QM textbooks; contrast, e.g., the first and second editions of~\cite{Weinberg_2015}.} by~\cite{zwieb}
\begin{eqnarray}
    \ket{\chi_N} &=& \ket{\chi_N^{\mathcal{B}}}
    - \xi  \bigg(
        \sum_{P \in \mathcal{S}_\perp}\,
        \frac{\langle V \rangle_{PN} }{\epsilon_P^{\mathcal{B}} - \epsilon_N^{\mathcal{B}}} \ket{P^{\mathcal{B}}}
        +
        \sum_{J \in \mathcal{S}_\mathrm{d}/\{N\}}\,
        \frac{\ket{J^{\mathcal{B}}}}{\epsilon_N^{(1)} - \epsilon_J^{(1)}}
        \;\sum_{P \in \mathcal{S}_\perp}\,
        \frac{\langle V \rangle_{PN} \,\langle V \rangle_{JP} }{\epsilon_P^{\mathcal{B}} - \epsilon_N^{\mathcal{B}}}
    \bigg) + \mathcal{O}(\xi ^2) \,, \nonumber
     \\
    \epsilon_N &=& \epsilon_N^{\mathcal{B}} + \xi \,  \epsilon^{(1)}_N
    - \xi ^2 \;\sum_{P \in \mathcal{S}_\perp}\,
    \frac{|\langle V \rangle_{PN} |^2}{\epsilon_P^{\mathcal{B}} - \epsilon_N^{\mathcal{B}}}
    + \mathcal{O}(\xi ^3)\,, \quad \epsilon^{(1)}_N \equiv    \langle V \rangle_{NN} \,, \quad \langle V \rangle_{PN} \equiv  \braket{P|V|\chi_N}_\mathcal{B} \label{eq:energy_correction}
\end{eqnarray}
where $\xi  \ll 1$ is a bookkeeping parameter.  \vskip 4pt

In practice, perturbations are not endowed with a manifest small parameter $\xi$, and one must instead assess the convergence of the expansion from the properties of the operator $V$. When the Hilbert space $\mathcal{S}$ is infinite-dimensional, the series is generically only asymptotic and a fully rigorous convergence analysis is subtle~\cite{Bender:2017fyz}. For our purposes, it suffices to control the leading corrections, for which the exactly solvable two-level problem, applied pairwise across $\mathcal{S}$, provides a useful diagnostic of perturbativity~\cite{zwieb,Bender:2017fyz}:
\begin{eqnarray} \label{eq:pert_crit}
\xi \sim \max_{P \in \mathcal{S}_\perp \,, \, N \in \mathcal{S}_\mathrm{d}} 
\frac{\sqrt{\left(\langle V\rangle_{NN} - \langle V\rangle_{PP}\right)^2 
+ 4\,\lvert\langle V\rangle_{NP}\rvert^2}}
{\lvert\epsilon^{\mathcal{B}}_N - \epsilon^{\mathcal{B}}_P \rvert} < 1 \,.
\end{eqnarray}
\vskip 4pt

We will classify the perturbations as \textit{internal}, $V_\mathrm{int}$, and \textit{external}, $V_\mathrm{ext}$. The former are the relativistic effects $V_\mathrm{R}$, proportional to powers of $\alpha$, as well as self-gravity $V_\mathrm{sg} \propto q_\mathrm{c}$, which formally never decouples (self-interactions are also of this type), while the latter originate from the presence of an external field (e.g.\ tidal perturbations from a binary companion). We will see explicitly in the following that $V_\mathrm{int}$ breaks the $SO(4)$ symmetry at first order. In analogy with the Zeeman and Stark effects in atomic physics, we distinguish between \textit{weak-field} and \textit{strong-field} regimes by comparing the external perturbation $V_\mathrm{ext}$ with the relevant internal one $V_\mathrm{int}$. In the weak-field regime, $|V_\mathrm{int}| \gg |V_\mathrm{ext}|$, the internal perturbation sets the good basis and the external field is treated as a perturbation on top of it; in the strong-field regime, $|V_\mathrm{ext}| \gg |V_\mathrm{int}|$, the roles are reversed. \vskip 4pt

Finally, let us note that in the following we will be particularly interested in the perturbative calculation of the energy splitting between different states at first order, $\Delta \epsilon^{(1)}_{NM}$. Firstly, these feature in the wavefunction correction in both the weak-field and the strong-field regimes [cf.~\eqref{eq:energy_correction}]. Secondly, as we discuss in Sec.~\ref{sec:pheno}, if $V_\mathrm{ext}$ induces a level transition $\ket{N} \to \ket{M}$, this mixing is resonantly enhanced when $\Omega_\mathrm{ext} \simeq \Delta \epsilon_{NM}/\Delta m_{NM}$. We can classify these transitions as being of the hyperfine ($\mathcal{H}$; $\Delta n_{NM} = \Delta \ell_{NM} = 0$, $\Delta m_{NM} \neq 0$), fine ($\mathcal{F}$; $\Delta n_{NM} = 0$, $\Delta \ell_{NM} \neq 0$), or Bohr ($\mathcal{B}$; $\Delta n_{NM} \neq 0$) type~\cite{Baumann:2018vus}. 

\subsection{Relativistic corrections}
\label{sec:relativistic}

The (hyper)fine corrections~\eqref{eq:S_micro} are diagonal in the standard basis $\ket{n \ell m}$, and thus ordinary PT can be used to find the first-order corrections to the spectrum~\cite{Baumann:2018vus,Baumann:2019eav}:
\begin{eqnarray} \label{eq:hf_spectrum}
 \epsilon^{\mathcal{F}}_N = \mu \left(- \frac{\alpha^4}{8n^4} + \frac{(2\ell-3n+1)\alpha^4}{n^4(\ell+1/2)} \right) \,, \quad  \epsilon^{\mathcal{H}}_N = \mu\frac{2 \tilde a m \alpha^5}{n^3\ell(\ell+1/2)(\ell+1)}  \,. 
\end{eqnarray}
In the context of the weak- vs.\ strong-field dichotomy, the relevant internal scale is set by the smallest splitting among the levels connected by non-zero matrix elements $\braket{P|V_\mathrm{ext}|N}_\mathcal{B}$. If $V_\mathrm{ext}$ mixes states within the same $n$-shell, this is the fine splitting, $V_\mathrm{R}=\mathcal{H}_\mathcal{F}$; if it mixes states within the same $\ell_N$-subshell, it is the hyperfine splitting, $V_\mathrm{R}=\mathcal{H}_\mathcal{H}$. \vskip 4pt

In order to estimate the regime of perturbative control, let us emphasise that, besides $\{\alpha,\tilde{a}\}$, the geometry of the state also controls the strength of the perturbation. Physically, as $n$ increases at fixed $\alpha$, the density of the state peaks at $r_N \sim n^2 r_\mathrm{c}$, further away from the BH, and thus the (hyper)fine corrections are progressively suppressed. It is straightforward to check from~\eqref{eq:pert_crit} that the perturbativity conditions for the fine and the hyperfine corrections are, respectively, given by
\begin{eqnarray} \label{eq:pert_crit_HF}
\alpha < \sqrt{n} \quad \left(\mathcal{F}\right) \,, \quad \quad \alpha < \frac{n^{2/3}}{\tilde{a}} \quad   \left(\mathcal{H}\right) \,,
\end{eqnarray}
where we have focused on the ``circular'' states $\ket{n \, (n-1) \, (n-1)}$, Rydberg limit $n \gg 1$ and ignored $\mathcal{O}(1)$ factors. We verify this trend explicitly by comparing the perturbative corrections with the full relativistic solution of the spectrum in App.~\ref{app:spectrum}. \vskip 4pt

Finally, as mentioned above, the ratio $\Delta \epsilon_{NM}/\Delta m_{NM}$ is of particular relevance when the atom is perturbed by an external field. However, at the order in~\eqref{eq:hf_spectrum}, the $m$-dependence in the resonance conditions cancels, and all transitions of the $\mathcal{H}$ type share the same value of this ratio. To break this degeneracy, we need to include corrections to the relativistic spectrum beyond $\mathcal{O}(\mu \tilde{a}\alpha^5)$, in particular those that are quadratic (or higher) in the magnetic quantum number of the state, $m^2$. As we show in App.~\ref{app:spectrum}, the leading-order contributions that break this degeneracy are of order $\mathcal{O}(\mu \tilde{a}^2 \alpha^6)$, leading to
\begin{eqnarray} \label{eq:hf_res_deg}
\frac{\Delta \epsilon_{NM}/\Delta m_{NM}}
     {\Delta \epsilon_{NP}/\Delta m_{NP}}
=
1
-\frac{3}{2}
\frac{m_M-m_P}{4\ell_N(\ell_N+1)-3}
\,\tilde a\alpha
+\mathcal{O}(\tilde a\alpha^2)\,.
\end{eqnarray}
\vskip 4pt

\subsection{Self-gravity} \label{sec:self_gravity}

Including self-gravity in the description of the bosonic clouds~\eqref{eq:S_micro} implies perturbing the background Kerr metric with the metric field $\mathfrak{h}_{\mu\nu}$ sourced by the bosonic field
\begin{eqnarray} \label{eq:self_grav_action}
S_\mathrm{micro} \supset S_\mathrm{sg} \equiv  S_{\mathrm{EH}+\mathrm{gf}}[\mathfrak{h}_{\mu \nu }] +  \int dt \int_{\bm{x}} \, \left[ \frac{1}{2} T^{\mathrm{c}}_{\mu \nu} \mathfrak{h}^{\mu \nu} \right] + \mathcal{O}(\mathfrak{h}^2) \,,
\end{eqnarray}
where the first term corresponds to the Einstein-Hilbert action and the gauge fixing term in linearized gravity. Performing a Post-Newtonian expansion in the cloud sector, to linear order in $q_\mathrm{c}$ and leading order in $\alpha$, the Einstein equations in de Donder gauge reduce to (e.g.~\cite{Weinberg:1972kfs}):
\begin{eqnarray}
\pd_k\pd^k \mathfrak{h}_{00} &=& - 8 \pi  (T^{00}_\mathrm{c})  \,,  \label{eq:self-gr_poisson}\\
\pd_k\pd^k \mathfrak{h}_{0i} &=& 16 \pi  (T^{0i}_\mathrm{c}) \,, \\
\pd_k\pd^k \mathfrak{h}_{ij} &=& -8 \pi  \delta_{ij} T^{00}_\mathrm{c} \,,
\end{eqnarray}
where $T^{\mu \nu}_\mathrm{c}$ is given by~\eqref{eq:T00}--\eqref{eq:T0b} (see also App.~\ref{app:nr_limit}). These can be easily inverted, allowing us to integrate out the self-gravity field $\mathfrak{h}_{\mu \nu}$. Focusing on the dominant Newtonian potential $\mathfrak{h}_{00}$, one finds that~\eqref{eq:self_grav_action} reduces to~\footnote{Note the following: (i) the on-shell action is one half of the coupling term; this is a standard result for the quadratic action, which can be checked explicitly by working in the Newtonian limit where $S_\mathrm{sg} = \int dt \int_{\bm{x}} \left[ \tfrac{1}{8\pi} \Phi_N \Delta \Phi_N - \Phi_N \mu \bar \psi \psi \right]$, and $\mathfrak{h}_{00} = - 2\Phi_N + \mathcal{O}(\Phi_N^2)$; (ii) the overlap integrals in~\cite{Kim:2025wwj} contain additional monopole and dipole terms. To the order we are working, $\mathcal{O}(\mu q_\mathrm{c}\alpha^2)$, and focusing on the self-energy shifts within the Bohr multiplet, these additional terms  do not produce any physical effect; this is further discussed in a separate study~\cite{Boskovic:2026boh}.}:
\begin{eqnarray}
 S_\mathrm{sg} = \frac{1}{2} \int dt \int_{\bm{x}} \left(- \bar \psi V^{\mathrm{n}}_\mathrm{sg} \psi  \right)  + \mathcal{O} \left(\mu q_\mathrm{c} \alpha^4 \right) \,, \quad
V^{\mathrm{n}}_\mathrm{sg} & \equiv&  - \mu^2 \int_{\bm{y}} \frac{\bar \psi \psi }{|\bm{x} - \bm{y}|} \label{eq:Vsg_newt} \,.
\end{eqnarray}
\vskip 4pt

We will denote by the \textit{dilute cloud} limit the regime where the relativistic corrections dominate self-gravity, and by the \textit{dense cloud} limit the opposite one. At the order-of-magnitude level, we can estimate the transition between these regimes by comparing~\eqref{eq:hf_spectrum} with~\eqref{eq:Vsg_newt}:
\begin{eqnarray} \label{eq:Vsg_estimate}
V^{\mathrm{n}}_\mathrm{sg} \gtrsim \left\{\mathcal{H}_\mathcal{F},\mathcal{H}_\mathcal{H} \right\}
\implies 
q_\mathrm{c} \gtrsim \left\{\alpha^{2}, \alpha^{4} \left(\frac{\tilde{a}}{\alpha}\right)\right\} \,.
\end{eqnarray}
In order to proceed with calculating the self-energy and the mixing corrections, we introduce the overlap integrals in the $\ket{n \ell m}$ basis
\begin{eqnarray}
    G_{MNBA} &\equiv& - \mu^2 \int_{\bm{x} \bm{y}} \frac{\bar \psi_M(\bm{x})\psi_N(\bm{x}) \psi_A(\bm{y}) \bar\psi_{B}(\bm{y})}{|\bm{x} - \bm{y}|} \label{eq:sg_newt} \,,
\end{eqnarray}
where the labels $AB$ correspond to the pairs of states that source the gravitational field and note that $N_\mathrm{c} G_{MNBA} \sim \mu q_\mathrm{c} \alpha^2 $. In the following we shall specialise to the case where a single state dominates the mass, $ \ket{a}$ (i.e. $I_N \ll 1$ for $N \neq a$), as is the case for SR-generated clouds with $\alpha \ll 1$~\cite{Arvanitaki:2010sy,Arvanitaki:2014wva}. Let us first, for simplicity, consider a two-state system, where the second state is weakly excited, and integrate out the radial structure, expressing $S_\mathrm{sg}$ in the phase space variables~\eqref{eq:aa_var}
\begin{eqnarray} \label{eq:direc_exch}
S_\mathrm{sg} &=& - \frac{N_\mathrm{c}^2}{2} \int dt \, \Big{\{} 
  G_{aaaa} I_a^2 + G_{NNNN} I_N^2 
  + 2 G_{NNaa} \, I_a I_N  \\
&& \quad 
  + \, 2 G_{NaaN} \, I_a I_N \nonumber \\
&& \quad 
  + \, 4 I_a^{3/2} I_N^{1/2} \, \mathrm{Re}\left[ e^{i \phi_N} G_{aaNa} \right]
  + 4 I_N^{3/2} I_a^{1/2} \, \mathrm{Re}\left[ e^{i \phi_N} G_{NNNa} \right] 
  + 2 I_a I_N \, \mathrm{Re}\left[ e^{2 i \phi_N} G_{NaNa} \right] \Big{\}} \,. \nonumber
\end{eqnarray}
We have three types of terms, the direct, exchange and mixing terms respectively, of which only the first has been considered in the previous literature~\cite{Baryakhtar:2020gao,Takahashi:2021yhy,May:2024npn,Kim:2025wwj}. Focusing just on the non-harmonic terms for the moment, the evolution of the phase $\phi_N$ that corrects the energy split follows
\begin{equation} \label{eq:phase_evo_sg}
\dot{\phi}_N = N_\mathrm{c} \left[ -  G_{aaaa} I_a +  \left( G_{NNaa} + G_{NaaN} \right) I_a +  \mathcal{O}(I_N)
\right]  \,.
\end{equation}
To arrive at the same result from standard perturbation theory, one needs to realise that different perturbing terms in the Schr\"odinger equation operate for the states $\sqrt{I_a} \psi_a$ and $\sqrt{I_N} \psi_N$. In the former case we just have a standard potential sourced by the cloud $a$ itself, while in the latter, in addition to $V_\mathrm{sg}[I_a] \, \sqrt{I_N} \psi_N$, one needs to incorporate also the perturbing term of the same size $V_\mathrm{sg}[\sqrt{I_a I_N}] \sqrt{I_a} \psi_a$. Setting $I_N \to 0$ from the start in the Poisson equation thus leads to an inconsistent enumeration of the self-gravity contributions when the occupancy is perturbed away from the dominant state $\ket{a}$. Inserting the full perturbing potential instead in the first-order energy shift~\eqref{eq:energy_correction} generates the correct result~\eqref{eq:phase_evo_sg}.

Using the multipolar expansion
\begin{equation} \label{eq:multipolar}
    \frac{1}{|\bm{x} - \bm{y}|} = \sum_{\ell m}\frac{4\pi}{2\ell +1} Y_{\ell m} (\bm{\hat{x}})Y^\ast_{\ell m}(\bm{\hat{y}}) \frac{r_{<}^\ell}{r_{>}^{\ell+1}} \,,
\end{equation}
with $r_{\{<,>\}}=\{\mathrm{min}(r_x,r_y),\mathrm{max}(r_x,r_y)\}$, we can read off the selection rules for the relevant kernels. For the direct term $G_{MNaa}$, from the $\bm{y}$ integral one finds $(\ell,m)=(2\mathbb{Z},0)$, which further implies, via the $\bm{x}$ integral, $m_M=m_N$ and $\ell_M+\ell_N \in 2\mathbb{Z}$, together with the triangular rule $|\ell_M-\ell_N| \leq \ell \leq \ell_M+\ell_N$. For the exchange kernel $G_{MaaN}$ $(N \neq a, M \neq a)$ we have $m_M = m_N$, $m = m_M - m_a$ and the standard parity and triangular rules for both sets $(\ell, \ell_M, \ell_a)$ and $(\ell, \ell_N, \ell_a)$.
Consider first the monopolar terms [see around~\eqref{eq:E_mult_moment} for the rescaled variables]
\begin{eqnarray} \label{eq:sg_mono}
G^{(\ell=0)}_{MNaa} &=& -  \mu^3 \alpha \left[\delta_{\ell_M \ell_N}  \delta_{m_M m_N} \int d \mathsf{r}_x  d \mathsf{r}_y  
\frac{\mathsf{r}_x^2 \mathsf{r}_y^2}{\mathsf{r}_{>}} \hat{\mathcal{R}}_{M} (\mathsf{r}_x )   \hat{\mathcal{R}}_{N}  (\mathsf{r}_x) 
 \hat{\mathcal{R}}^2_{a}   (\mathsf{r}_y )  \right] \,, \\
G^{(\ell=0)}_{MaaN} &=& -  \mu^3 \alpha  \left[\delta_{\ell_a \ell_N} \delta_{\ell_M \ell_N}  \delta_{m_a m_N} \delta_{m_M m_N} \int d \mathsf{r}_x  d \mathsf{r}_y  
\frac{\mathsf{r}_x^2 \mathsf{r}_y^2}{\mathsf{r}_{>}} \hat{\mathcal{R}}_{M}  (\mathsf{r}_x )
\hat{\mathcal{R}}_{a} (\mathsf{r}_x )     \hat{\mathcal{R}}_{a} (\mathsf{r}_y )   \hat{\mathcal{R}}_{N}  (\mathsf{r}_y ) \right] \,. \nonumber
\end{eqnarray}
For the diagonal splits in the (hyper)fine multiplet of the dominant state $n_M=n_N = n_a$, the direct term breaks the $\ell_N$-degeneracy of the Bohr levels  
$\Delta \epsilon^{\mathrm{sg}, (\ell =0)}_{aN} = G^{(\ell=0)}_{NNaa}  - G^{(\ell=0)}_{aaaa} \geq 0$, as the radial self-overlap $\hat{\mathcal{R}}_{a}^4$ dominates the term proportional to $\hat{\mathcal{R}}_{a}^2 \hat{\mathcal{R}}_{N}^2$. In contrast, the exchange kernel contribution vanishes at this order.  It thus follows that the off-diagonal terms (within the $n_N$-shell) are at minimum quadrupolar (dipolar) via the direct (exchange) kernels, making the use of ordinary PT approximately correct even in the dense cloud regime~\footnote{In more detail, the off-diagonal overlaps via the direct kernel, inside the (hyper)fine multiplet, are possible only between states that share the same magnetic number but whose angular numbers differ by (at least) $2$; this first occurs for $n_N \geq 3$, e.g. $\ket{M}=\ket{300}$, $\ket{N}=\ket{320}$. Regarding the exchange kernel, the dipolar coupling requires $\{ \ell_M , \ell_N \} = \{ \ell_a - 1 , \ell_a + 1 \}$ and $|m_M - m_a| \leq 1$, e.g. $\ket{a}=\ket{310}$, $\ket{M}=\ket{300}$, $\ket{N}=\ket{320}$. Such coupling is therefore not possible within the multiplet of a circular state, while the quadrupolar one starts at $n_N \geq 3$, as in the direct kernel case.} \vskip 4pt

In order to break the $m_N$-degeneracy, we include the quadrupolar term in the $\ell_N$-subshell, where, using
\begin{equation} \label{eq:quadr_int}
c^{(2)}_{\ell m} \equiv \int |Y_{\ell m}|^2\,Y_{20}\,d\Omega=\sqrt{\frac{5}{4\pi}}\;\frac{\ell(\ell+1)-3m^2}{(2\ell-1)(2\ell+3)} \,,
\end{equation}
we find (for the direct term)
\begin{eqnarray} \label{eq:hf_sg_newt}
G^{(\ell=2)}_{NNaa} &=& - \frac{4\pi}{5} \,  \mu^3 \alpha  \left[ c^{(2)}_{\ell_a m_a} \, c^{(2)}_{\ell_N m_N} \int d \mathsf{r}_x  d \mathsf{r}_y  \mathsf{r}_x^2 \mathsf{r}_y^2
\frac{\mathsf{r}_{<}^2}{\mathsf{r}_{>}^{3}}
\hat{\mathcal{R}}^2_{a} (\mathsf{r}_x) \hat{\mathcal{R}}^2_{N}   (\mathsf{r}_y ) \right] \,.
\end{eqnarray}
If $c^{(2)}_{\ell_a m_a} < 0$, one finds $G^{(\ell=2)}_{NNaa} - G^{(\ell=2)}_{MMaa} > 0$ whenever $m_M^2 > m_N^2$. Both conditions are satisfied if $a=M$ is an SR-generated state. For the exchange term let us return to the non-expanded form and perform the Fourier transform, showing
\begin{eqnarray} \label{eq:sg_exch_ft}
   G_{NaaN} = -\mu^3 \alpha \int \frac{d^3 k}{(2\pi)^3} \, \frac{4\pi}{k^2} \, \Big| \left[\bar\psi_N \psi_a \right]_{\bm{k}}  \Big|^2 < 0 \,,
\end{eqnarray}
thus working in the \textit{opposite} direction from the direct term. Within the $\ell_a$-subshell the two quadrupolar contributions are of the same order, and the overall sign is given by the direct term, barring two special cases. The first is $m_N = - m_a$, where due to the discrete symmetry $\bm{S}_\mathrm{c} \to - \bm{S}_\mathrm{c}$ the direct terms of both states are identical, and the energy split is therefore due entirely to the exchange term, $\Delta \epsilon^\mathrm{sg}_{aN} < 0$. The second case is $|\Delta m_{aN}| = 1$. These states are connected by the action of the ladder operators, $(\hat{S}_\mathrm{c})_\pm \ket{n_a \ell_a m_a} \propto \ket{n_a \ell_a\, m_a \pm 1}$, which together with $(\hat{S}_\mathrm{c})_z$ form a basis of the Lie algebra of $SO(3)$. As the self-gravity interactions are invariant under rotations, they cannot generate an energy split between states with $|\Delta m_{aN}| = 1$. The self-gravity-induced splitting can thus occur only due to the spin of the Kerr BH, which breaks the rotational symmetry explicitly, at the order $\Delta \epsilon^\mathrm{sg}_{aN} \sim \mu q_\mathrm{c} \tilde{a} \alpha^5$, that we do not consider explicitly here. Finally, note that, modulo the two previously discussed cases, there is an interplay of the partial cancellation of direct terms and the (single) exchange term in the  determination of the sign shift.  Inside the Bohr multiplet we find, performing calculations for several relevant state splits, that for  $|\Delta \ell_{aN}| =1 $, $|\Delta m_{aN}| =1 $ fine splits the exchange term dominates, thus leading to $\Delta \epsilon^\mathrm{sg}_{aN} < 0$, while for hyperfine and fine splits with $|\Delta m_{aN}| \geq 2$ the direct terms dominate leading to $\Delta \epsilon^\mathrm{sg}_{aN} > 0$. Considering the Bohr splits, as $|n_N - n_a|$ grows, the cancellation among the direct terms becomes less effective while the exchange progressively gets suppressed; both trends lead to the overall correction being dominated by the direct kernel. We summarise how self-gravity breaks the Bohr degeneracy, and contrast with the relativistic corrections, in Table~\ref{tab:degeneracy}.  \vskip 4pt

\begin{table}[t]
\centering
\begin{tabular}{c || c | c}
 & Relativistic corrections & Self-gravity  \\
\hline\hline
$n_a$-shell ($\ell_a$ degeneracy) & $\mathcal{O}( \mu \alpha^4 )$ & $\mathcal{O}(  \mu  q_\mathrm{c} \alpha^2) \times  \int_{\bm{x} \bm{y}}  \frac{1}{r_{>}}  $  \\[6pt]
$\ell_a$-subshell ($m_a$ degeneracy) & $\mathcal{O}(  \mu \alpha^5 \tilde{a} )$ & $\mathcal{O}(  \mu  q_\mathrm{c} \alpha^2) \times  \int_{\bm{x} \bm{y}}  \frac{r_{<}^2}{r_{>}^{3}}  $   \\[6pt]
$\ell_a$-subshell ($m_N = m_a \pm 1$ residual) & no degeneracy & $ \mathcal{O} (\mu q_\mathrm{c} \tilde{a} \alpha^5)$  \\
\end{tabular}
\caption{Hierarchy of spectral degeneracy breaking via the relativistic corrections and self-gravity.}
\label{tab:degeneracy}
\end{table}

\subsubsection{Perturbativity and interplay of perturbations}

The strength of self-gravity is controlled not only by $\{\alpha,q_\mathrm{c}\}$, but also the geometry of the state $n$. As $n$ grows for fixed $\{\alpha,q_\mathrm{c}\}$, the cloud is diluted and the absolute effect of self-gravity gets smaller. This however needs to be compared to the smallest energy gap in~\eqref{eq:pert_crit} of the neighbouring state (consistent with the selection rules) $\ket{M} = \ket{(n_N+1) \, \ell_N m_N}$. To first estimate the numerator of~\eqref{eq:pert_crit}, consider the monopolar term from the direct kernel~\eqref{eq:sg_mono} and note that the external $\mathsf{r}_y$ integral is dominated by the peak of the state $\ket{N}$ at $\mathsf{r}_{n_N}$:
\begin{eqnarray}
\mathsf{F}(\mathsf{r}_y) \equiv \int^\infty_0 d \mathsf{r}_x  
\frac{\mathsf{r}_x^2}{\mathsf{r}_{>}}  \hat{\mathcal{R}}^2_{a}(\mathsf{r}_x  ) \,, \quad \mathsf{F}(\mathsf{r}_y) = \mathsf{F}(\mathsf{r}_{n_N})  - \left(\mathsf{r}_y - \mathsf{r}_{n_N} \right) \frac{1}{\mathsf{r}_{n_N}^2} \int^{\mathsf{r}_{n_N}}_0 d \mathsf{r}_x  \mathsf{r}_x^2  \hat{\mathcal{R}}^2_{a} + \dots \,,
\end{eqnarray}
where, for $n_N \gtrsim n_a$ we have $\int^{\mathsf{r}_{n_N}}_0 d \mathsf{r}_x  \mathsf{r}_x^2  \hat{\mathcal{R}}^2_{a} \simeq 1$. Focusing again on circular  $\ket{N}=\ket{n_N \, (n_N-1) \, (n_N-1)}$  and Rydberg  $n_N \gg 1$ states  we find that the off-diagonal contribution $G_{MNaa} \propto n^{-5/2}_N$ dominates $|G_{NNaa}-G_{MMaa}| \propto n^{-3}_N$ and the perturbative condition for $V^\mathrm{n}_\mathrm{sg}$ is
\begin{eqnarray} \label{eq:pert_crit_sg}
q_\mathrm{c} < \frac{1}{\sqrt{n_N}} \,,
\end{eqnarray}
up to $\mathcal{O}(1)$ factors, thus becoming more stringent with increasing $n_N$.

We finish this subsection with two further comments on the interplay between internal and external corrections. Firstly, in contrast to the relativistic corrections, where the leading-order hyperfine term~\eqref{eq:hf_spectrum} is linear in the magnetic quantum number and thus leads to a degeneracy in the resonant-condition ratio $\Delta \epsilon_{NM}/\Delta m_{NM}$, the contributions~\eqref{eq:hf_sg_newt} to the hyperfine energy split are quadratic in the magnetic numbers (for $|m_N| \neq |m_M|$), and thus this degeneracy is not present. Secondly, the relativistic corrections and self-gravity in many instances pull the energy splits in different directions, pointing to the possibility of \textit{level crossing}. In Rayleigh--Schr\"odinger PT, the presence of off-diagonal terms in the (hyper)fine multiplet tends to produce avoided crossings (von Neumann--Wigner theorem; e.g.~\cite{zwieb}). Here, however, in many instances the off-diagonal terms exactly vanish (e.g.\ in the $\ell_N$-subshell) and would thus seem to allow level crossing to take place. Crucially, in order to probe the GA's structure, and thus for the level crossing to have a physical consequence, we need to apply an external field, which will generically induce off-diagonal terms. We will therefore discuss the question of level crossing in the time-independent case in Sec.~\ref{sec:love} and in the time-dependent case in Sec.~\ref{sec:pheno}, separately.

\section{Love Numbers} \label{sec:love}

In the following, we will compute the (gravitational) Love numbers of GAs within linear response theory. In Sec.~\ref{sec:pheno} we will model the external tidal field $H_{\mu \nu}$ as sourced by the orbital companion of the GA and contextualise our results. \vskip 4pt

\subsection{Love numbers in the WEFT} \label{sec:love_weft}

The induced multipoles depend on $C_N(t)$, which can evolve on the scales of the external driving $\Omega_\mathrm{ext}$ (e.g.\ during a resonance). Provided we expand the multipoles not only in $E_{i_L}$ but also in powers of $C_N$, the coefficients in the expansion are governed only by the degrees of freedom evolving much faster than the tidal field, resulting in a relation that is local in time (see App.~\ref{app:bottom})\footnote{\label{fn:inst}Here we focus on the leading contribution, coming from the instantaneous limit of the response. Going beyond it corresponds to including contributions $Q^E_{i_L} \propto \pd_{\tau} E^{j_{L'}}$, which are further suppressed by the ratio of the frequencies at which the tidal field and the microphysics evolve (see also Sec.~\ref{sec:love_dyn}).}:
\begin{eqnarray} 
       \left(\mathcal{Q}^E_{i_L}\right)_{\text{ind}} &=& -2 \sum_{\ell'}\left(\lambda^{(0)}_{i_L j_{L'}}  + \sum_{MN} C_N \bar{C}_M \lambda^{MN}_{i_L j_{L'}} + \mathcal{O}(C^4)\right) E^{j_{L'}} \,, \nonumber \\ 
     \lambda^{MN}_{i_{L} j_{L'}} &=& \sum_{m, m'} \mathsf{N}_\ell  \mathsf{N}_{\ell '} (\mathcal{Y}_{i_L}^{\ell m})^\ast \mathcal{Y}_{j_{L'}}^{\ell' m'} \lambda^{MN}_{(\ell m) (\ell' m')} \,, \label{eq:Love_projections}
\end{eqnarray}
where we include the leading contribution from $C_N$ compatible with the $U(1)$ symmetry. Furthermore, we perform the decomposition in the appropriate hydrogenic basis, to be defined in the following, and in the STF basis, respectively. Several comments are in order. The first coefficient is the only one surviving in the $C_N \to 0$ limit, and thus corresponds to the BH response, while the higher-order terms in the occupancy $\mathcal{O}(C^4)$ are generated via scalar self-coupling [this follows explicitly from the energy-momentum tensor~\eqref{eq:T00}--\eqref{eq:T0b}]. As we focus here on the limit of vanishing self-interactions, the self-coupling is mediated by self-gravity and thus suppressed by $q_\mathrm{c}$. In the regime of validity of the WEFT we are working on the scales $\Omega_\mathrm{ext} < \Delta \epsilon^\mathcal{B}_{NM} \ll M^{-1}$.  Let us keep a finite number of states $C_N$ explicitly, while the rest is integrated out. The dynamics of these integrated-out degrees of freedom, as well as the internal BH dynamics (quasi-normal modes), is dormant and thus captured in the constant coefficients $\{\lambda^{(0)}_{i_L j_{L'}}$, $\lambda^{MN}_{i_L j_{L'}}\}$. Famously, the (static) Kerr BH Love numbers vanish, $\lambda^{(0)}_{i_L j_{L'}}=0$~\cite{Chia:2020yla,Rodriguez:2026iot}, and thus the response of the GA coincides with that of the cloud, at the level we are working at. Finally, notice that these Love numbers, defined here per state pair, $\lambda^{MN}_{(\ell m) (\ell' m')}$, have a much richer structure than those of non-rotating objects. A generic state of the GA breaks spherical symmetry, allowing a response with $SO(3)$ quantum numbers different from those of the tidal field, and introducing a dependence on the magnetic numbers. \vskip 4pt

Instead of working with the induced multipoles as intermediate objects, one can capture the static tidally induced effects directly in the action
\begin{equation}
    S_{\rm tidal} \approx S_{\rm perm} + S_{\mathrm{Love}}
\end{equation}
where the first term contains the contributions from the permanent moments and the second one is (we focus on the dominant electric Love number)
\begin{equation} \label{eq:love_action}
    S_{\mathrm{Love}}= \int d \tau \sum_{\substack{\ell \geq 2 \, \ell' \geq 2}} \sum_{M N} C_N \bar{C}_M \lambda^{MN}_{i_L j_{L'}} E^{i_L} E^{j_{L'}} \,.
\end{equation}
The tidal field $H_{\mu \nu}$ induces a response $h_{\mu \nu}$ in the GA, so that the total metric perturbation in $S_\mathrm{EH}$ is $\delta g_{\mu \nu} = H_{\mu \nu} + h_{\mu \nu}$. For $H_{\mu \nu}$ we take a solution of the Einstein equations in the zero-frequency limit that is regular at the origin (see e.g.~\cite{Iteanu:2024dvx}). Working in de Donder gauge\footnote{The final result that we find for the cloud Love numbers will not depend on this choice. This is manifest from their EFT definition as Wilson coefficients in front of gauge-invariant operators; thus, the matching calculation can be done in any convenient gauge.} and using $E_{i_L} = -\frac{1}{2}\pd_{i_L} H_{00} + \mathcal{O}(H^2, \pd_t)$, we find the linear correction to the tidal potential $H_{00}$ by varying $S_\mathrm{EH}+ S_\mathrm{Love}$ w.r.t.\ $h_{00}$ (see e.g.~\cite{Rodriguez:2026iot}):
\begin{equation}
    \nabla^2 h_{00} = - \sum_{\ell'}\sum_{MN} C_N \bar C_M (-1)^{\ell'} \, 8\pi G \, \ell! \, A^{i_L}  \lambda^{MN}_{i_L j_{L'}} \pd^{j_{L'}} \delta^{(3)}(\bm{x}) \,,
\end{equation}
where we parametrise the tidal field as $H_{00} = A_{i_L} x^{i_L}$ and discard the subleading $\pd_t^2 h_{00}$ term consistent with the instantaneous response limit~\footref{fn:inst}. We solve this equation by going to momentum space and using the following Fourier integral for the STF tensors~\cite{Rodriguez:2026iot}
\begin{equation}
    (-i)^\ell \int \frac{d^d\bm{p}}{(2\pi)^d} e^{i \bm{p} \cdot \bm{x}} \frac{p^{j_L}}{|\bm{p}|^2} = (-1)^\ell \frac{\Gamma(d/2 - 1)\Gamma(2-d/2)}{2^\ell (4\pi)^{d/2}\Gamma(2-\ell-d/2)} x^{j_L} \left(\frac{\bm{x}^2}{4}\right)^{1-d/2-\ell} \,,
\end{equation}
resulting in
\begin{equation}
    h_{00} = 16\pi G \sum_{\ell'}\sum_{MN} C_N \bar C_M \, \ell! \, \frac{(2\ell'-1)!!}{8\pi}  \, A^{i_L}  \lambda^{MN}_{i_L j_{L'}} \frac{n_x^{j_{L'}}}{r^{\ell'+1}_x} \,,
\end{equation}
where $\Gamma(1/2) = \sqrt{\pi}$ and $\Gamma(1/2 - \ell) = (-1)^\ell \frac{2^\ell}{(2\ell-1)!!}\sqrt{\pi}$. By unpacking the STF tensors, we then find
\begin{equation} \label{eq:h00_EFT}
    h^{\rm WEFT}_{00} = 2 G \sum_{MN} \sum_{(\ell' m')(\ell m)} A_{\ell m} \frac{Y^{\ell' m'}(\hat{\bm{x}})}{r^{\ell'+1}_x} C_N \bar C_M \left[ \lambda^{MN}_{(\ell m)(\ell' m')} \, \ell! \, (2\ell' - 1)!! \,  \mathsf{N}_\ell  \mathsf{N}_{\ell '} \right] \,.
\end{equation}
\vskip 4pt

 Finally, let us introduce the ``effective Love number of the GA'' (which carries canonical dimensions) and can be compared with the usual Love numbers of astrophysical compact objects, like neutron stars~\footnote{In the limit where the time dependence induced by the evolution of $C_N(t)$ can be neglected, this object quantifies the effect of the induced response on the waveform. In the time-dependent case, additional care is needed when determining the contribution to the waveform; this is beyond the scope of the present work.}
\begin{equation} \label{eq:love_eff}
   \lambda^{\mathrm{eff}}_{i_L j_{L'}} = \sum_{MN} C_N \bar{C}_M \lambda^{MN}_{i_L j_{L'}} \,.
\end{equation}
\vskip 4pt

\subsection{Cloud in the external field \& matching}

The previous result~\eqref{eq:h00_EFT} was calculated in the WEFT. Let us now calculate the same object in the UV theory, $h^\mathrm{UV}_{00}$, in order to perform the matching:
\begin{equation} \label{eq:poisson_pert_uv}
    \nabla^2 h^{\mathrm{UV}}_{00} = - 8\pi G \delta T_{00}   \,, \quad \delta T_{00} \approx \delta \rho_\mathrm{c} \big|_\mathrm{ext} \,, 
\end{equation}
where $\delta \rho_\mathrm{c} \big|_\mathrm{ext}$ denotes the density perturbation of the cloud due to the tidal perturbation $V_\mathrm{ext} = -\frac{1}{2} \mu H_{00}$. In order to avoid the resonances within the (hyper)fine multiplet, we take $\Omega_\mathrm{ext} \ll \Delta \epsilon^\mathcal{H}$ and further assume a cloud dominated by a single state $\ket{a}$, with amplitude $C_a$, while all the other states $N$ are populated at most perturbatively by the tide (in Secs.~\ref{sec:love_spin} and~\ref{sec:love_dyn} we make this criterion more precise and generalise the discussion). As the  coefficient $\lambda^{MN}_{i_L j_{L'}}$ describes the leading, instantaneous part of the response, the matching can be done in the static limit, i.e., we track the perturbation of the ``good basis'' states (cf. Sec.~\ref{sec:pert})
\begin{eqnarray} \label{eq:pi_state}
 \sqrt{N_\mathrm{c}(t_0)I_N (t_0)} e^{-i \phi^\mathcal{B}_N }  \ket{\chi_N}  \to   \sqrt{N_\mathrm{c}(t_0)I_N (t_0)}   e^{-i \phi_N } \left( \ket{\chi_N} +  \ket{\chi_N^{(1)}} \right) \,,
\end{eqnarray}
with $\phi_N =\left( \Delta \epsilon_{aN}^\mathcal{B} + \Delta  \epsilon_{aN}^{(1)} \right)  t$, where the corrections to the energies and the states follow from~\eqref{eq:energy_correction}. \vskip 4pt

Let us first assume the validity of ordinary PT, i.e.\ $\mathcal{S}_\mathrm{d}=\{\ket{N}\}$. The solution of~\eqref{eq:poisson_pert_uv}, via~\eqref{eq:pi_state}, is, to leading order in the non-relativistic expansion,
\begin{eqnarray} \label{eq:uv_tidal}
     h^\mathrm{UV}_{00} &=&  \sum_{MN}\sum_{(\ell' m')(\ell m)} \frac{Y^{\ell'm'}(\hat{\bm{x}})}{r^{\ell'+1}_x}  \frac{4\pi G \mu^2 }{2\ell'+1}  C_N \bar C_M A_{\ell m} \mathsf{N}_\ell  \mathcal{K}^{MN}_{(\ell m)(\ell' m')} \,, \\
    \mathcal{K}^{MN}_{(\ell m)(\ell' m')} &=&  \sum_{\ket{P} \in \mathcal{S}_\perp} \left[\frac{\bra{P}r^\ell_z Y_{\ell m} (\bm{\hat{z}})\ket{N}\bra{M}r^{\ell'}_y Y_{\ell'm'}^\ast(\bm{\hat{y}}) \ket{P}}{\epsilon_P - \epsilon_N} + \frac{\bra{P} r^{\ell'}_y Y_{\ell'm'}^\ast(\bm{\hat{y}}) \ket{N} \bra{M}r^\ell_z Y_{\ell m}(\bm{\hat{z}}) \ket{P} }{\epsilon_P - \epsilon_M}  \right]  \,, \nonumber
\end{eqnarray}
where we have used the reality of the perturbation $V_\mathrm{ext}$ and the multipolar expansion~\eqref{eq:multipolar}, and considered only the particular solution ``outside'' the cloud. The selection rules for the first term in $\mathcal{K}^{MN}_{(\ell m)(\ell' m')}$ constrain $m_P = m_N +m = m_M + m'$, while the second term contributes only if $m_P = m_N -m' = m_M - m$. \vskip 4pt

Matching~\eqref{eq:uv_tidal} with the WEFT calculation~\eqref{eq:h00_EFT}, we find the explicit expression for the cloud, and thus the GA, Love numbers:
\begin{eqnarray}
    \lambda^{MN}_{(\ell m) (\ell' m')} &=& \frac{2\pi\mu^2}{\mathsf{N}_{\ell '} (2\ell'+1)!! \, \ell! } \,  \mathcal{K}^{MN}_{(\ell m)(\ell' m')} \,.
\label{eq:LoveAB}
\end{eqnarray}

\vskip 4pt

\subsection{Love numbers of the ground state} \label{sec:love_spherical}

Here we consider a spherical atom in the ground state, $\ket{a}=\ket{100}$. As the ground state has a trivial degenerate subspace, ordinary PT is applicable. The selection rules further reduce to $\ell=\ell'=\ell_P$, $m=m'$, and $m_P=m$ (first term) or $m_P=-m$ (second term), implying
\begin{eqnarray}
        \lambda^{aa}_{(\ell m)(\ell' m')} &=& \frac{2 \pi \mu^2 r_\mathrm{c}^{2\ell}  }{\mathsf{N}_\ell  (2\ell+1)!! \, \ell! } \delta_{\ell \ell'} \delta_{m m'} \frac{1}{4\pi} \sum_{\ket{P} \in \mathcal{S}_\perp} \frac{\left(\mathcal{I}^{aP|\ell}_r \right)^2 \delta_{\ell_P \ell}}{\epsilon_P - \epsilon_a}  \left(\delta_{m_P,m} + \delta_{m_P,-m} \right) \,.
    \label{eq:Love_sph}
\end{eqnarray}
The result is clearly $m$-independent, as expected for a spherically symmetric object. Furthermore, the result is positive, in agreement with the general expectations for weakly-gravitating spherically symmetric systems in the ground state~\cite{Creminelli:2026xxx}. \vskip 4pt

It follows that for the one-level atom the ``effective Love numbers'' scale as \\ $  \lambda^{\mathrm{eff}}_{(\ell m)(\ell' m')} \sim (M_\mathrm{c}/M) \, r^{\ell+\ell'+1}_\mathrm{c}$, in agreement with the estimate in~\cite{Baumann:2019ztm} and with the Newtonian calculation of the Love numbers in~\cite{Arana:2024kaz}. Let us compare our results with those of~\cite{Arana:2024kaz} in more detail.  We calculate the sum in~\eqref{eq:Love_sph} using the Dalgarno-Lewis method~\cite{1955RSPSA.233...70D} in App.~\ref{app:DL}, obtaining, via~\eqref{eq:Love_projections},
\begin{eqnarray} \label{eq:love_100}
    \lambda^\mathrm{eff}_{i_L j_{L'}} = \delta_{i_L j_{L'}} \lambda^\mathrm{eff}_\ell \,, \quad  \lambda^\mathrm{eff}_2 = \frac{5}{4} \frac{M_\mathrm{c}}{M} \frac{M^5}{\alpha^{10}} \,,
\end{eqnarray}
in agreement with the result presented in~\cite{Arana:2024kaz}\,\footnote{By comparing the respective definitions, we find that the $k^{a}_{\ell m}$ defined in~\cite{Arana:2024kaz} is related to our convention by \\ $k^{a}_{2 m} = 6 \mathsf{N}_2  (N_\mathrm{c}/M^5) \lambda^{aa}_{(2m)(2m)} = (15/2) \alpha^{-10} \,(M_\mathrm{c}/M)$.}. \vskip 4pt

\enlargethispage{-\baselineskip}
Note that the sum $\sum_{\ket{P} \in \mathcal{S}_\perp}$ in~\eqref{eq:LoveAB} runs over both the bound and the continuum states\footnote{The convergence of the sums is in general not guaranteed, and possible IR and UV divergences are to be addressed with the standard methods of QFT.}. This can be checked by performing the sum above explicitly. We find that adding up the bound states up to $n_P=50$ gives only $\simeq 36 \%$ of the full answer, while including the unbound contributions yields the rest of the result, in agreement with the Dalgarno-Lewis method, which automatically resums the continuum contribution. \vskip 4pt

 \vskip 4pt

\subsection{Love numbers of excited states} \label{sec:love_spin}

Let us now turn to the more subtle case of excited states, including the SR-generated clouds with spin, where~\cite{Arana:2024kaz} only succeeded in calculating the Love numbers originating from $m=0$ perturbations, while obtaining divergent results for $m \neq 0$. The issue lies in the starting point of~\cite{Arana:2024kaz}, where only the Bohr-level contributions to the eigenfrequencies are considered. The non-trivial degenerate subspace, $\mathcal{S}_\mathrm{d} \neq \{\ket{N}\}$, however, requires the use of DPT~\footnote{This is in general also the case for $m=0$ perturbations. However, if the starting point is the circular state (as in~\cite{Arana:2024kaz}), the selection rules in~\eqref{eq:uv_tidal} for $m=0$ perturbations imply $m_P=m_N$, and the circular state is the only one in its $n$-shell with that magnetic number. This forbids any off-diagonal overlap within the degenerate subspace of $\ket{N}$, allowing ordinary PT to be used.\label{fn:m0}}. As discussed in Sec.~\ref{sec:pert}, one needs to diagonalise the total perturbation, including both the ``internal'' effects (relativistic and self-gravity) and the external field. In order to highlight the main conceptual aspects, we consider the regimes of dilute vs.\ dense clouds, where the relativistic effects dominate self-gravity and vice versa, respectively. In each case we further consider a weak- vs.\ strong-field regime, where the corresponding internal effect dominates the external field or the opposite, respectively. \vskip 4pt

\subsubsection{Dilute clouds}

Starting with the weak-field regime, the (hyper)fine degeneracies have already been lifted by the relativistic corrections. Thus, one can apply ordinary PT to the relativistically-corrected energies and states ($\ket{P^\mathcal{B}} \to \ket{P^{\mathcal{B}\mathrm{R}}}$, etc., where R denotes that the appropriate (hyper)fine corrections have been incorporated in the standard $\ket{n \ell m}$ basis) [via~\eqref{eq:energy_correction} and $\mathcal{S}_\mathrm{d}=\{\ket{N}\}$]:
\begin{eqnarray} \label{eq:state_weak_f}
 \ket{N^{(\mathrm{ext}, 1)}} = -\left(  \sum_{n_P \neq n_N}  +\sum_{\mathcal{F}/\mathcal{H}} \right) \frac{\braket{P^{\mathcal{B}\mathrm{R}}|V_\mathrm{ext} |N^{\mathcal{B}\mathrm{R} }}}{\epsilon_P^{\mathcal{B}\mathrm{R}} - \epsilon_N^{\mathcal{B}\mathrm{R}}} \ket{P^{\mathcal{B}\mathrm{R}}} \,.
\end{eqnarray}  
We have split the sum into two parts. The first runs over different Bohr levels (and the continuum states), where the (hyper)fine contributions to the energy splitting in the denominator are subleading. There, we can approximate the energies and the states by their leading-order (Bohr) values (i.e.\ $\ket{P^{\mathcal{B}\mathrm{R}}} \approx \ket{P^{\mathcal{B}}}$, etc.), for small-to-moderate $\alpha < \sqrt{n_N}$. In addition, we need to explicitly calculate the finite second sum. We apply this decomposition in~\eqref{eq:uv_tidal}, where for the intra-Bohr (and continuum) contributions we again use the DL technique (App.~\ref{app:DL}) to evaluate the infinite sum. It is already clear that the contributions from inside the multiplet will be \textit{enhanced} (if the overlap $\braket{V_\mathrm{ext}}^{\mathcal{B}\mathrm{R}}_{PN}$ is non-zero), compared to the DL-resummed contribution, by relative factors of $(\Delta \epsilon^{\mathcal{B}\mathrm{R}}_{NP} / \Delta \epsilon^{\mathcal{B}}_{NP})^{-1} \sim \{\alpha^{-2},\alpha^{-3}/\tilde{a}\}$, corresponding to the fine and hyperfine splittings, respectively [cf.~\eqref{eq:hf_spectrum}]. This will then lead to Love numbers enhanced\footnote{Roughly, at the saturation of SR $\tilde{a} \sim \alpha$, giving a maximal enhancement of $\alpha^{-4}$, compared to the scaling of the ground state~\eqref{eq:love_100}.} with respect to the baseline scaling established in Sec.~\ref{sec:love_spherical}. In more detail, consider the $\ell'=\ell$, $m'=m$ case, where [Eq.~\eqref{eq:uv_tidal}]
\begin{eqnarray}
 \mathcal{K}^{aa}_{(\ell,m),{(\ell,m)}} \bigg|_{\substack{\mathrm{weak} \\ \mathrm{field}}} \supset  \sum_{\mathcal{F}/\mathcal{H}}  \frac{|\braket{P^{\mathcal{B}\mathrm{R}}| r^{\ell} Y_{\ell m} |N^{\mathcal{B}\mathrm{R} }}|^2 +|\braket{P^{\mathcal{B}\mathrm{R}}| r^{\ell} Y_{\ell m}^\ast |N^{\mathcal{B}\mathrm{R} }}|^2}{\epsilon_P^{\mathcal{B}\mathrm{R}} - \epsilon_N^{\mathcal{B}\mathrm{R}}} \,.
\end{eqnarray}
Thus, if the initial state is circular (i.e.\ $\ket{n_N \,(n_N-1) \,(n_N-1)}$), we have $\epsilon_P < \epsilon_N$ in the (hyper)fine multiplet, leading to \textit{negative} Love numbers. \vskip 4pt

In contrast, in the strong-field regime, we can ignore the relativistic corrections to the spectrum and apply DPT directly at the Bohr level, after finding the ``good basis'' states. As the sum in~\eqref{eq:uv_tidal} now runs only over the Bohr and continuum states, the scaling is in agreement with the one in Sec.~\ref{sec:love_spherical}, while the sign depends on the interplay between the infinitely many higher-energy states and the finite number of lower-energy ones. \vskip 4pt

Let us now focus on the example of the SR-generated $\ket{211}$ state, with the equatorial perturbation $\ell=2$, $m \in \{-2,0,2\}$ (see Sec.~\ref{sec:pheno}). The $m = 0$ perturbation is diagonal inside the (hyper)fine multiplet\footref{fn:m0}, while $m=\pm 2$ leads to off-diagonal overlaps between $\{\ket{211},\ket{21{-}1}\}$. Consequently, $\mathcal{S}_\mathrm{d}=\mathrm{span}\left(\ket{211},\ket{21{-}1}\right)$, and the overlap matrix that we need to diagonalise is (using the reality of the perturbation)
\begin{eqnarray} \label{eq:eta_matrix}
\hat{\mathcal{P}}_\mathrm{d} \left( V_\mathrm{R} + V_\mathrm{ext} \right) \hat{\mathcal{P}}_\mathrm{d} = \begin{pmatrix}
\left( \epsilon^\mathcal{F}_{\ket{211}} +\eta_0 \right) + \epsilon^\mathcal{H}_{\ket{211}} & \eta_2 \, e^{i \Pi } \\[6pt]
\eta_2\, e^{-i\Pi } & \left(\epsilon^\mathcal{F}_{\ket{211}} +\eta_0 \right) - \epsilon^\mathcal{H}_{\ket{211}}
\end{pmatrix} \,,
\end{eqnarray}
where $\hat{\mathcal{P}}_\mathrm{d}$ is the projection operator onto $\mathcal{S}_\mathrm{d}$, $\epsilon^\mathcal{F}_{\ket{211}}=\epsilon^\mathcal{F}_{\ket{21{-}1}}$, $\epsilon^\mathcal{H}_{\ket{211}}=-\epsilon^\mathcal{H}_{\ket{21{-}1}}$, and $\{\eta_{0},\eta_{2}\} \in \mathbb{R}$ correspond to the $m=0$ and $|m|=2$ external perturbations, respectively. Solving the eigenvalue problem for this matrix, we find the ``good basis'' states $\ket{\pm}$:
\begin{eqnarray} \label{eq:eigen_val_vec}
\tilde{\varepsilon}^{(1)}_\pm &=& \epsilon^\mathcal{F}_{\ket{211}} + \eta_0 \pm \sqrt{\left(\epsilon^\mathcal{H}_{\ket{211}}\right)^2 + \eta_2^2} \,, \quad \ket{\pm} = \cos{\varrho} \ket{2 1 \pm 1}  \pm e^{ \mp i \Pi}  \ket{ 2 1 \mp 1 }  \sin{\varrho} \,, \\
\varrho &=& \frac{1}{2} \arctan{\left(\frac{\eta_2}{\epsilon^\mathcal{H}_{\ket{211}}}\right)} \,.
\end{eqnarray}
Already from $\tilde{\varepsilon}^{(1)}_\pm$ it is clear that the asymptotic expansion for $\eta_2 \to 0$ depends on the ratio $\epsilon^\mathcal{H}_{\ket{211}}/\eta_2$: 
\begin{eqnarray}
 \quad \tilde{\varepsilon}^{(1)}_\pm &\simeq& \eta_0 + \epsilon^\mathcal{F}_{\ket{211}} \pm \left( \epsilon^\mathcal{H}_{\ket{211}} + \frac{\eta_2^2}{2 \epsilon^\mathcal{H}_{\ket{211}}} \right) \,, \quad \eta_2 \ll \epsilon^\mathcal{H}_{\ket{211}} \,, \label{eq:en_weak} \\
  \tilde{\varepsilon}^{(1)}_\pm &\simeq&\eta_0 +  \epsilon^\mathcal{F}_{\ket{211}} \pm \eta_2 \,,\quad  \eta_2 \gg \epsilon^\mathcal{H}_{\ket{211}} \,, \label{eq:en_strong}
\end{eqnarray}
corresponding, respectively, to the weak- and the strong-field regimes. \vskip 4pt

We now compute the Love numbers in these two limits. %
In the weak-field limit, $\varrho \to 0$ and the standard basis is the good one, as expected. Focusing on the dominant second term in~\eqref{eq:state_weak_f}, one finds the Love number
\begin{eqnarray} \label{eq:love_weak}
N_\mathrm{c} \lambda^{\braket{211|211}}_{(2,m),{(2,m')}} \bigg|_{\eta_2 \ll \epsilon^\mathcal{H}_{\ket{211}}} = - \frac{405}{2 \pi} \frac{M_\mathrm{c}}{M} \frac{M^5}{\alpha^{13} \tilde{a}} \left(\delta_{m,2} \delta_{m',2} + \delta_{m,-2} \delta_{m',-2}  \right) + \mathcal{O}(\alpha^{-12}) \,,
\end{eqnarray}
where the subleading terms include the beyond-hyperfine contribution to the dominant term (App.~\ref{app:spectrum}), as well as the contributions from the sum over Bohr and continuum states for both the $m=\pm 2$ and $m=0$ tidal components, which start at $\mathcal{O}(\alpha^{-10})$ (App.~\ref{app:DL}).
\vskip 4pt

In the strong-field limit, $\varrho \to \pi/4$, the hyperfine partner $\ket{21{-}1}$ can no longer be integrated out. Now the good-basis states $\ket{\pm}$ constitute the set of active states that determine the permanent multipoles, while the contribution to the Love numbers comes from the sum over the Bohr and continuum states in~\eqref{eq:uv_tidal}. The effective Love number of the atom~\eqref{eq:love_eff} is then found from the Love numbers of the $\ket{\pm}$ states. We perform this sum via the DL technique\footnote{The contribution due to virtual second-order transitions vanishes in this particular case.} (see App.~\ref{app:DL})
\begin{eqnarray} \label{eq:love_strong}
N_\mathrm{c} \lambda^{\braket{\pm|\pm}}_{(2,m),{(2,m')}} \bigg|_{\eta_2 \gg \epsilon^\mathcal{H}_{\ket{211}}} &=& \frac{15}{\pi} \frac{M_\mathrm{c}}{M} \frac{M^5}{\alpha^{10}} \Big[68 \left(\delta_{m,2} \delta_{m',2} + \delta_{m,-2} \delta_{m',-2}  \right) \nonumber \\ && + 40\,\delta_{m,0}\delta_{m',0} \pm 7\sqrt{6}\left(\delta_{m,0}\delta_{|m'|,2}+\delta_{|m|,2}\delta_{m',0}\right)\Big]\,, \\
N_\mathrm{c} \lambda^{\braket{+|-}}_{(2,m),{(2,m')}} \bigg|_{\eta_2 \gg \epsilon^\mathcal{H}_{\ket{211}}} &=& - N_\mathrm{c} \lambda^{\braket{-|+}}_{(2,m),{(2,m')}} \bigg|_{\eta_2 \gg \epsilon^\mathcal{H}_{\ket{211}}} \nonumber \\
&=& \frac{15}{\pi} \frac{M_\mathrm{c}}{M} \frac{M^5}{\alpha^{10}} \nonumber \\ && \times 7\sqrt{6} \left(\delta_{m,2}\delta_{m',0}+\delta_{m,0}\delta_{m',-2}-\delta_{m,-2}\delta_{m',0}-\delta_{m,0}\delta_{m',2}\right) \,, \label{eq:love_strong_pm}
\end{eqnarray}
where we set $\Pi=0$ without loss of generality. As expected, these Love numbers follow the Newtonian scaling, instead of the enhanced one.  We will contextualise these regimes, and comment further on the meaning of negative Love numbers, in Sec.~\ref{sec:pheno} and Sec.~\ref{sec:conc}, respectively. \vskip 4pt

Although the example we have considered is a spinning state, an enhancement of the Love numbers can also occur for spherical states with $n\geq3$, due to the non-zero overlap in the fine sector, e.g.\ $\braket{322|r^2 Y_{22}|300}\neq0$; as these partners lie above $\ket{n00}$, the enhancement, $\propto \alpha^{-2}$ relative to the baseline scaling, is positive. \vskip 4pt

 \vskip 4pt

\subsubsection{Dense clouds} \label{sec:love_dense}

The general discussion for dense clouds requires diagonalising $V_\mathrm{int}+V_\mathrm{ext}$ simultaneously, where $V_\mathrm{int} = V_\mathrm{R} + V_\mathrm{sg}$. Let us first revisit the previous example, i.e.\ the $\ell=2$, $m=\{0,\pm 2\}$ perturbations applied to $\mathcal{S}_\mathrm{d}=\mathrm{span}\left(\ket{211},\ket{21{-}1}\right)$. Here the discussion is completely parallel to the dilute-cloud limit: the strong-field regime is by construction blind, at leading order, to the post-Bohr corrections in the microphysics, and we thus find the same $\lambda^{\braket{\pm|\pm}}_{(2,m),{(2,m')}}$ as in the dilute-cloud limit. In contrast, in the weak-field regime ordinary PT can be applied, with the ``background'' energies and states corrected via self-gravity on top of the Bohr description, $\ket{P^\mathcal{B}} \to \ket{P^{\mathcal{B}\, \mathrm{sg}}}$, instead of via the (hyper)fine corrections. We would again find negative and parametrically enhanced $m=\pm 2$ Love numbers (compared to the baseline scaling, which applies to the $m=0$ Love numbers in this case), although the enhancement is now controlled by $1/q_\mathrm{c}$. \vskip 4pt

In general, the weak/strong-field asymptotics are more subtle, and the full overlap matrix needs to be considered. In particular, if a level crossing occurs in the \textit{diabatic} basis, which would naively make the state corrections~\eqref{eq:state_weak_f} diverge, the matching needs to be performed in the ``good basis'', where the level crossing is avoided. To see this explicitly, take the $\mathcal{S}_\mathrm{d}=\mathrm{span}\left(\ket{322},\ket{320},\ket{300}\right)$ coupled via $(\ell,m)=(2,\pm 2)$. Focusing only on the hyperfine split for simplicity, the overlap matrix  is
\begin{eqnarray} 
\hat{\mathcal{P}}_\mathrm{d} \left( V_\mathrm{int} + V_\mathrm{ext} \right) \hat{\mathcal{P}}_\mathrm{d} = \begin{pmatrix}
\epsilon^\mathcal{F}_{\ket{322}}  + \epsilon^\mathcal{H}_{\ket{322}}  - |\epsilon^\mathrm{sg}_{\ket{322}}| & \eta_2 \, e^{i \Pi } \\[6pt]
\eta_2\, e^{-i\Pi } & \epsilon^\mathcal{F}_{\ket{322}}  + |\epsilon^\mathrm{sg}_{\ket{320}}| 
\end{pmatrix} \,,
\end{eqnarray}
with eigenvalues
\begin{eqnarray} \label{eq:eigen_val_vec_210}
\tilde{\varepsilon}^{(1)}_\pm &=& \epsilon^\mathcal{F}_{\ket{322}} + \Delta_+ \pm \sqrt{\Delta^2_- + \eta_2^2} \,,  \quad \Delta_\pm  \equiv  \frac{\left( \epsilon^\mathcal{H}_{\ket{322}} - |\epsilon^\mathrm{sg}_{\ket{322}}| \right) \pm  |\epsilon^\mathrm{sg}_{\ket{320}}| }{2} \,.
\end{eqnarray}
The diabatic level crossing occurs at $\Delta_-=0$, where in fact the weak-field approximation~\eqref{eq:state_weak_f} breaks down. The ``good basis'' states, however, avoid crossing, $\Delta \tilde{\varepsilon}_{-+} = 2\sqrt{\Delta_-^2 + \eta_2^2} \to 2|\eta_2| > 0$. In other words, as in the dilute-cloud example, here also, in the strong-field regime, the hyperfine partner can no longer be integrated out, and the good-basis states have to be promoted to active degrees of freedom. \vskip 4pt

\section{Gravitational Atom in a Binary} \label{sec:pheno}

In this section, we consider a specific choice for the external field, due to an orbital companion in a binary with mass $M_\star$ and mass ratio $q = M_\star/ \mathcal{M}$. The full WEFT action~\eqref{eq:action} is now extended to include the terms corresponding to the companion 
\begin{equation} \label{eq:action_star}
    S_{\rm WEFT} \supset  \left( S_{\rm pp} + S_{\rm tidal} + S_{\rm micro} \right)_\star \,. 
\end{equation}
To reduce the parameter space, we will assume that the companion is simply a point particle described by its mass. \vskip 4pt

\subsection{Dynamics in the cloud-orbit phase space}

To track the relativistic corrections in the inspiral dynamics, one introduces a post-Newtonian (PN) counting $v_\star^2 \sim \mathcal{M}(1+q)/R$. Considering only point-particle terms and $S_\mathrm{EH}$ in~\eqref{eq:action},~\eqref{eq:action_star}, to leading order in the PN expansion, one can integrate out the gravitational field and obtain a Newtonian binding potential for the binary (e.g.~\cite{Porto:2016pyg}). The latter can, after going to the centre-of-mass frame, be written as a Kepler Hamiltonian. We now proceed to integrate out the cloud spatial structure in $S_\mathrm{micro}$ (Sec.~\ref{subsecmicro}), together with the internal perturbations (Sec.~\ref{sec:pert}). The expansion~\eqref{eq:ampdecomp} allows us in principle to project the scalar field at any moment onto the hydrogenic basis and to track the corrections induced by the perturbations only in terms of the flow of the occupancies and the relative phases~\eqref{eq:aa_var}. This perspective, however, obscures the fact that the expansion coefficients in~\eqref{eq:ampdecomp} change once the perturbations are turned on. Thus, we will take the basis to correspond to kets shifted by the effect of the internal perturbations. To this order all internal perturbations that we consider are diagonal~\footref{ft:sg_mix}, and we thus obtain a single Hamiltonian acting on the full cloud-binary phase space [cf.~\eqref{eq:aa_var}]
\begin{eqnarray} \label{eq:unperturbed}
\overline{H} = N_\mathrm{c} \sum_{N \neq a} \Delta \epsilon_{aN}I_N - \frac{q^3 \mathcal{M}^5 }{2 (1+q) \Lambda^2} \,,  \quad  \Delta \epsilon_{aN} = \Delta \epsilon_{aN}^\mathcal{B} + \Delta \epsilon_{aN}^{\mathrm{int}} + \dots 
\end{eqnarray}
where $(\vartheta,\Lambda)$ are the angle-action Delaunay variables~\cite{Tremaine_Dynamics} for the orbit: the mean anomaly $\vartheta$ and $\Lambda \equiv \frac{\mathcal{M}^{3/2}q}{\sqrt{1+q}} \sqrt{a}$ ($a$ is the semi-major axis). As $\overline{H}$ depends only on the action variables $\vec{J}=(\Lambda, N_\mathrm{c} I_N)$, only the angle variables $\vec{\theta}=(\vartheta, \phi_N)$, with $N \neq a$, will have a flow
\begin{eqnarray} \label{eq:angles_unprt2}
\dot{\phi}_N &=&  \Delta \epsilon_{aN} \,, \\
\dot{\vartheta} &=& \frac{q^3 \mathcal{M}^5 }{(1+q) \Lambda^3}  = \sqrt{\frac{\mathcal{M}(1+q)}{a^3}} \equiv \Omega \,, \quad 
\end{eqnarray}
where $\Omega$ is the Keplerian orbital frequency.
\vskip 4pt

The tidal field acting on the GA, which we introduced in Sec.~\ref{sec:love_weft}, is generated by the companion via $H_{00} = 2 M_\star/R  \left[ 1 + \mathcal{O}(v_\star^2) \right] $. The interaction of the GA with the tidal field is described in~\eqref{tidalaction} and~\eqref{eq:linresp}, corresponding, to leading order, to permanent and (linear) induced multipoles, where the active states contribute to the former, while the states that are integrated out contribute to the latter.  In the rest of this Section we focus on SR-generated, i.e.\ spinning, clouds and consider the impact of the dominant permanent multipoles on the phase space dynamics, taking into account several active states, and thus generalising the discussion of~\cite{Boskovic:2025ixx} to multi-state atoms [via~\eqref{tidalaction}]
\begin{eqnarray}
H' &\supset & \sum_{\ell \geq 2} (\mathcal{Q}^{\mathrm{ perm}})_E^{i_L}  E_{i_L} \approx - M_\star \sum_{\ell \geq 2} \sum^{\ell}_{m=-\ell} \ell! \frac{ 4 \pi }{2\ell+1} \frac{Y^\ast_{\ell m}(\bm{\hat{R}})}{R^{\ell+1}}  (Q^\mathrm{perm}_\mathrm{c})^E_{\ell m} \nonumber \\
& =& \sum_\ell \, \frac{Y_{\ell 0}(\Theta,0)}{2} \left(  \mathcal{E}^{aa}_{\ell 0} D^2 + \sum_{N \neq a} \mathcal{E}^{NN}_{\ell 0} I_N \right) + \sum_{\ell,m \geq 0,N \neq a} \mathcal{E}^{aN}_{\ell m} Y_{\ell m}(\Theta,0) D \sqrt{I_N}    \cos{\left(m \Phi + \phi_N \right)} \nonumber \\
&&  + \sum_{\ell, m \geq 0, \diamond} \mathcal{E}^{MN}_{\ell m} Y_{\ell m}(\Theta,0)  \sqrt{I_M I_N} \cos{\left( m \Phi + \phi_N - \phi_M\right)} \,, \quad \diamond \equiv {\substack{M \neq N \\ M \neq a \\ N \neq a}} \,, \label{eq:H_mix} 
\end{eqnarray}
\begin{eqnarray}
\hspace{-3em} \mathcal{E}^{MN}_{\ell m} Y_{\ell m } \left(\frac{\pi}{2},0 \right) \equiv 2 \eta^{MN}_{\ell m} N_\mathrm{c} \,, \quad \eta^{MN}_{\ell m} \equiv -  
\frac{\mu M_\star}{r_\mathrm{c}} \frac{4 \pi}{2 \ell  + 1} \left(\frac{r_\mathrm{c}}{R} \right)^{\ell + 1} Y_{\ell m } \left(\frac{\pi}{2},0 \right) (\mathcal{I}_r \mathcal{I}_{\Omega,E})^{(MN|\ell m)} \,, \label{eq:etaN}
\end{eqnarray}
where $(\Theta,\Phi)$ are the spherical coordinates of the companion and we have assumed that the selection rules~\eqref{eq:selection_electric} are satisfied for the overlap integrals (which should include the correction via the internal perturbations) $\{\mathcal{I}_r^{(MN|\ell m)}, \mathcal{I}_{\Omega,E}^{(MN|\ell m)}\}$. In the first term of the second row, we used the result from~\eqref{eq:selection_electric} that the diagonal terms are non-zero only for $m=0$, while in the mixing terms we made the reality of the Hamiltonian explicit via the trigonometric representation. There are two types of mixing terms: those coupling the reference state $a$ directly to the other states (second term in the second row), and those coupling the other states among themselves (third row). Note that the potential above carries a $(v_\star^2)^\ell$ PN counting, i.e.\ starting at $2\mathrm{PN}$ for the leading quadrupole. The associated coefficient $(Q^\mathrm{perm}_\mathrm{c})^E_{\ell m} \propto q_\mathrm{c} r_\mathrm{c}^\ell$, however, is enhanced by powers of the cloud's extent, which for clouds of moderate-to-high density partially compensates for the PN suppression.  \vskip 4pt

Let us now further specialise to circular, equatorial orbits, where $\Theta = \pi/2$ and $\Phi = s \vartheta$, with $s=\pm 1$ corresponding to co- and counter-rotating orbits, respectively. The diagonal terms then depend only on the action variables $\vec{J}$ and thus can induce precession in the angle-variable dynamics. On the other hand, the interference terms also depend on $\vec{\theta}$ and thus lead to changes in the actions: mixing of atomic energy levels and changes in the size and shape of the orbit. As the interference terms have the structure
\begin{eqnarray}
 H'  \supset  h(\vec{J}) \times \{\cos{\left(  \phi_N + g \vartheta \right)},   \cos{\left(  \phi_N - \phi_M + g \vartheta \right)}\} \,, \quad g = m s \,,
\end{eqnarray}
this constitutes a paradigmatic example of a Hamiltonian for a nearly integrable system, where resonances occur when the argument of the cosine varies slowly with time (cf.~\cite{Tremaine_Dynamics}), i.e.\ $g\dot{\vartheta}_g = - \dot{\phi}_N$ and $g\dot{\vartheta}_g = - (\dot{\phi}_N-\dot{\phi}_M)$, respectively. Using the selection rules~\eqref{eq:selection_electric}, it follows (taking $M=a$ corresponds to a direct coupling term)
\begin{eqnarray} \label{eq:res}
\dot{\vartheta}^{MN}_g = \frac{s}{\Delta m_{MN}} \left[ \Delta \epsilon_{MN} + \sum_\ell \left( \eta^{NN}_{\ell 0} - \eta^{MM}_{\ell 0} \right) + \dots \right] \,,
\end{eqnarray}
where the dots correspond to the contributions from the mixing terms in $H'$. As $H'  \ll \overline{H}$, we have $\dot{\vec{\theta}} \approx \partial_{\vec{J}} \overline{H}$, i.e.\ the internal perturbations dominate the resonant position, and thus we recover the resonant condition of~\cite{Baumann:2018vus} at leading order, $\Omega^{MN}_g  \approx \tfrac{\Delta \epsilon_{MN}}{\Delta m_{MN}}$.\,\footnote{Allowing for eccentricity $e$ and spin-orbit misalignment $\beta$ on the one hand enlarges the orbital phase subspace with the action variables connected to these two parameters and the conjugate angle variables~\cite{Tremaine_Dynamics}. On the other hand, in order to express $(\Theta,\Phi)$ via the mean anomaly $\vartheta$, one needs to expand $Y_{\ell m}(\Theta,\Phi)$ in two sets of overtones $(g,k)$, where the circular equatorial limit is supported at $(sm,0)$. This ultimately leads to two towers of resonances (at leading order, $\dot{\vec{\theta}} \approx \partial_{\vec{J}} \overline{H}$),
$ 
\Omega^{MN}_{(g,k)} = \tfrac{m}{g-k} \tfrac{\Delta \epsilon_{MN}}{\Delta m_{MN}} \,,
$
as the precession dynamics of the remaining two orbital angles starts at $H'$~\cite{Boskovic:2025ixx}. Finally, to track the off-equatorial dynamics one needs to include $S_\mathrm{spin}$ in~\eqref{eq:action_pp} explicitly, although to leading order in the radiation reaction the total spin of the system (orbit + GA) is conserved and the dynamics of the spin angles can be expressed via the evolution of the orbital elements~\cite{Boskovic:2025ixx}.\label{eq:res_general}} \vskip 4pt

In the WEFT framework one can systematically include higher-order PN corrections to the gravitational field~\cite{Porto:2016pyg}, which drive the precession dynamics in the orbital phase subspace, as well as higher-order finite-size contributions, starting with the permanent magnetic multipoles that we briefly discuss in App.~\ref{app:magnetic}. The latter impact both the precession and the mixing dynamics (with a different set of selection rules), although in practice not as a dominant driver. So far we have included only conservative effects (at the level of the full phase space). However, dissipative effects, both due to the BH horizon and the GW emission of the binary (and the cloud itself), are also relevant. Although both can be described formally within the WEFT~\cite{Goldberger:2005cd,Porto:2007qi,Galley:2009px,Porto:2016pyg,Goldberger:2020fot}, for our purposes we will include them in an adiabatic manner: the former gives rise to the growth/decay of the occupancies (App.~\ref{sec:dis}),
\begin{eqnarray}
\dot{I}_N  \Big|_\mathrm{dis} = - 2 \Gamma_N I_N - \frac{\dot{N}_\mathrm{c}}{N_\mathrm{c}} I_N \,,
\end{eqnarray}
while the latter is included via energy-momentum fluxes, leading to Peters' evolution of the semi-major axis $a$ and the orbital eccentricity $e$~\cite{Peters:1963ux,Peters:1964zz}. In this way, radiation reaction (RR) drives the orbital evolution from lower orbital frequencies towards the resonance(s)~\eqref{eq:res}. The non-linear feedback between the orbit, the cloud, and the BH can then, at the resonance, lead to \\ \textit{(quasi-)floating} dynamics for $\Delta \epsilon_{aN} <0$ (the orbit gains energy at the expense of the cloud), where the orbital evolution is slowed down compared to the evolution in vacuum (in the extreme limit, stalling), and \textit{sinking} dynamics when $\Delta \epsilon_{aN} >0$, where the inspiral significantly accelerates over the resonance timescale (the orbit loses energy both to the cloud and to GW emission). Crucially, while the former presents a positive-feedback loop between the efficacy of the level mixing and the orbital impact, generically completing the level transitions (and significantly depleting the cloud for $\ell_a \sim \mathcal{O}(1)$ states), the latter is a negative-feedback loop, usually only leading to incomplete level mixing~\cite{Baumann:2019ztm,Boskovic:2024fga,Tomaselli:2024bdd,Boskovic:2025ixx}. \vskip 4pt

\subsection{Chronology of level mixing} \label{sec:chrono}

In both the dilute and the dense cloud limit, there is a hierarchy between the Bohr $(\mathcal{B})$, fine $(\mathcal{F})$, and hyperfine $(\mathcal{H})$ splits of the levels (Sec.~\ref{sec:pert}), implying, via~\eqref{eq:res}, a separation in the inspiral stages where the resonances occur, with $\Omega^\mathcal{H}_g \ll \Omega^\mathcal{F}_g \ll \Omega^\mathcal{B}_g$. On the other hand, the decay widths complicate the story by broadening the transitions, leading to overlaps between the different types in part of the parameter space. Let us address these effects in turn, starting from an SR-generated cloud in the circular state $\ket{n_a \, (n_a-1) \, (n_a-1)}$, $n_a \geq 2$. \vskip 4pt

\subsubsection{Narrow resonances}

In most of the literature so far, the position of the resonance~\eqref{eq:res} has been estimated using only\footnote{The effect of self-gravity on the resonances that can occur on co-rotating equatorial orbits has been discussed for $\ket{a}=\ket{211}$ in~\cite{Takahashi:2021yhy} and for $\ket{a}=\ket{322}$ in~\cite{Kim:2025wwj}, albeit both missing the crucial effect of the exchange kernel~\eqref{eq:phase_evo_sg}. } the leading-order relativistic corrections~\eqref{eq:hf_spectrum}, which we denote $(\Omega^{aM}_g)^{(\mathrm{R},1)}$. For the $\mathcal{H}/\mathcal{F}$ transitions we always have $\Delta \epsilon_{aM}^{(\mathrm{R},1)} < 0$, and thus one expects, via~\eqref{eq:res}, floating-type resonances to be supported on co-rotating orbits and no support for resonances on counter-rotating orbits. Formally, the $\mathcal{B}$ regime, which would allow both types of resonance on both co- and counter-rotating orbits, is not within the regime of validity of the WEFT. Note, however, that the selection rules always truncate the multipolar expansion~\eqref{eq:etaN} to a finite sum, and thus the finite-size contributions~\eqref{tidalaction} do not diverge even in the $\mathcal{B}$ regime, while now also giving support for tidal mixing with the continuum (ionization)~\cite{Baumann:2021fkf,Tomaselli:2024bdd}. Rather, the obstruction is the breakdown of the separation of scales: once $R \lesssim r_\mathrm{c}$, the multipolar coupling no longer captures the full interaction, and a new class of ``contact'' processes describing the direct coupling of the companion to the scalar cloud, such as scalar absorption by the companion~\cite{Baumann:2021fkf}, must be accounted for. These are beyond the scope of this work, but we will consider early-$\mathcal{B}$ transitions, mediated by higher multipoles $\ell \gg 1$, towards circular states, e.g.\ $\ket{(\ell+2) \, (\ell+1) \, (\ell+1)}$, where the large denominator $\Delta m_{aM} \simeq \ell$ in~\eqref{eq:res} can lower the resonant frequency to earlier stages of the inspiral, even coinciding with the $\mathcal{F}$ regime~\cite{Boskovic:2025ixx}. \vskip 4pt

\begin{figure}[p]
\centering
\begin{minipage}{0.48\textwidth}
    \centering
    \includegraphics[width=0.76\linewidth]{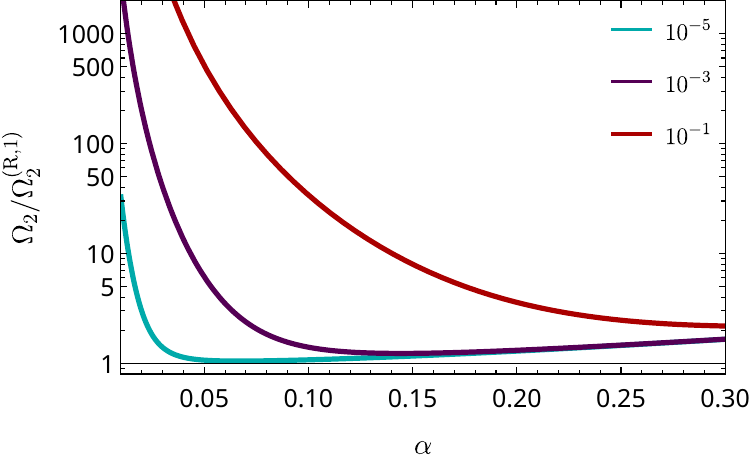}
\end{minipage}
\hfill
\begin{minipage}{0.48\textwidth}
    \centering
    \includegraphics[width=0.76\linewidth]{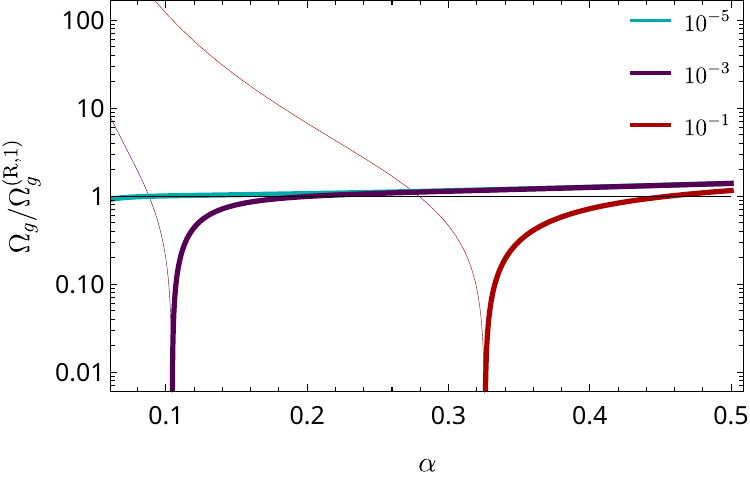}
\end{minipage}
\caption{\small [\textit{Hyperfine transitions}] Resonant frequency (including internal perturbations) $\Omega^{MN}_g$, relative to the value determined by including only (hyper)fine corrections $(\Omega^{MN}_g)^{(\mathrm{R},1)}$, for the $\ket{211}\to\ket{21{-}1}$ (left) and $\ket{322}\to\ket{320}$ (right) transitions, and $q_\mathrm{c} = \{\,10^{-5},\,10^{-3},\,10^{-1}\,\}$. Thick (thin) lines correspond to $\Delta \epsilon_{MN}<0$ ($\Delta \epsilon_{MN}>0$).}
\label{fig:res_hf}
\begin{minipage}{0.48\textwidth}
    \centering
    \includegraphics[width=0.76\linewidth]{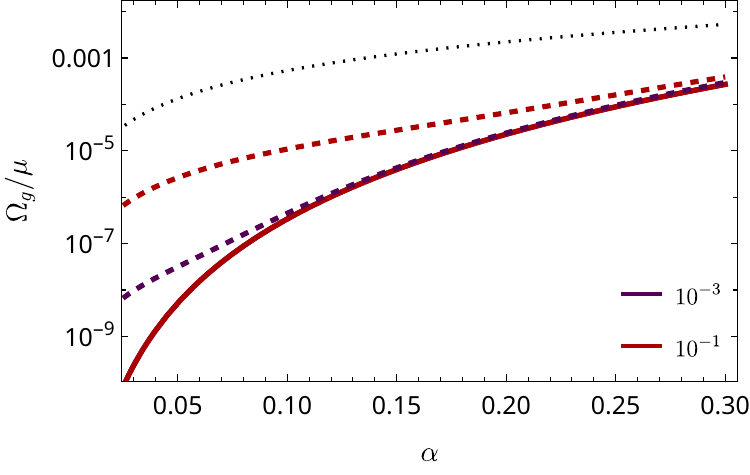}
\end{minipage}
\hfill
\begin{minipage}{0.48\textwidth}
    \centering
    \includegraphics[width=0.76\linewidth]{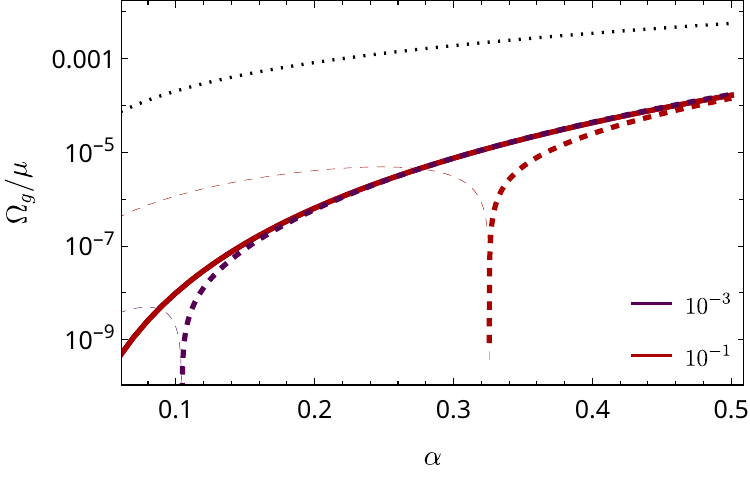}
\end{minipage}
\caption{\small [\textit{Hyperfine transitions}] Resonant frequency (including internal perturbations) $\Omega^{MN}_g/\mu$ of two different transitions in the hyperfine regime, for $q_\mathrm{c} = \{\,10^{-3},\,10^{-1}\,\}$. Left: $\ket{211}\to\ket{210}$ (solid lines) and $\ket{211}\to\ket{21{-}1}$ (dashed lines). Right: $\ket{322}\to\ket{321}$ (solid lines) and $\ket{322}\to\ket{320}$ (dashed lines). The dotted curve indicates the onset of the deep-Bohr regime. Thick (thin) lines correspond to $\Delta \epsilon_{MN}<0$ ($\Delta \epsilon_{MN}>0$).}
\label{fig:res_hf_abs}
\begin{minipage}{0.48\textwidth}
    \centering
    \includegraphics[width=0.76\linewidth]{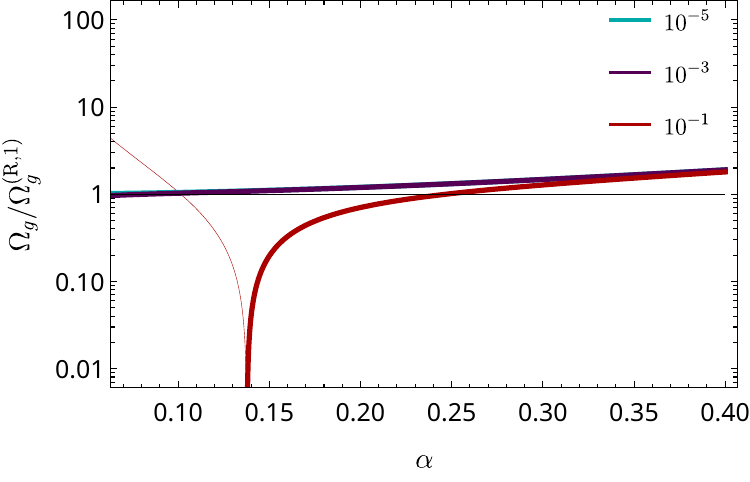}
\end{minipage}
\hfill
\begin{minipage}{0.48\textwidth}
    \centering
    \includegraphics[width=0.76\linewidth]{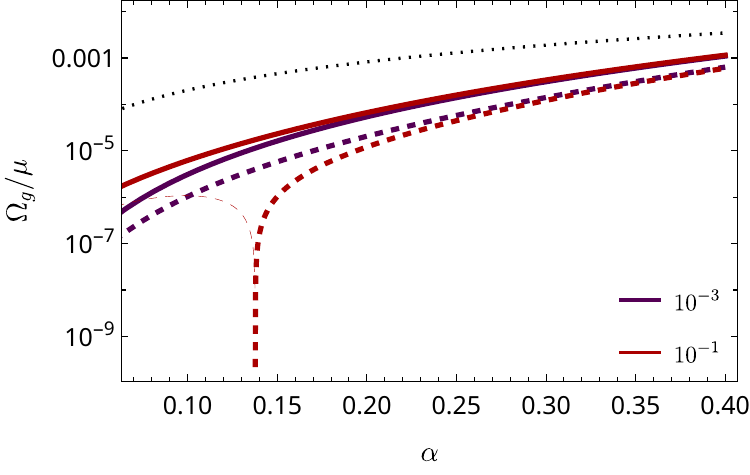}
\end{minipage}
\caption{\small [\textit{Fine transitions}] (Left) Same as Fig.~\ref{fig:res_hf} for $\ket{322}\to\ket{31{-}1}$. (Right) Same as Fig.~\ref{fig:res_hf_abs} for $\ket{322}\to\ket{311}$ (solid lines) and $\ket{322}\to\ket{31{-}1}$ (dashed lines).}
\label{fig:res_fine}
\begin{minipage}{0.48\textwidth}
    \centering
    \includegraphics[width=0.76\linewidth]{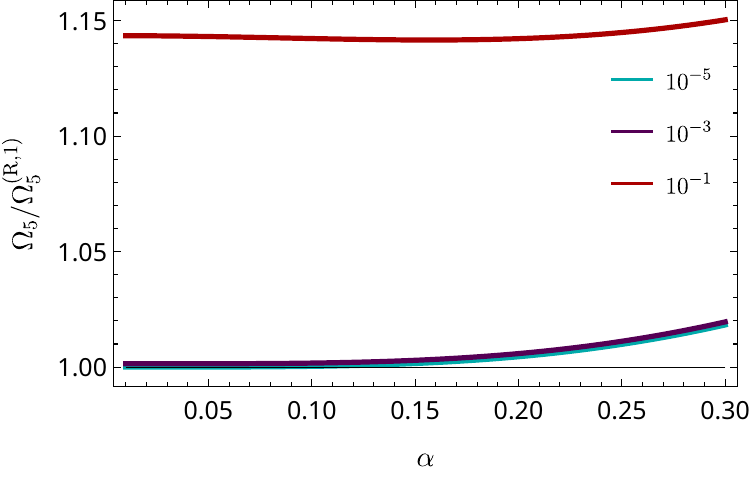}
\end{minipage}
\hfill
\begin{minipage}{0.48\textwidth}
    \centering
    \includegraphics[width=0.76\linewidth]{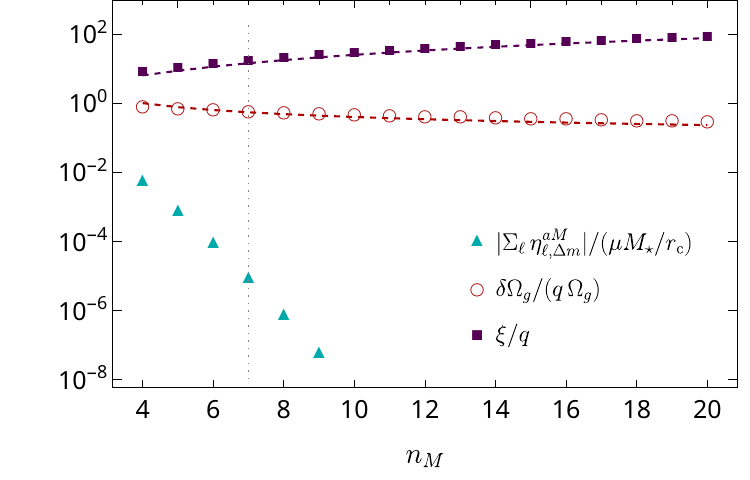}
\end{minipage}
\caption{\small [\textit{Early Bohr transitions}] (Left) Same as Fig.~\ref{fig:res_hf} for $\ket{211}\to\ket{766}$. (Right) Mixing overlap $|\sum_\ell\eta^{aM}_{\ell,\Delta m}|$ [cf.~\eqref{eq:etaN}; triangles], in units of $\mu M_\star/r_\mathrm{c}$, and resonance shift $\delta\Omega_g/\Omega_g$ [circles, with the monopolar estimate~\eqref{eq:dOm_B} dashed] and perturbativity $\xi$ [cf.~\eqref{eq:pert_crit}; squares, with the dipolar estimate~\eqref{eq:xi_B} dashed], in units of $q$ (suppressing the $(1+q)$ factors), for early-$\mathcal{B}$ transitions from $\ket{211}$ to the circular states $\ket{n_M\,(n_M{-}1)\,(n_M{-}1)}$. The dotted line marks $\ket{211}\to\ket{766}$.}
\label{fig:res_bohr}
\end{figure}

Let us assess the impact of the internal perturbations (Sec.~\ref{sec:pert}) relative to $(\Omega^{aM}_g)^{(\mathrm{R},1)}$ across the parameter space\footnote{By specifying $\alpha$ and the BH ``age'' $t_\mathrm{age}$ (the timescale over which the growth of the cloud can proceed unimpeded by environmental effects, cf.~\cite{Caputo:2025oap}), one can calculate $q_\mathrm{c}=q_\mathrm{c}(\alpha,t_\mathrm{age}/M)$~\cite{Baryakhtar:2020gao,Khalaf:2024nwc}. In~\cite{Kim:2025wwj}, a specific choice of $t_\mathrm{age}$ is assumed in all the results; however, models of binary BH formation allow for a range of $t_\mathrm{age}$, and environmental accretion onto the BH can even boost $q_\mathrm{c}^\mathrm{max} \to 0.3$~\cite{Brito:2014wla,Hui:2022sri}. To get the broadest view of the parameter space, we treat $q_\mathrm{c}$ as a free parameter.\label{ft:cloud_evo1}} $\{\alpha,q_\mathrm{c}\}$. From~\eqref{eq:Vsg_estimate} we see that, at fixed $q_\mathrm{c}$, the lower-$\alpha$ regime is more susceptible to self-gravity, while in the higher-$\alpha$ regime we expect the beyond-hyperfine relativistic corrections to make an impact. The latter always work in the direction of further widening the energy gap and thus pushing the resonance to higher values of the orbital frequency, up to relative $\mathcal{O}(1)$ factors in the relevant higher-$\alpha$ range for $\mathcal{H}/\mathcal{F}$ resonances. For the former, in addition to the absolute scaling, the relative importance of the direct and exchange kernels [cf.~\eqref{eq:phase_evo_sg}] is important in determining the overall direction of the energy shift. 
Consider first the $\mathcal{H}$ transitions, most susceptible to the impact of self-gravity. Based on the discussion in Sec.~\ref{sec:self_gravity} we find three different scenarios (Figs.~\ref{fig:res_hf}--\ref{fig:res_hf_abs}):
\begin{itemize}
\item[$\mathrm{(i)}_\mathcal{H}$] $\Delta m_{aN}= -1$ (e.g.\ $\ket{211} \to \ket{210}$, $\ket{322} \to \ket{321}$; possible on off-equatorial orbits), where the leading-order effects of self-gravity vanish, inducing only a correction parametrically weaker than the hyperfine relativistic term;
\item[$\mathrm{(ii)}_\mathcal{H}$] $\Delta m_{aN}= - 2 m_a$ (e.g.\ $\ket{211} \to \ket{21{-}1}$, $\ket{322} \to \ket{32{-}2}$; mediated by $\ell =  2 \ell_a$), where the contribution of the direct kernel exactly cancels and only the exchange kernel operates (self-gravity is mediated by the same multipole $\ell =  2 \ell_a$), in consonance with the relativistic corrections, thereby shifting the resonance to higher orbital frequencies; this shift can be parametric in the dense cloud regime;
\item[$\mathrm{(iii)}_\mathcal{H}$] other transitions (e.g.\ the strongest equatorial for the $\ket{322}$ state: $\ket{322} \to \ket{320}$), where the direct self-gravity kernel overtakes the exchange term, operating in the opposite direction to the relativistic corrections. This can ultimately lead, for denser clouds, to $\Delta \epsilon_{aM}^{(1)} > 0$, thus implying sinking-type resonances on counter-rotating orbits instead of the expectation based solely on relativistic effects. 
\end{itemize}

For $\mathcal{F}$ resonances we instead find (Fig.~\ref{fig:res_fine}):
\begin{itemize}
\item[$\mathrm{(i)}_\mathcal{F}$] $\Delta m_{aN}= -1$ (e.g.\ $\ket{322} \to \ket{311}$), where the dipolar contribution of the exchange kernel overtakes the \textit{difference} of the monopolar contributions of the direct kernels; this works in the direction of shifting the corresponding resonances to higher orbital frequencies;
\item[$\mathrm{(ii)}_\mathcal{F}$] other transitions (e.g.\ $\ket{322} \to \ket{31{-}1}$), where the direct kernel wins, and analogously to scenario (iii) for the $\mathcal{H}$ resonances, shifts the resonance to lower orbital frequencies, ultimately, for dense clouds and the smaller-$\alpha$ range, changing the nature of the resonance from floating-type on co-rotating orbits to sinking-type on counter-rotating orbits. 
\end{itemize}
Note that in both scenarios the effect of self-gravity is quantitatively weaker and consequently the parameter subspace where the level crossing occurs is significantly smaller compared to the $\mathcal{H}$ transitions. Finally, for early-$\mathcal{B}$ resonances the effect of the internal perturbations is more modest and primarily due to self-gravity, amounting to up to $\mathcal{O}(15\%)$ shifts relative to $(\Omega^{aM}_g)^{(\mathrm{R},1)}$ (Fig.~\ref{fig:res_bohr}, left). \vskip 4pt

Tidal coupling affects both the relative phase~\eqref{eq:res} and the orbital precession [cf.~\eqref{eq:H_mix}], while the higher-order conservative PN terms in the orbital sector affect only the latter. The estimate of these effects can be found in~\cite{Boskovic:2025ixx}, and in the regime of validity of the WEFT they amount to at most $\mathcal{O}(1\%)$ corrections to the resonance frequency, with the exception of the tidal shift of the spectrum in the early-$\mathcal{B}$ regime, which we now address. 
Although the target states of early-$\mathcal{B}$ transitions lie outside the orbit at the resonance, $R_g/r_M\simeq(2n_a^2\Delta m)^{2/3}/n_M^2\lesssim1$ [cf.~\eqref{eq:res}], these transitions are all of the sinking type and typically ineffective, while the companion still lies outside the dominant state, making ionization and contact interactions inefficient. However, when calculating the tidal overlap, the expansion~\eqref{eq:etaN} does not apply to the target; outside the orbit, the leading terms of $V_\mathrm{ext}$ are instead the monopole and the dipole~\cite{Brito:2023pyl,Tomaselli:2024bdd}
\begin{equation} \label{eq:outer_kernel}
V_\mathrm{ext}^{(\ell=0)}=\mu M_\star\left(\frac1R-\frac1r\right)\theta(r-R)\,,\quad
V_\mathrm{ext}^{(\ell=1)}=-\mu M_\star\,\frac{4\pi}{3}\sum_{m}Y^\ast_{1m}(\bm{\hat{R}})\,Y_{1m}(\bm{\hat{r}})\left(\frac{R}{r^2}-\frac{r}{R^2}\right)\theta(r-R)\,,
\end{equation}
while higher-order terms have contributions from both ``inside'' and ``outside'' the orbit. For the self-energy shift, the dipole vanishes by parity and the monopolar contribution $\eta^{MM}\sim\mu M_\star/R_g$ dominates, thus pushing the resonance towards higher orbital frequencies:
\begin{equation} \label{eq:dOm_B}
\frac{\delta\Omega_g}{\Omega_g}\sim \frac{q}{(1+q)^{1/3}}\left(\frac{n_a}{\Delta m}\right)^{2/3}\,.
\end{equation}
\vskip 4pt

As the shift grows with $q$, let us then check if the corrections are perturbative~\eqref{eq:pert_crit}. First, we go to the dressed frame~\cite{Baumann:2019ztm}, which co-rotates with the companion, so that the static perturbation theory applies. Next, we recognise that the strongest mixing between neighbouring levels is dipolar:
\begin{equation} \label{eq:xi_B}
\xi\simeq\frac{\mu M_\star r_M}{R_g^2\,\Omega_g}\simeq \frac{q}{(1+q)^{2/3}}\,n_M^2\left(2n_a^2\Delta m\right)^{-1/3}\,,
\end{equation}
e.g.\ for $\ket{211}\to\ket{766}$, it gives $\xi\simeq14\,q$, i.e.\ the treatment is under perturbative control for $q\lesssim0.07$. Note that although, for fixed $q$, perturbativity worsens as $n_M$ grows, the mixing is mediated by the highest-$\ell$ multipoles and will thus decrease with $n_M$, making these transitions increasingly suppressed. These trends are illustrated in Fig.~\ref{fig:res_bohr} (right). Physically, the breakdown of perturbativity signals that the binding potential of the scalars is no longer dominated by $M$, and the system is more akin to a gravitational molecule~\cite{Ikeda:2020xvt,Guo:2025ckp}; in that part of the parameter space, one needs to use molecular wavefunctions as the basis instead. \vskip 4pt

\subsubsection{Decay width and cloud patterns} \label{sec:cloud_patter}

\enlargethispage{-\baselineskip}
The presence of the decay width broadens the transitions, particularly for transitions to spherical states $\ket{n_M 0 0}$, where the mixing can take place over decades of orbital frequency prior to $\Omega^{aM}_g$~\cite{Tong:2022bbl,Tomaselli:2025jfo,Boskovic:2025ixx}. These states are also the ones for which the relativistic correction of the spectrum is the strongest, as the support of the wavefunction is pushed towards the horizon. On the other hand, precisely because of the non-resonant nature of the mixing~\footnote{ In addition, self-gravity-induced mixing can operate via the kernel $G_{Maaa}$, which requires $m_M =   m_a$. Within the (hyper)fine multiplet this mixing never connects the SR-generated circular state to any other state of the multiplet. Self-gravity-induced mixing with strongly decaying states can, however, be relevant in the overall dynamics; this is the subject of a separate study~\cite{Boskovic:2026boh}.\label{ft:sg_mix}}, the precise determination of the resonant frequency is of lesser importance. \vskip 4pt

The broad history of the cloud evolution was described in~\cite{Boskovic:2025ixx}, a priori applicable in the dilute cloud limit. Let us now briefly revisit this history for dense clouds, in light of the shifted resonant frequencies, leaving a more quantitative analysis for future studies. The $\ket{211}$ cloud\footnote{Given a set of parameters describing the cloud-BH(-environment) co-evolution~\footref{ft:cloud_evo1}, together with dissipative processes such as axion annihilation into GWs~\footref{ft:gw}, the succession of clouds $\ket{211} \to \ket{322} \to \ket{433} \to \dots$ can be tracked explicitly~\cite{Arvanitaki:2014wva,Baryakhtar:2020gao,Khalaf:2024nwc}, assuming that the formation takes place outside the strong tidal field. Here we do not model this succession, and simply identify, for a rough interval in $\alpha$ (evaluated at the SR saturation of a given state), the longest-lived state, $t_a \gtrsim t_\mathrm{age}$, for BH masses $10 \lesssim M/M_\odot \lesssim 10^5$.} is long-lived for $0.01 \lesssim \alpha \lesssim 0.15$. The earliest (i.e.\ lowest in orbital frequency) equatorially supported $\mathcal{H}$ resonance $\ket{211} \to \ket{21{-}1}$ [scenario $\mathrm{(ii)}_\mathcal{H}$] can be significantly shifted to higher orbital frequencies for dense clouds compared to dilute ones (Fig.~\ref{fig:res_hf}), even reaching the LISA band in the part of the parameter space ($M \lesssim 100 M_\odot$, $\alpha \gtrsim 0.1$, $q_\mathrm{c} \gtrsim 0.05$). On the other hand, the off-equatorially supported transition $\ket{211} \to \ket{210}$ is virtually intact by self-gravity [scenario $\mathrm{(i)}_\mathcal{H}$]. As the only $\mathcal{F}$-type transition $\ket{211} \to \ket{200}$ is dipolar and thus forbidden\footnote{In the deep-$\mathcal{B}$ regime, as the companion enters the density support of the cloud, dipolar transitions also become allowed~\cite{Brito:2023pyl}. Dipolar mixing can gain support at higher $\alpha$ for $\mathcal{F}$ and early-$\mathcal{B}$ transitions as well~\cite{Tomaselli:2024bdd}, although the $\ket{211}$ cloud is short-lived in that range.}, GAs with surviving clouds (formed on $\Omega > \Omega_{\mathcal{H}}$) can proceed towards the $\mathcal{B}$ regime. \vskip 4pt

In the case of the $\ket{322}$ cloud, supported for $0.1 \lesssim \alpha \lesssim 0.35$, all three scenarios operate (Figs.~\ref{fig:res_hf}--\ref{fig:res_hf_abs}): the off-equatorially supported resonance $\ket{322} \to \ket{321}$ as in the dilute cloud regime [$\mathrm{(i)}_\mathcal{H}$]; while the equatorially supported and quadrupole driven $\ket{322} \to \ket{320}$ is either shifted to lower orbital frequencies or is not of the floating type in a significant part of the parameter space [$\mathrm{(iii)}_\mathcal{H}$]; in a further subspace of the previous scenario the $(\ell=4)$-mediated $\ket{322} \to \ket{32{-}2}$ resonance is pushed towards sufficiently higher orbital frequencies in order to be adiabatic [$\mathrm{(ii)}_\mathcal{H}$]. Overall, the survival probability of the cloud in the $\mathcal{H}$ regime and the equatorial plane is somewhat increased in the dense cloud regime, compared to the dilute one. The cloud that survived can then proceed towards the $\mathcal{F}$ regime, where the floating nature of the transitions $\ket{322} \to \ket{31 m_M}$ is more robust even for denser clouds (Fig.~\ref{fig:res_fine}), en route  being slowly depleted via the non-resonant mixing $\ket{322} \to \ket{300}$. The history of higher-$n_a$ states ($\ket{433},\dots$), supported for $\alpha \gtrsim 0.3$, in the dense cloud regime is similar. Note that scenario $\mathrm{(ii)}_\mathcal{H}$ that protects the floating nature of the equatorial resonant transitions $\to \ket{n_a \, (n_a-1) \, {-}(n_a-1)}$ is mediated via $\ell=2(n_a-1)$, and thus increasingly irrelevant. On the other hand, the $\ell=(n_a-1)$-mediated non-resonant mixing with the spherical state $\ket{n_a 0 0}$ also gets progressively weaker~\cite{Boskovic:2025ixx}. Overall this means that the cloud is more likely to survive both the $\mathcal{H}$ regime and the wide mixing as $n_a$ grows. In most of the parameter space, the $\mathcal{H}$ resonances do not occur in the band of future GW detectors, while the imprint of the cloud in the higher-than-expected binary eccentricity, which these transitions can lead to when floating is possible, is mostly washed away by the subsequent RR dynamics before the binary reaches the band~\cite{Boskovic:2024fga,Boskovic:2025ixx}. Consequently, the impact of self-gravity in the equatorial plane for $n_a \geq 3$ is observationally favourable: it allows binaries formed in the early inspiral to proceed to the $\mathcal{F}$ regime, which is typically either in the band~\cite{Kim:2025wwj} or closer to it, thus allowing off-band imprints in the binary eccentricity distribution to survive to the band~\cite{Boskovic:2024fga,Boskovic:2025ixx}. On the other hand, the dominant off-equatorial ($\ell=2$)-mediated resonance $\to \ket{n_a \, (n_a-1)  \,  (n_a-2) }$ is not perturbed by self-gravity. Precisely these allow for off-equatorial fixed points in the spin-orbit misalignment, a cloud footprint that is not washed away by the RR-driven evolution~\cite{Boskovic:2025ixx}. \vskip 4pt

\subsection{Love numbers in binaries} \label{sec:love_dyn}

In the non-rotating case (as for, e.g., dark matter-induced clouds~\cite{Hui:2019aqm,Hui:2022sri,Budker:2023sex}), Love numbers represent the leading-order imprint of the object's structure in the GW waveform, and thus the large Love numbers expected for BHs dressed by a bosonic cloud have been discussed as a smoking-gun signature~\cite{Baumann:2018vus,Baumann:2019ztm,DeLuca:2021ite,Chia:2023tle}. It should be emphasised, however, that SR-generated clouds are \textit{by construction} spinning objects, and thus the permanent quadrupole term will dominate the induced one in the waveform due to the relative PN suppression ($2\mathrm{PN}$ vs.\ $5\mathrm{PN}$ at leading order), let alone the orbital consequences of the mixing dynamics discussed above, generated by the permanent quadrupole. This dynamics is, as argued, disruptive for the cloud, and one may ask -- even if the cloud is so dilute that the effect on the orbital dynamics is vanishing, can the presence of the bosonic overdensity be inferred from the conservative effect of the finite-size terms alone? \vskip 4pt

The same coupling inside the (hyper)fine multiplet that is responsible for the enhanced scaling and (in some cases) the negative values of the Love numbers (Sec.~\ref{sec:love_spin}) is also responsible for driving the level mixing and, ultimately, the resonant dynamics. In Sec.~\ref{sec:love_spin} the tide was taken in the strictly static limit. Let us here instead consider a scenario where, as the binary inspirals via RR, the tidal field inherits the orbital frequency, $\Pi = -ms\vartheta(t)$. In the dressed frame~\cite{Baumann:2019ztm}, which co-rotates with the tidal field, the overlap matrix evolves adiabatically and can be diagonalised as in Sec.~\ref{sec:love_spin}, with the splitting $\Delta\epsilon_{aN}$ of the diagonal entries replaced by the detuning $\Delta\epsilon_{aN}+ms\Omega$. The weak-field regime, where the (hyper)fine partner $\ket{N}$ can be integrated out and treated perturbatively, thus requires $|\eta^{aN}_{\ell m}| \ll |\Delta\epsilon_{aN}+ms\Omega|$. Three regimes can therefore be delineated: in the very early inspiral, $|ms\Omega| \ll |\Delta\epsilon_{aN}|$, the static Love numbers derived in Sec.~\ref{sec:love_spin} provide the appropriate description of the induced multipoles; further in the inspiral, dynamical corrections need to be included; and close to the resonance, the (hyper)fine partner $\ket{N}$ in general needs to be promoted to an active degree of freedom, contributing to the permanent multipoles instead, while the Bohr and continuum states generate the Love number contribution. 

\vskip 4pt

We can instead start by considering two active states $\{\ket{a},\ket{N}\}$ in the (hyper)fine multiplet and the mixing term of the permanent multipole, where [cf.~\eqref{eq:H_mix} and~\cite{Boskovic:2025ixx}]
\begin{eqnarray} \label{eq:pert_mix}
(Q^\mathrm{perm}_\mathrm{c})_E^{i_L} E_{i_L} \Big|_{aN} &=& 2 N_\mathrm{c}\, \eta^{aN}_{\ell m} D \sqrt{I_N} \cos\left(m s \vartheta + \phi_N\right) \nonumber \\
&\approx& - 2 N_\mathrm{c} \frac{\left( \eta^{aN}_{\ell m} \right)^2 \left(m s \dot{\vartheta} + \Delta \epsilon_{aN}\right)}{\left(m s \dot{\vartheta} + \Delta \epsilon_{aN} \right)^2 + \left(\Gamma_a - \Gamma_N \right)^2} \,,
\end{eqnarray}
and in the second equality we assumed that the relative mixing of the two states is perturbative and adiabatic (this can be appropriate even at the resonance for states of moderate decay width, $\ell_a,\ell_N \simeq 1$; cf.~\cite{Boskovic:2025ixx}). By substituting the perturbative mixing result above, we have integrated out\footnote{Here we perform the matching directly in the hydrogenic basis. In contrast, in Sec.~\ref{sec:love} we work in the perturbed basis $\ket{\chi_N}+\ket{\chi^{(1)}_N}$ [cf.~\eqref{eq:pi_state}], and thus the occupancies in the perturbed basis in general differ from the ones used here, $\tilde{C}_N\neq C_N$. However, as we have assumed there that the occupancies of all states $N\neq a$ are perturbative, the matching to the Love numbers of Sec.~\ref{sec:love_weft} can proceed straightforwardly.} the state $\ket{N}$ and thus generated a contribution to the induced response of $\ket{a}$, which in the weak-field and static limit reads $\lambda^{aa}_{i_L j_{L'}} E^{i_L} E^{j_{L'}} \supset (\eta^{aN}_{\ell m})^2/\Delta\epsilon_{aN}$ [cf.~\eqref{tidalaction} and~\eqref{eq:love_weak}] (on top of those from the Bohr and continuum states). The result~\eqref{eq:pert_mix}, however, also resums dynamical and dissipative contributions to Love numbers, which we do not discuss individually further here.
\vskip 4pt

Returning to the static limit (allowing these terms to be mapped from the action to the waveform), let us now estimate the PN scaling of the permanent- [cf.~\eqref{eq:E_mult_moment},~\eqref{eq:H_mix}] and the induced- [cf.~\eqref{eq:love_action},~\eqref{eq:uv_tidal}] multipole contributions to the finite-size terms in the action:
\begin{eqnarray}
\sum_{\ell} (Q^{\mathrm{ perm}})_E^{i_L}  E_{i_L} &\sim& q M v_\star^2 \bigg[ (v_\star^2)^2 \tilde{a}^2   + (v_\star^2)^4 \tilde{a}^4 + \dots \bigg] \,,  \nonumber \\
\sum_{\ell} (Q^{\mathrm{ perm}}_\mathrm{c})_E^{i_L}  E_{i_L} &\sim&  q M v_\star^2 \bigg[  \underbrace{ q_\mathrm{c}  \left(v_\star^2 \right)^2 \left(\frac{r_\mathrm{c}}{M}\right)^2 + q_\mathrm{c}  \left(v_\star^2 \right)^4 \left(\frac{r_\mathrm{c}}{M}\right)^4 + \dots}_{aa} \bigg] \,,  \nonumber \\
\sum_{(\ell m) (\ell' m')}  \lambda^{\mathrm{eff}}_{i_L j_{L'}} E^{i_L} E^{j_{L'}}  &\sim&  q M v_\star^2  \bigg[  \underbrace{  \, q_\mathrm{c}  q \left(v_\star^2 \right)^5 \left(\frac{r_\mathrm{c}}{M}\right)^4 \left( \frac{|\Delta \epsilon_{aN}|}{\mu} \right)^{-1} }_{aa} + \dots \bigg] \,, \label{eq:2state}
\end{eqnarray}
where the first two rows correspond to the $\ell=2,4,\dots$ hierarchy of the BH and the initial cloud permanent multipoles (see App.~\ref{app:magnetic} for the magnetic multipoles), while the third row corresponds to the leading-order, quadrupolar enhanced Love number. In the spirit of our question, we consider the scenario where the BH finite-size contribution dominates the diagonal contribution of the cloud up to $5\mathrm{PN}$, i.e.\ $|(Q^\mathrm{perm})^{E,B}_{\ell m}| \gtrsim |(Q^\mathrm{perm}_\mathrm{c})^{E,B}_{\ell m}(q_\mathrm{c})|_{aa}$ for $\ell < 5$.  %
As the BH Love number vanishes, the cloud Love number is the dominant $5\mathrm{PN}$ contribution~\footnote{There will also be a contribution at $5\mathrm{PN}$ from the $aa$ cloud sector and the BH part of $\mathcal{Q}^\mathrm{perm}_{2m}$, due to the PN correction of $E_{i_L}$ (see~\cite{Porto:2016pyg}), but the corresponding finite-size enhancement $\propto (r_\mathrm{c}/M)^2$ in the former and the spin suppression $\propto \tilde{a}^2$ in the latter will be subdominant to the Love number.}. Given the additional powers of the cloud radius and the enhancement factor, it can be sizeable even for dilute clouds, as we shall see in the following. %
\vskip 4pt

 \vskip 4pt

\subsubsection{Enhanced multipoles in the band} \label{sec:love_band}

Although a detailed study of the observability of permanent and induced multipoles is beyond the scope of this work (see e.g.~\cite{Shterenberg:2024tmo} for the inference of finite-size parameters with future detectors), let us describe the main features of the dynamical context, focusing, as motivated above, on the dilute cloud limit~\footnote{The dilute regime can be reached due to the combination of initial conditions (lower initial BH spin), the GW emission from the cloud while inspiralling to the band, and the decay-induced off-resonant mixing. Note that even floating resonances never completely deplete the cloud, as the floating is interrupted when the occupancy drops below the critical value~\cite{Boskovic:2025ixx}, allowing for the possibility of non-trivial Love numbers post-resonance.}. Consider first a $\ket{211}$ cloud on an equatorial orbit. As the tidal partner $\ket{21{-}1}$ has a large decay width in a sizable part of the parameter space, its occupancy is perturbative and the state can thus be integrated out. 

\begin{figure}
\begin{minipage}{0.48\textwidth}
    \centering
    \includegraphics[width=0.9\linewidth]{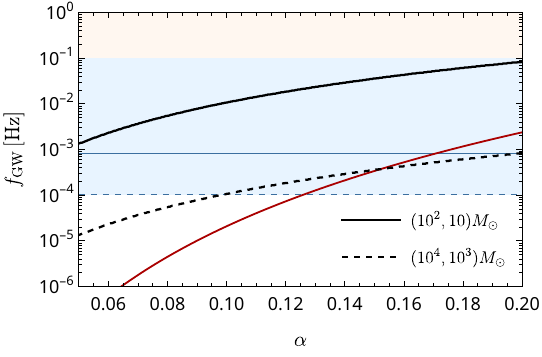}
\end{minipage}
\hfill
\begin{minipage}{0.48\textwidth}
    \centering
    \includegraphics[width=0.9\linewidth]{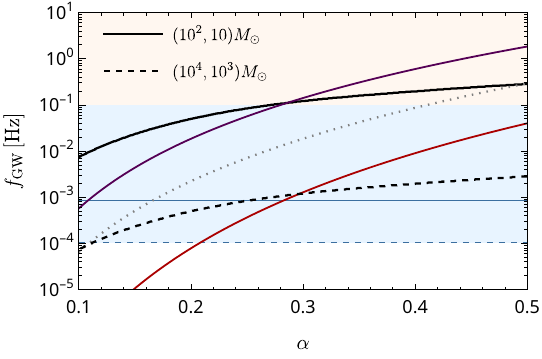}
\end{minipage}
\caption{(Left) The black curves delimit the region (below them) where the Love number on counter-rotating orbits is negative, i.e.\ where the hyperfine part, with the frequency dependence of~\eqref{eq:pert_mix}, dominates the positive $\mathcal{O}(\alpha^{-10})$ terms and the weak-field regime holds, shown for a stellar-mass $(10^2+10)\,M_\odot$ (solid) and an intermediate-mass black hole $(10^4+10^3)\,M_\odot$ (dashed) binary. The corresponding red curve indicates the hyperfine resonance $\ket{211} \to \ket{21{-}1}$. The light blue and orange shaded regions correspond to the LISA and (deci-Hz) DECIGO bands, and the horizontal blue lines, for each binary, to the onset of chirping (the frequency drifts by more than ten bins of width $1/T_\mathrm{obs}$ over $T_\mathrm{obs}=4\,$yr). (Right) \textit{Mutatis mutandis} for the $\ket{322}$ state, where the red, dotted gray and purple curves denote the  resonances $\ket{322} \to \ket{320}$, $\ket{322} \to \ket{31{-}1}$ and $\ket{322} \to \ket{300}$, respectively. In both cases we show the resonances only for the stellar-mass case, as for the intermediate-mass black hole example the hyperfine resonance always lies outside the band.}
\label{fig:love_freq}
\end{figure}

 In Fig.~\ref{fig:love_freq} (left) we show, in terms of the GW frequency $f_\mathrm{GW} = \Omega/\pi$ and for two types of binaries, part of the parameter space where the enhancement of (negative) Love numbers is relevant. On co-rotating orbits, this is the case prior to the $\mathcal{H}$ resonance (in red), which can resonantly decay the cloud. On counter-rotating orbits, the dynamical factor $(1+\Omega/\Omega^{\braket{211|21{-}1}}_2)^{-1}$ [cf.~\eqref{eq:pert_mix}] eventually offsets the relative enhancement in $\alpha$ compared to the $\mathcal{O}(\alpha^{-10})$ scaling of the Love numbers (black curves). Thus, the plots can be read off as delineating the portion of the band where enhanced Love numbers can serve as a handle on the dilute cloud: for the $(10^2+10)\,M_\odot$ binary, between the blue and red lines on co-rotating orbits and between the blue and black lines on counter-rotating orbits (and, for the $(10^4+10^3)\,M_\odot$ binary, between the dashed blue and dashed black lines on counter-rotating orbits). The effective Love number, including its dynamical frequency dependence (away from the resonance, $|\Delta\epsilon_{aN}+ms\Omega| \gg |\Gamma_a-\Gamma_N|$, where the decay widths can be dropped), then reads
\begin{eqnarray}
\frac{ \lambda^{\mathrm{eff}}_{(2, \pm 2),{(2, \pm 2)}}(\Omega)}{M^5} = -6 \cdot 10^{7} \, \left(\frac{\alpha}{0.15} \right)^{-14} \left(\frac{q_\mathrm{c}}{10^{-5}} \right) \left(1 - s\,\frac{\Omega}{\Omega^{\braket{211|21{-}1}}_2}\right)^{-1} \,,
\end{eqnarray}
where we take $\tilde{a}$ to correspond to the saturation of SR, $\tilde{a}_\mathrm{sat} = 4 m_a \alpha/(m_a^2 + 4 \alpha^2)$ [via~\eqref{eq:detw}] (for the reference values chosen above, the BH permanent quadrupole dominates the diagonal component of the cloud's permanent quadrupole, $|(Q^\mathrm{perm}_\mathrm{c})^E_{20}/(Q^\mathrm{perm})^E_{20}| = 6q_\mathrm{c}/(\alpha^4\tilde{a}^2) \simeq 0.4$, while the cloud's diagonal hexadecapole is zero by the selection rules). \vskip 4pt

A similar analysis can be performed for higher-$n_a$ states. For the $\ket{322}$ cloud, the enhancement of the Love numbers in the weak-field regime receives contributions from the overlap with $\{\ket{320},\ket{300}\}$. The same overlap again leads to level mixing, the former being an $\mathcal{H}$ resonance and the latter a strong decay-induced off-resonant mixing. In general, the scaling of the enhancement, the overall sign of the Love number, and the nature of the resonance depend on the interplay between $\{\alpha, q_\mathrm{c}\}$ (cf.\ Sec.~\ref{sec:love_dense},~Sec.~\ref{sec:chrono}).   In contrast to the $\ket{211}$ cloud, the size of the parameter space where the hyperfine tidal partner can be integrated out is smaller, and thus the two-state system $\{\ket{322},\ket{320}\}$ needs to be evolved close to the resonance. Similarly to the previous case, we delineate in Fig.~\ref{fig:love_freq} (right) the part of the parameter space where the negative and enhanced multipolar term at $5\mathrm{PN}$ is observationally relevant, i.e.\ prior to the $\ket{322}\to\ket{320}$ resonance on co-rotating orbits. Note that the Bohr multiplet has a richer structure in this example: we indicate the $\ket{322}\to \ket{320}$ resonance in red, the $\ket{322}\to\ket{300}$ resonance in purple,  and the octupole-mediated $\mathcal{F}$ resonances with $\ket{31 \pm 1}$ in dotted gray.\footnote{ The latter overlap is responsible for the enhancement of the octupolar Love numbers, but the resonance indirectly leads to a change in the quadrupolar Love numbers, by depleting the initial cloud. The efficiency of these resonances, as well as the wide mixing of $\ket{322}$ and $\ket{300}$, depends on $q_\mathrm{c}$; we leave a more detailed exploration of the parameter space for future work.} In both cases, the onset of the deep-$\mathcal{B}$ regime takes place at frequencies a few times higher than those of the black curves. For reference, at the onset of chirping on counter-rotating orbits, the negative multipolar term at $5\mathrm{PN}$ of the stellar-mass $(10^2+10)\,M_\odot$ binary exceeds in magnitude its Bohr and continuum part by a factor of $\simeq 11$ for $\ket{211}$ at $\alpha=0.1$ and $\simeq 18$ for $\ket{322}$ at $\alpha=0.2$ (rising to $\simeq 40$ and $50$, respectively, for a $(10+1)\,M_\odot$ binary). On co-rotating orbits, the dynamical enhancement grows further as the resonance is approached, with the mixing still perturbative: e.g.\ for $\ket{322}$ at $\alpha=0.3$ and the stellar-mass binary, the factor grows from $\simeq 10^2$ at the onset of chirping to $\simeq 8\times10^2$ at $\Omega \simeq 0.95\,\Omega^{\braket{322|320}}_2$. \vskip 4pt

\section{Conclusions} \label{sec:conc}

In this work, in the framework of WEFT, we have extended the description of GAs beyond the permanent multipoles, to include the induced response to tidal perturbations. At the level of the microphysics, we have used Rayleigh-Schr\"odinger perturbation theory to take into account the relevant competing perturbations, including (internal) relativistic corrections and self-gravity, thereby breaking all the degeneracies of the Bohr spectrum. The mixing dynamics is described at the level of a single cloud-binary phase space and is applicable to generic orbits, by extending the orbital subspace as in~\cite{Boskovic:2025ixx}. The framework that we set up, building on previous work~\cite{Baumann:2018vus,Baumann:2019ztm,Boskovic:2025ixx}, allows for a systematic inclusion of higher-order relativistic effects, both at the level of the inspiral (higher-order PN corrections) and of the microphysics (higher-order corrections in $\{\alpha,q_\mathrm{c}\}$).  Let us now emphasise the most important contributions and some possibilities for future studies:  \vskip 4pt

\begin{itemize}
\item[\bul{\faFingerprint}] \textbf{Smoking-gun signatures of Gravitational Atoms:} For SR clouds of moderate to high density, the permanent multipoles are the dominant finite-size effect. Particularly when a resonance is triggered, they can lead to distinct features in the orbital dynamics if the resonance occurs in band, i.e.\ (quasi-)floating and sinking behaviour, or to imprints in the distribution of binary orbital elements if it occurs off band~\cite{Baumann:2019ztm,Boskovic:2024fga,Tomaselli:2024bdd,Boskovic:2025ixx}. By  including self-gravity corrections as well as  beyond-hyperfine corrections in the relativistic spectrum, we can now predict, across a much wider parameter space than before, when a resonance occurs and what its nature is. In particular, for the former type our consistent treatment corrects the previous statements in the literature, even qualitatively:  the exchange contribution to the self-gravity kernel arises at the same order as the direct one and acts against it, so that several transitions previously expected to turn sinking for dense clouds in fact remain of the floating type. Crucially, once the permanent multipoles of the cloud are subdominant to those of the BH (e.g.\ after a resonance), the induced ones (Love numbers) can still be sizeable and remain a clear indicator of the bosonic overdensity. We have calculated the Love numbers of spinning clouds for the first time, showing how their scaling evolves across the parameter space and how, in the phenomenologically relevant regions, they acquire negative values and an enhancement of up to $\alpha^{-4}$ compared to the baseline estimate $\propto r_\mathrm{c}^5$. Furthermore, we demonstrate that the evolution of the induced response can be tracked even during the resonance. Let us emphasise that previous studies of the observational prospects for GA Love numbers~\cite{DeLuca:2021ite,Chia:2023tle} did not include the proper dynamical context and underestimate their size, for both non-spinning and spinning clouds, by a factor $\alpha^{-2} - \alpha^{-6}$. A reassessment of these forecasts is thus strongly motivated. \vskip 4pt

\item[\bul{\faSearch}] \textbf{Precision gravitational collider physics:} The dilute regime is therefore of phenomenological relevance --- when tracking the dynamical evolution of GAs in binaries across the parameter space, the cloud must be followed also at occupancies well below the SR-saturated values. This motivates assessing whether the Love numbers could be detectable past the full set of (hyper)fine resonances, in the band of future ground-based detectors such as ET. Another instance where tracking of low occupancies is of particular importance is when axion self-interactions are non-negligible. As their strength grows, the extraction of BH spin slows down and consequently the cloud mass is more modest~\cite{Baryakhtar:2020gao,Witte:2024drg}. Spin-based constraints on ultra-light bosons thereby lose their power, while a possible detection based on the permanent multipoles is no longer feasible. Precisely in this part of the axion parameter space, the Love numbers can be a useful handle. \vskip 4pt

\item[\bul{\faAtom}] \textbf{Towards the deep Bohr regime:} Our results apply to frequencies below the (deep) Bohr scale; above it, ionization~\cite{Baumann:2021fkf,Tomaselli:2024bdd} and contact interactions also need to be taken into account. Here, in particular, synergies with recent advances in the self-force approach would be welcome in the (intermediate/extreme) mass-ratio binary limit~\cite{Brito:2023pyl,Dyson:2025dlj,Keijzer:2026vul,Xu:2026cky,Xu:2026aic}. On the other hand, in the comparable-mass limit the system is more appropriately described as a gravitational molecule, as we have argued, for which several numerical studies have recently been performed~\cite{Ikeda:2020xvt,Guo:2025ckp,Guo:2025pea,Roy:2025qaa,Cheng:2025wac}; these could again serve as a useful benchmark in developing a controlled analytical framework. \vskip 4pt

\item[\bul{\faHeart}] \textbf{Signs of Love:} The sign of the Love numbers is of interest from the perspective of the positivity bounds program, where consistency of the underlying microphysics typically requires the leading Wilson coefficients to be positive~\cite{Adams:2006sv}. A non-trivial background (such as the GA) can, however, obstruct these expectations~\cite{Creminelli:2022onn,Hui:2023pxc,Creminelli:2023kze,Creminelli:2024lhd,Hui:2025aja}. In particular, the positivity of the Love numbers has been tied both to the passivity of the medium (the requirement that it only absorb energy from an external source)~\cite{Creminelli:2024lhd} and to the stability of classical weakly-gravitating systems~\cite{Creminelli:2026xxx}. In agreement with the first identification, we find that the negative Love numbers arise for a state $\ket{N}$ coupled, via the external field, to lower-energy states within its (hyper)fine splitting. The same coupling that induces the conservative response in the adiabatic limit also drives the level mixing between $\ket{N}$ and these lower-energy states. For GAs we can go further and track the non-linear evolution of this mixing: in the active setup it leads to the decay of the bulk of the initial cloud, whereas in the passive setup the mixing remains incomplete and the initial state is approximately stable. This suggests that GAs may serve, more broadly, as a useful setting in which to illustrate and test the theoretical aspects of response theory in GW physics. \vskip 4pt

\end{itemize}

\vskip 4pt {\bf Acknowledgements.} We would like to thank Paolo Creminelli, Riccardo Della Monica, Rafael Porto, Borna Salehian, Luca Santoni and Filippo Vernizzi for helpful discussions. We also thank Rafael Porto and the anonymous referee for feedback on a draft. The work of MB was supported in part by the Deutsche Forschungsgemeinschaft (DFG, German Research Foundation) under Germany's Excellence
Strategy – EXC 2121 ``Quantum Universe" – 390833306.

\appendix

\section{Non-relativistic effective theory for the scalar} \label{app:nr_limit}

In this appendix we follow~\cite{Salehian:2020bon} in order to set up a well-defined procedure for obtaining the non-relativistic limit of the real scalar field. First, we introduce an auxiliary field $X$ satisfying the constraint $X = \dot\Psi$. The description in terms of $\mathcal{L}[g,\Psi,\dot{\Psi}]$ is then equivalent to the one given by the following Lagrangian:
\begin{equation} \label{eq:nreft1}
    \tilde{\mathcal{L}}[g,\Psi,X,\pi] = \mathcal{L}[g,\Psi,X] + \pi(\dot\Psi - X) \,,
\end{equation}
where we have introduced the Lagrange multiplier $\pi$ enforcing the constraint. Since $\pi$ is not dynamical, one can eliminate it by solving its equation of motion. The resulting action depends only on $\{\Psi, \dot\Psi,X\}$ and their spatial derivatives. Next, one performs a field redefinition trading the pair of scalar fields for the complex scalar $\psi$:
\begin{equation}
    \Psi \rightarrow \frac{1}{\sqrt{2\mu}}\left(\psi e^{-i\mu t} + \bar{\psi} e^{i\mu t}\right), \hspace{0.5cm} X \rightarrow -i\sqrt{\frac{\mu}{2}}\left(\psi e^{-i\mu t} - \bar{\psi} e^{i\mu t}\right) \,,
\label{eq:NRfieldredef}
\end{equation}
resulting in the following Lagrangian:
\begin{eqnarray} 
    \tilde{\mathcal{L}} &=& -\bigg[i g^{0\mu}(\Bar\psi \pd_\mu \psi - \psi \pd_\mu \Bar \psi) + \frac{1}{\mu} g^{0 a}(\pd_t \psi \pd_a \Bar\psi + \pd_a \psi \pd_t \Bar\psi) + \frac{1}{\mu}g^{ab}\pd_a \psi \pd_b \Bar\psi + \mu(g^{00}+1)\psi \Bar\psi + \frac{1}{2\mu}e^{-2 i\mu t} \times  \nonumber \\
    &&\Big\{ g^{0 a}(\dot \psi \pd_a \psi + \pd_a \psi \dot\psi) + g^{ab}\pd_a \psi \pd_b\psi + \mu^2(g^{00}+1)\psi \psi + \frac{i \mu}{\sqrt{-g}}\pd_{\mu}(\sqrt{-g}g^{0\mu})\psi^2 + \mathrm{c.c.} \Big\}\bigg ] \,.
\end{eqnarray}
A couple of comments are in order. First, because the field redefinition~\eqref{eq:NRfieldredef} is invertible, the description in terms of $\psi$ is equivalent to that of the real scalar $\Psi$ we started from. Second, because~\eqref{eq:NRfieldredef} is non-local in time, so is the resulting Lagrangian for $\psi$. We will see that systematically integrating out the fast modes results in a manifestly local description. Finally, in the absence of gravity and upon dropping the oscillating factors, we recover the Lagrangian of the Schr\"odinger field as the leading non-relativistic approximation. \vskip 4pt 

We now specialise to the situation of interest, a non-relativistic scalar around the (deformed) Kerr black hole:
\begin{equation}
    g_{\mu\nu} = \bar g_{\mu\nu} + \mathfrak{h}_{\mu\nu} \,,
\end{equation}
where $\bar g_{\mu\nu}$ and $\mathfrak{h}_{\mu\nu}$ denote the Kerr background and the perturbation due to self-gravity $\sim q_\mathrm{c}$, respectively. Furthermore, let us fix the background de Donder gauge
\begin{equation}
    S_{\rm GF} = -\int d^4x \sqrt{-\bar g}\bar g^{\alpha\beta}\bar{\Gamma}_\alpha \bar \Gamma_\beta \,, \hspace{0.5cm} \bar\Gamma_{\mu} \equiv \bar g^{\alpha\beta}\bar\nabla_\alpha \mathfrak{h}_{\beta\mu} - \frac{1}{2}\bar g^{\alpha\beta}\bar\nabla_\mu \mathfrak{h}_{\alpha\beta} \,,
\end{equation}
with the covariant derivative $\bar\nabla_\mu$ defined with respect to the background metric. We will use the Boyer--Lindquist coordinates, in which the Kerr line element is
\begin{eqnarray}
    \bar g_{\mu\nu}dx^\mu dx^\nu &=& -\left(1-\frac{r_s r}{\Sigma}\right)dt^2 + \frac{\Sigma}{r^2 -r_s r + a^2}dr^2 + \Sigma d\theta^2 + \nonumber \\
    && \left(r^2 + a^2 + \frac{r_s a^2 r}{\Sigma}\right)\sin^2\theta d\phi^2 - \frac{2 r_s a r}{\Sigma}\sin^2\theta dt d\phi
\end{eqnarray}
with $r_s \equiv 2 M $, $\Sigma \equiv r^2 + a^2\cos^2 \theta$, and $a=\tilde{a} M$. Varying the action with respect to $\psi$ and $\mathfrak{h}_{\mu\nu}$ gives the equations of motion, where we organise the perturbative expansion in $\alpha \ll 1$
\begin{equation}
    (i \pd_t - \mathcal{H}_\mathcal{B})\psi  - e^{2i\mu t} \mathcal{H}_\mathcal{B} \bar \psi + \ldots  = 0 \,, \quad \mathcal{H}_\mathcal{B} \equiv -\frac{\nabla^2}{2\mu} - \frac{\alpha}{r} \,.
\label{eq:osceq}
\end{equation}
Here and in the following we aim to explain how to deal with the oscillatory contributions, and thus in the previous equation only the leading-order post-Bohr oscillatory contribution has been considered, while the next-to-leading-order ones as well as the non-harmonic contributions are ``hidden'' in $\dots$. One can estimate the impact of the oscillatory term by averaging over the Bohr scale, $\braket{e^{2i\mu t}\left\{ \ldots \right\}} \sim \alpha^2 \left\{ \ldots \right\}$, which thus contributes already at the fine order. \vskip 4pt

In order to proceed with the next-order corrections, we integrate out the ``fast'' modes order by order in $\alpha$, analogously to~\cite{Salehian:2020bon}
\begin{equation} \label{eq:psi_slow}
    \psi = \underbrace{\psi_{\rm s}}_{\text{slow}} + \underbrace{e^{2 i \mu t}\psi_{2} + \ldots}_{\text{fast modes}} \,,
\end{equation}
where $\psi_{\rm s}$ and $\psi_2$ evolve on the Bohr scales, and we have suppressed the components oscillating at even higher frequencies\footnote{Beyond the leading relativistic correction, the dynamics will induce more modes oscillating at frequencies that are multiples of $\mu$. Furthermore, there will also be ``slow'' and ``fast'' components of $\mathfrak{h}_{\mu\nu}$. These modes do not contribute at the order we focus on but are to be handled in a similar way~\cite{Salehian:2020bon}. Thus, $\mathfrak{h}_{\mu\nu}$ in our considerations varies at most on the Bohr scale.\label{ft:sg}}. Using the fact that, at leading order in $\alpha$, the slow component $\psi_{\rm s}$ satisfies $(i\pd_t - \mathcal{H}_\mathcal{B} + \frac{1}{2}\mu 
\mathfrak{h}^{00})\psi_{\rm s} = 0$, we find
\begin{equation}
    (i\pd_t - \mathcal{H}_\mathcal{B} )\left(\psi_{2} e^{2i\mu t}\right) = +e^{2i\mu t}\mathcal{H}_\mathcal{B}\bar\psi_{\rm s} \,.
\end{equation}
The leading-order term, in $\mathfrak{h}$ and $\alpha$, on the l.h.s. is $i \psi_2 \pd_t e^{2i\mu t}$, allowing us to solve for
\begin{equation} \label{eq:psi_slow2}
    \psi_2 = -\frac{\mathcal{H}_\mathcal{B}}{2\mu}\bar\psi_{\rm s} \,.
\end{equation}
Substituting~\eqref{eq:psi_slow} and~\eqref{eq:psi_slow2} in Eq.~\eqref{eq:osceq}, we find the dynamics of $\psi_{\rm s}$ to the fine-order relativistic correction
\begin{equation}
    (i \pd_t - \mathcal{H}_\mathcal{B})\psi_{\rm s} +\frac{1}{2\mu}\mathcal{H}_\mathcal{B}^2\psi_{\rm s} + \ldots  = 0 \,,
\end{equation}
where again we consider only the contribution from the oscillatory terms. Including also the non-harmonic contributions at this order, one finds $\mathcal{H}_\mathcal{F}$ in~\eqref{eq:S_micro}. Note that in the treatment of the complex scalar non-relativistic theory in~\cite{Baumann:2018vus}, a term proportional to $\partial_t^2$ contributes at the fine order. In our formalism, such a term is not present, by construction of the theory~\eqref{eq:nreft1}-\eqref{eq:NRfieldredef}. Instead, the term $-\frac{1}{2\mu}\mathcal{H}_\mathcal{B}^2$ yields exactly the same expectation value. \vskip 4pt

Similarly, one can find the equation of motion for~\footnote{Here we do not include processes such as axion annihilation $a + a \to g$ , which is forbidden for complex scalars but occurs for real ones~\cite{Arvanitaki:2010sy,Yoshino:2013ofa,Brito:2017zvb,Yang:2023vwm}. This is a high-energy process taking place above the $\mu$ scale and is thus not formally described by the low-energy EFT, although the corresponding decay channel can be added post hoc in an adiabatic manner (App.~\ref{sec:dis}).\label{ft:gw}}
 $\mathfrak{h}_{\mu\nu}$:
\begin{equation}
    \bar\nabla_\rho \bar\nabla^{\rho} \left( \mathfrak{h}_{\alpha \beta} - \frac{1}{2}\bar g_{\alpha \beta} \bar g^{\mu\nu}\mathfrak{h}_{\mu\nu}\right) = -16\pi G (T_\mathrm{c})_{\alpha\beta} \,,
\label{eqhsource}
\end{equation}
with the energy-momentum tensor $(T_\mathrm{c})_{\alpha\beta}$ containing the oscillatory contribution $\sim \psi^2_{\rm s} e^{-2i \mu t} + \mathrm{c.c.}$. Using the decomposition into slow and fast modes and the solution for $\psi_2$, we find the local expression for $(T_\mathrm{c})_{\alpha\beta}$, bilinear in $\psi_{\rm s}$ and $\bar\psi_{\rm s}$, given by
\begin{eqnarray}
    T_\mathrm{c}^{00} &=& \mu\Bar\psi_{\rm s} \psi_{\rm s} + \bigg( \frac{2\alpha}{r} - \frac{\nabla^2}{4\mu}\bigg)\Bar{\psi}_{\rm s}\psi_{\rm s} +\frac{1}{\mu}\pd_a\bar\psi_{\rm s} \pd^a\psi_{\rm s} -\frac{i}{2}(\Bar\psi_{\rm s} \dot\psi_{\rm s} - \psi_{\rm s} \dot{\Bar\psi}_{\rm s}) + \mathcal{O}( \mu \alpha^4) \,, \\
    T_\mathrm{c}^{0a} &=& -\frac{i}{2}(\Bar\psi_{\rm s} \pd^a \psi_{\rm s} - \psi_{\rm s} \pd^a \Bar\psi_{\rm s})  + \mathcal{O}(\mu \alpha^3) \,, \\
    T_\mathrm{c}^{ab} &=& \frac{1}{\mu} \pd^{\langle a}\Bar\psi_{\rm s} \pd^{b\rangle}\psi_{\rm s}+ \frac{1}{2}\delta^{ab}\left[ i(\bar \psi_{\rm s} \dot \psi_{\rm s} -\psi_{\rm s} \dot{\bar\psi}_{\rm s}) + \frac{2\alpha}{r}\psi_{\rm s} \bar\psi_{\rm s} - \frac{1}{3\mu}\pd_a\psi_{\rm s} \pd^a\bar\psi_{\rm s} \right]  + \mathcal{O}(\mu \alpha^4) \,.
\end{eqnarray} 
Note that the isospectrality between the real and complex scalar is broken~\footnote{Differences between the real and complex scalar self-gravitating solutions have been most thoroughly explored in the limit of vanishing horizon - oscillatons and boson stars, respectively, via numerical construction of the solution; cf.~\cite{Brito:2015yfh, Boskovic:2018rub,Salehian:2021khb}.} at $\mathcal{O}(\mu q_\mathrm{c} \alpha^4)$ (see the separate study~\cite{Boskovic:2026boh} for details). 

\section{Worldline EFT for the occupancies} \label{app:bottom}

In the following, we will construct the EFT with additional degrees of freedom (DoF) attached to the worldline, in the \textit{bottom-up} spirit, by taking a minimal set of assumptions on these DoF and discussing how symmetries constrain the dynamics. Throughout the exposition we illustrate the formalism with the example of the GA, and at the end we apply it to the tidally driven oscillations of compact objects. On top of the point-particle and finite-size terms and the Einstein-Hilbert action, explained in Sec.~\ref{sec:weft}, we introduce the scalar functions of the affine parameter (which we take to be proper time, in order to make the reparametrisation invariance manifest) $C_N(\tau)$, describing the additional DoF (``occupancies'' or ``modes''). The label $N$ runs over the modes and in general carries indices of any symmetry group under which they transform. To make direct contact with the discussion of GAs in the main body (where $N$ labels the atom states $\ket{n \ell m}$), we will take $C_N$ to be a complex scalar. \vskip 4pt

At leading order in derivatives and powers of $C_N$ we have the quadratic action, including the kinetic term, which can be brought to the normal-mode form provided it is positive-definite~\cite{Stone:bose_bogoliubov}:
\begin{equation}
    S^{(C)}_{\rm{min}} = \int d \tau \left[ i \bar C_N \pd_\tau C_N - \epsilon_{N}\, \bar C_N C_N \right]  \,.
\end{equation}
This provides a first-order differential equation for $C_N$ (or, equivalently, a second-order equation for its real and imaginary components). As a consequence, higher-order terms in the action involving $\pd_\tau C_N$ can always be reduced, by an appropriate field redefinition, to terms without derivatives on $C_N$.

One can introduce terms with more powers of $C_N$, corresponding to interactions among these degrees of freedom (e.g.\ self-gravity):
\begin{multline}
\label{selfint}
    S^{(C)}_{\rm{self\text{-}int}} = \int d \tau \sum_{NMP}\Big( \xi_1^{NMP} \left( C_N C_M C_P + \mathrm{c.c.} \right) + i  \xi_2^{NMP} \left( C_N C_M  C_P - \mathrm{c.c.} \right) + \\+  \xi_3^{NMP} \left( C_N C_M \bar C_P + \mathrm{c.c.} \right) + i  \xi_4^{NMP} \left( C_N C_M  \bar C_P - \mathrm{c.c.} \right)\Big) + \mathcal{O}(C^4) \,,
\end{multline}
where $\xi_i^{NMP} \in \mathbb{R}$ and $\mathrm{c.c.}$ denotes complex conjugation, enforcing the reality of the action. We now describe the composite operators corresponding to the mass, spin and multipoles of the point particle. In general, these depend both on the active degrees of freedom (i.e.\ $C_N$ and the long-wavelength gravitational field) and on the sectors of the microphysics $\mathcal{U}$ (for a GA this corresponds to the BH horizon and the dynamics associated with the deep-$\mathcal{B}$ regime). Integrating out $\mathcal{U}$ in the case of  the permanent multipoles  results in (suppressing indices for the moment)
\begin{multline}
    Q^{\mathrm{perm}}[C_N(t), \mathcal{U}] = \int dt' \left( G^{(1)}_{N}(t,t')\, C_N(t') + \bar G^{(1)}_{N}(t,t')\, \bar C_N(t') \right) + \\
    + \int dt'\, dt'' \left( G^{(2)}_{NM}(t,t',t'')\, C_N(t') C_{M}(t'') + \ldots \right) + \mathcal{O}(C^3) \,,
\end{multline}
with $G^{(i)}_{N \dots}$ the Green's functions corresponding to the dynamics of $\mathcal{U}$, $(\ldots)$ in the second row denoting the other possible combinations of conjugations and we imposed the reality of $Q^{\mathrm{perm}}$. If $\mathcal{U}$ evolves on scales much faster than the tidal field or the $C_N$, $G^{(i)}_{N \dots}$ functions reduce to products of delta functions, rendering the relation instantaneous. In this limit (and restoring the indices), one finds
\begin{multline}
    (Q^{\rm perm})_{\mu_L} =  (Q^{\rm {perm}}_0)_{\mu_L} + \sum_N(Q^{\rm {perm}}_1)^N_{\mu_L} (C_N + \bar C_N)+ \sum_N(Q^{\rm {perm}}_2)^N_{\mu_L}\, i(C_N - \bar C_N) +\\+
    \sum_{NM} \Big((Q^{\rm {perm}}_1)^{NM}_{\mu_L}(C_N C_M + \mathrm{c.c.}) +(Q^{\rm {perm}}_2)^{NM}_{\mu_L}\, i(C_N C_M - \mathrm{c.c.})+\\+(Q^{\rm {perm}}_3)^{NM}_{\mu_L}(C_N \bar C_M + \mathrm{c.c.}) +(Q^{\rm {perm}}_4)^{NM}_{\mu_L}\, i(C_N \bar C_M - \mathrm{c.c.}) \Big) + \mathcal{O}(C^3) \,, \label{eq:permQ_gen}
\end{multline}
with the Wilson coefficients $(Q^{\rm{perm}}_k)^{N \dots}_{\mu_L}$, which can be matched for a concrete object. Including the tidal field $E_{i_L}$ works in exactly the same way, where again the separation of scales enforces a local relation, giving the induced multipoles in terms of $E_{i_L}$ and $C_N(t)$. The composite operator $\lambda_{i_L j_{L'}}$ has the same structure as~\eqref{eq:permQ_gen}, with the substitution $(Q^{\rm{perm}}_k)^{N \dots}_{\mu_L} \to (\lambda_k)^{N \dots}_{i_L j_{L'}}$.

The discussion so far maps to the one in the main body for GAs if the global $U(1)$ symmetry $C_N \to e^{i \theta} C_N$, $\theta \in \mathbb{R}$, is imposed, forbidding terms with different powers of $C_N$ and $\bar C_N$ in the action (e.g.\ the linear couplings $(Q^{\rm perm}_1)^N_{\mu_L}$). This symmetry emerges from the non-relativistic limit of the full scalar field, and corresponds to the conservation of the number of particles, as the continuum is not excited (App.~\ref{app:nr_limit}). \vskip 4pt

Let us now apply the formalism to the case of fluid compact objects, whose normal modes can be parametrised by the fluid displacement vector and whose spectrum can also be expressed in the $N=(\ell m)$ basis (e.g.~\cite{Flanagan:2007ix,Chakrabarti:2013xza}). Once the radial structure of these modes is integrated out, one is left with the scalar $C_N$, depending on time only, corresponding to the amplitude of a particular mode. Tidal driving of the compact object's mode dynamics has been discussed extensively in the context of neutron stars, mostly at the level of the equations of motion, but also in the WEFT framework~\cite{Chakrabarti:2013xza,HegadeKR:2026kku,Martinez-Rodriguez:2026omk}. For example, the strongest mode self-couplings~\eqref{selfint} correspond to the cubic mode interactions considered for neutron-star oscillations in~\cite{Kwon:2025zbc}. In contrast to the GA, no global $U(1)$ acts on the stellar mode amplitudes, so these odd-power terms are not forbidden. \vskip 4pt

Finally, let us illustrate some parallels between the tidal driving of neutron stars and GAs. The strongest modes of neutron stars are the $f$-modes, with resonances at $\mathcal{O}(1\,\mathrm{kHz})$, and the $g$-modes, at $\mathcal{O}(100\,\mathrm{Hz})$. Since the merger regime in the coalescence of neutron stars starts roughly at $\mathcal{O}(1\,\mathrm{kHz})$, the $f$-modes are only adiabatically excited, and to leading order they can be integrated out (corresponding to the sector $\mathcal{U}$ above); the neutron star's Love numbers are then dominated by the $f$-mode contribution~\cite{Chan_2014,Pitre_2024}. The $g$-modes, on the other hand, must be kept as active DoF, and correspond to the modes $C_N$. In this case, the induced multipole $(\lambda_1)^N_{i_L j_{L'}} C_N E^{j_{L'}}$ is interpreted as the coupling of the (dormant) $f$- and (active) $g$-modes to the tidal field. Following the $g$-mode dynamics explicitly allows one to keep track of the induced response also during the resonance. In parallel with the GA, the $f$-modes then correspond to the deep-Bohr dynamics, while the $g$-modes correspond to the dynamics associated with the (hyper)fine and early-Bohr overlaps. Note that the response to resonant tides in compact objects is modelled with a Lorentzian~\cite{Flanagan:2007ix,Chakrabarti:2013xza}, the same structure we find explicitly for the GA in the perturbative mixing limit~\eqref{eq:pert_mix}.
\vskip 4pt

\section{Magnetic multipoles of Gravitational Atoms} \label{app:magnetic}

Permanent (spin-induced) magnetic multipoles~\eqref{eq:J_L} are suppressed relative to the electric ones~\eqref{eq:I_L} by relativistic corrections, both in the cloud microphysics, $T^{0i}_\mathrm{c} \sim T^{00}_\mathrm{c}/(\mu r_\mathrm{c})$, and in the orbital dynamics of the bound orbit, $B_{i_L} \sim v_\star E_{i_L}$ [cf.~\eqref{eq:2state}]:
\begin{eqnarray} \label{eq:magnetic_scaling0}
 \mathcal{Q}^B_{i_L}  B^{i_L} \sim \alpha v_\star \left(  \mathcal{Q}_E^{i_L} E_{i_L} \right)  \sim q\, q_\mathrm{c}\, M \alpha^{1-2\ell} (v_\star^2)^{\ell+3/2}  \,.
\end{eqnarray}
As in the electric sector, for dense clouds and $\alpha \ll 1$, the large extent of the cloud causes the magnetic multipoles of the cloud to dominate over the BH ones $(Q^\mathrm{perm})^B_{\ell m} \simeq M^{\ell + 1} \tilde{a}^\ell \delta_{\ell, 2\mathbb{Z}+1}$, while the corresponding potential in the bound orbit is enhanced relative to its 2.5$\mathrm{PN}$ suppression. \vskip 4pt

Taking the leading-order contributions and working in the spherical-harmonic basis, we find [in parallel to~\eqref{eq:E_mult_moment}]
\begin{eqnarray}
Q^B_{\ell m} &\approx&  M_\mathrm{c} r_\mathrm{c}^{\ell-1} \sum_{MN} c^B_{\ell \ell_N} \mathcal{I}^{(MN|\ell-1 \, m)}_r \int d\Omega_{\hat{r}} \, \mathrm{Im}\left( Y_M^\ast\, \bm{Y}^E_N\, \frac{\bar{C}_M C_N}{N_\mathrm{c}} \right) \cdot \bm{Y}^B_{\ell m} \,, \\
c^B_{\ell \ell_N} &\equiv&  -\frac{2\sqrt{\ell}}{\ell!\sqrt{\ell+1}} \ \sqrt{\ell_N(\ell_N+1)} \,, \nonumber
\end{eqnarray}
where we introduced the vector spherical-harmonic components
\begin{eqnarray}
\bm{Y}^r_{\ell m} = Y_{\ell m} \hat{r} \,, \quad \bm{Y}^E_{\ell m} = \frac{1}{\sqrt{\ell(\ell+1)}} r \bm{\nabla} Y_{\ell m} \,, \quad  \bm{Y}^B_{\ell m}= \hat{r} \times \bm{Y}^E_{\ell m} \,.
\end{eqnarray}
The selection rules then follow from the angular overlap integral
\begin{eqnarray}
 \mathcal{I}^{(MN|\ell m)}_{\Omega,B} =  \int d\Omega_{\hat{r}} \, Y^\ast_M  \bm{Y}^{E}_N \cdot \bm{Y}^{B}_{\ell m} \,,
\end{eqnarray}
implying
\begin{eqnarray}
&& m = -\Delta m_{MN} \,, \nonumber \\
&&  \ell+\ell_M+\ell_N \in 2\mathbb{Z} + 1 \,, \nonumber \\
&& |\ell_M-\ell_N| \leq \ell \leq \ell_M+\ell_N \,, \label{eq:selection_magneticc}
\end{eqnarray}
i.e.\ only the second rule differs from that in the electric sector~\eqref{eq:selection_electric}. Focusing first on the diagonal case $M=N$, one finds that the lowest allowed multipole is $\ell=3$. Thus, the $\ell_M=1$ states do not carry a permanent magnetic octupole, whereas the higher $\ell_M > 1$ states do. \vskip 4pt

Including the magnetic multipoles in the cloud-orbit Hamiltonian~\eqref{eq:H_mix} generates analogous diagonal and mixing terms. Whenever the magnetic-mediated overlap $(\ell,m)$ is allowed, so is the electric one $(\ell+1,m)$. Reflection symmetry of the $E$ and $B$ tensors through the equatorial plane requires that, on equatorial orbits, the electric (magnetic) tidal components with odd (even) $\ell+m$ vanish. Thus, the previous condition implies that when one mixing is forbidden on equatorial orbits, the other is as well. Let us compare the strength of these terms at the resonance $M \Omega^{MN}_g \sim \alpha^p$, where $p=\{3,5,7\}$ for $\{\mathcal{B},\mathcal{F},\mathcal{H}\}$, respectively (assuming $\tilde{a} \sim \alpha$), allowing us to parametrise their ratio solely in terms of $\alpha$ and $p$. Taking, e.g.\ $p=5$, one can explicitly check that the $\ell=2$ magnetic-mediated tidal overlap competes with the $\ell=4$ electric one, together with the $\alpha^2$-suppressed relativistic corrections to the $\ell=2$ electric multipoles [cf.~\eqref{eq:J_L},~\eqref{eq:T00}]. Thus, the magnetic-mediated contribution must be included for consistency when next-to-leading-order corrections to~\eqref{eq:etaN} are considered. \vskip 4pt

\section{Relativistic corrections to the spectrum} \label{app:spectrum}

As explained in App.~\ref{app:nr_limit}, in the decoupling limit $q_\mathrm{c} \to 0$, the spectrum of real and complex scalars in the Kerr background is the same. Thus, we can work with a simpler non-relativistic theory for complex scalars, find the higher-order corrections in $q_\mathrm{c}^0 \times \{\alpha,\tilde{a}\}$, and use them for real scalars. The main motivation for doing this, beyond establishing the domain of validity of the perturbative calculation, is to break the degeneracy of the $\Delta \epsilon^{MN}/\Delta m^{MN}$ ratio in the $\mathcal{H}$ regime (see Sec.~\ref{sec:relativistic}). In going beyond the hyperfine order, we provide not only the next-to-leading-order corrections for these transitions, but also the consistent corrections to the fine transitions. Namely, when $\tilde{a} \sim \alpha$ (at the saturation of SR), the first hyperfine correction is of order $\tilde{a}\alpha^5 \sim \alpha^6$ [cf.~\eqref{eq:hf_spectrum}]. But this is not the only correction at $\mathcal{O}(\alpha^6)$, and thus the rest need to be computed in order to obtain consistent subleading terms. \vskip 4pt

The procedure is straightforward, albeit tedious. We start from the quadratic action for the non-relativistic complex scalar~\cite{Baumann:2018vus}
\begin{equation}
  \mathcal{L}_\psi =  \sqrt{-g}\left(-\frac{1}{2\mu}\left[
      \nabla_a\psi^*\nabla^a\psi
      + i\mu\,g^{0a}\!\left(\psi^*\nabla_a\psi-\psi\nabla_a\psi^*\right)
      + \mu^2\!\left(g^{00}+1\right)\psi^*\psi
    \right]\right),
  \label{eq:app-action}
\end{equation}
and expand in $\{\alpha,\tilde{a}\}$ using the power-counting rules that can be inferred from the Bohr-level description (Sec.~\ref{subsecmicro}), i.e.\ $r^{-1} \sim r^{-1}_\mathrm{c} \sim \mu \alpha$, $\partial_t \sim \epsilon^\mathcal{B} \sim \mu \alpha^2$. We report here the relevant Kerr metric expansions:
\begin{eqnarray}
\frac{\sqrt{-g}}{r^2 d\Omega_{\hat{r}}}\,\left(g^{00}+1\right) &=&
-\frac{2M}{r}-\frac{4M^{2}}{r^{2}}-\frac{8M^{3}}{r^{3}}-\frac{16M^{4}}{r^{4}}-\frac{32M^{5}}{r^{5}}
+\tilde{a}^{2}M^{2}\left(\frac{4M^{2}}{r^{4}}+\frac{16M^{3}}{r^{5}}\right)
+\mathcal{O}\left(r^{-6}\right)\,, \nonumber \\
\frac{\sqrt{-g}}{r^2 d\Omega_{\hat{r}}}\,g^{tt} &=&
-1-\frac{2M}{r}-\frac{4M^{2}}{r^{2}}-\frac{8M^{3}}{r^{3}}-\frac{16M^{4}}{r^{4}}-\frac{32M^{5}}{r^{5}}
 \\
&& + \tilde{a}^{2}M^{2}\left(-\frac{\cos^{2}\theta}{r^{2}}+\frac{4M^{2}}{r^{4}}+\frac{16M^{3}}{r^{5}}\right)
+\mathcal{O}\left(r^{-6}\right)\,, \nonumber \\
\frac{\sqrt{-g}}{r^2 d\Omega_{\hat{r}}}\,g^{rr} &=&
1-\frac{2M}{r}+\frac{\tilde{a}^{2}M^{2}}{r^{2}}\,, \nonumber \\
\frac{\sqrt{-g}}{r^2 d\Omega_{\hat{r}}}\,g^{\phi\phi} &=&
\frac{1}{r^{2}\sin^{2}\theta}
+\tilde{a}^{2}M^{2}\left(-\frac{1}{r^{4}}-\frac{2M}{r^{5}}\right)
+\mathcal{O}\left(r^{-6}\right)\,, \nonumber  \\
\frac{\sqrt{-g}}{r^2 d\Omega_{\hat{r}}}\,g^{t\phi} &=&
\tilde{a}M\left(-\frac{2M}{r^{3}}-\frac{4M^{2}}{r^{4}}-\frac{8M^{3}}{r^{5}}\right)
+\tilde{a}^{3}M^{3}\left(\frac{2M}{r^{5}}\right)
+\mathcal{O}\left(r^{-6}\right)\,,
\end{eqnarray}
where, e.g., $M^5/r^5 \sim \alpha^{10}$, etc. When evaluated on the hydrogenic wavefunction, the $\mathcal{O}(r^{-4})$ term in $\mu^{-1} g^{\phi \phi} \partial_\phi \partial_\phi \sim \mu m^2 \tilde{a}^2 \alpha^6$, together with the $\mathcal{O}(\tilde{a}^2r^{-2})$ term in $g^{tt} \partial_t \sim \mu m^2 \tilde{a}^2 \alpha^6$,\footnote{Here, the $m^2$ dependence enters via the pure quadrupole $Y_{20}$ in $g^{tt}$ [cf.~\eqref{eq:quadr_int}].} will break the resonant-position degeneracy for $\mathcal{H}$ transitions, leading to~\eqref{eq:hf_res_deg}. Using the above, we calculate all corrections up to $\mu \{\tilde{a}^0,\tilde{a}^1,\tilde{a}^2\} \times \alpha^8$. On top of this, for consistency, one needs to include the second-order eigenfrequency corrections [via~\eqref{eq:energy_correction}] from $\mathcal{O}(\mathcal{H}_\mathcal{F}^2) \sim \mu \alpha^6$, $\mathcal{O}(\mathcal{H}_\mathcal{F}\mathcal{H}_\mathcal{H}) \sim \mu m \tilde{a} \alpha^7$, and $\mathcal{O}(\mathcal{H}_\mathcal{H}^2) \sim \mu m^2 \tilde{a}^2 \alpha^8$. In the course of performing these calculations, one relies on the averages $\braket{r^k}_{NN}^\mathcal{B}$, $k \in \mathbb{Z}$, where $k \geq - 2 \ell_N - 2$ in order for the radial integral to converge. When the inequality is not satisfied, one needs to regularise the integral and renormalise the self-energy corrections. Here we instead simply drop these terms, in practice obtaining a weaker precision for the $\ell_N = \{0,1\}$ states. \vskip 4pt

\begin{figure}
\begin{minipage}{0.48\textwidth}
    \centering
    \includegraphics[width=0.9\linewidth]{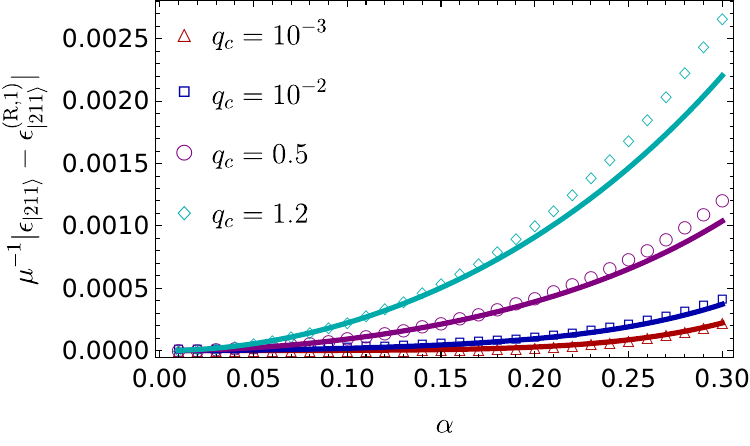}
\end{minipage}
\hfill
\begin{minipage}{0.48\textwidth}
    \centering
    \includegraphics[width=0.9\linewidth]{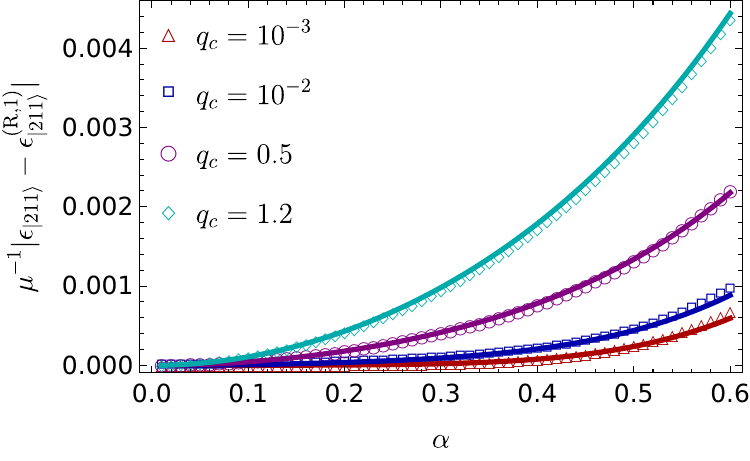}
\end{minipage}
\caption{Comparison of the numerical determination of the relativistic spectra for a self-gravitating cloud in~\cite{May:2024npn} (points) with our beyond-hyperfine relativistic and $\mathcal{O}(q_\mathrm{c} \alpha^2)$ self-gravity calculation (full lines), obtained by truncating the Bohr-to-hyperfine contributions $\epsilon^{(\mathrm{R},1)}_N$, for the $\ket{211}$ state (left) and the $\ket{322}$ state (right).}
\label{fig:num1}

\vspace{1.5em}

\begin{minipage}{0.48\textwidth}
    \centering
    \includegraphics[width=0.9\linewidth]{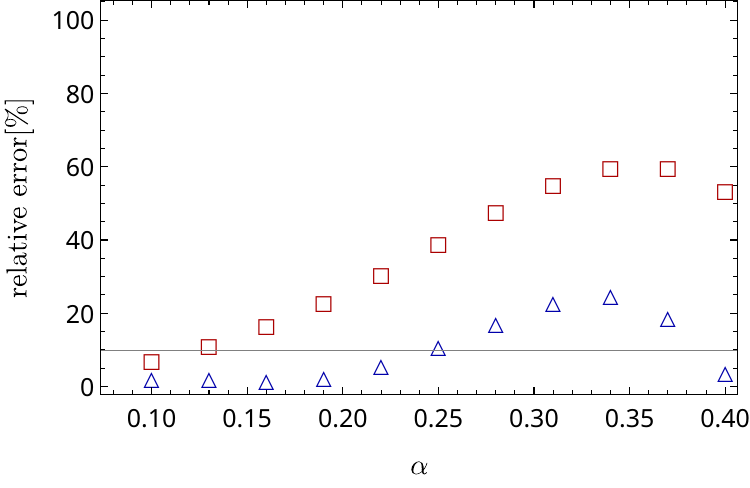}
\end{minipage}
\hfill
\begin{minipage}{0.48\textwidth}
    \centering
    \includegraphics[width=0.9\linewidth]{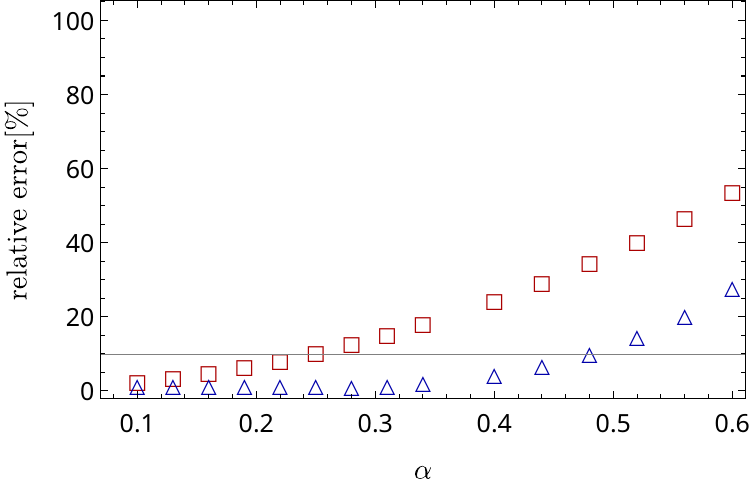}
\end{minipage}
\caption{Relative error of the analytical determination of $\Delta \epsilon^{NM}$, compared to the numerical determination of the relativistic spectrum, using only the leading-order hyperfine corrections~\eqref{eq:hf_spectrum} (red squares) and including the higher-order terms explained in App.~\ref{app:spectrum} (blue triangles). The transitions considered are $\ket{211} \to \ket{21{-}1}$ (left) and $\ket{322} \to \ket{320}$ (right).}
\label{fig:num2}
\end{figure}

Finally, we compare our results for the relativistic corrections with the numerical determination of the spectrum using the Leaver method~\cite{Dolan:2007mj}. We use the public code of~\cite{May:2024npn}, which provides the relativistic spectra of a self-gravitating (complex) scalar for the $\{\ket{211},\ket{322}\}$ states, as a benchmark on two fronts: in the $q_\mathrm{c}\to0$ limit it validates our own numerical calculation, while at finite $q_\mathrm{c}$ it tests our analytic self-gravity approximation. In Fig.~\ref{fig:num1} we subtract the leading-order relativistic result~\eqref{eq:hf_spectrum} from both the numerics of~\cite{May:2024npn} and our analytical result, which includes the beyond-hyperfine and $\mathcal{O}(\mu q_\mathrm{c}\alpha^2)$ self-gravity corrections, and compare the residuals, finding very good agreement. In Fig.~\ref{fig:num2} we perform a more stringent test of the relativistic calculation in the $q_\mathrm{c}\to0$ limit, comparing with our numerical results for the energy splittings $\Delta \epsilon^{\braket{211|21{-}1}}$ and $\Delta \epsilon^{\braket{322|320}}$ that enter the resonance condition~\eqref{eq:res}. These indicate that the first order (improved) relativistic calculation provides a good description, $\mathcal{O}(\lesssim 10\%)$, up to $\alpha \lesssim 0.12$ ($\alpha \lesssim 0.25$) for the $\ket{211}$ $\mathcal{H}$ resonances and up to $\alpha \lesssim 0.25$ ($\alpha \lesssim 0.5$) for the $\ket{322}$ $\mathcal{H}$ resonances. In these cases, and others that we have checked—including $\ket{322}$ $\mathcal{F}$ transitions and $\ket{433}$ $\mathcal{H}/\mathcal{F}$ resonances—we find agreement with the discussion around~\eqref{eq:pert_crit_HF}: for a given $\alpha$, the error of the analytical approximation decreases as $n_N,n_M$ increase.

\section{The Dalgarno - Lewis summation} \label{app:DL}

In perturbative calculations in Sec.~\ref{sec:pert} and Sec.~\ref{sec:love}, we have used the resummation technique developed by Dalgarno and Lewis (DL)~\cite{1955RSPSA.233...70D} for calculating the polarisability of the hydrogen atom. Let us briefly review the procedure and illustrate it for spherical states. The DL method trades an infinite sum over states for the solution of certain differential equation(s). One seeks a state $\ket {F^V_N}$ such that:
\begin{equation} \label{eq:F_implicit}
    (\mathcal{H}_\mathcal{B} - \epsilon_N)\ket {F^V_N} = \left[ V - \braket{V}_N \right] \, \ket N \,,
\end{equation}
where $V$ can, in principle, be a non-Hermitian operator. In position space, $F^V_N(\bm{x}) \equiv \braket{\bm{x}|F^V_N}$, the previous equation becomes
\begin{equation} \label{eq:F_scaled}
    \left(-\frac{\nabla^2_\mathsf{r}}{2}  -\frac{1}{\mathsf{r}} - \hat{\epsilon}_N  \right) \hat{F}^V_N(\bm{x}) = [\hat{V}(\bm{x}) - \braket{\hat{V}}_N ] \hat{\psi}_N(\bm{x}) \,,
\end{equation}
where we introduced the rescaling used in~\eqref{eq:E_mult_moment}, with $\hat{\psi} \equiv r_\mathrm{c}^{3/2} \psi$, $\hat{\epsilon}_N  \equiv \epsilon_N \, r_\mathrm{c} /  \alpha $, $\hat{V} \equiv V \, r_\mathrm{c} / (  \alpha \kappa)$, $\hat{F}^V_N \equiv F^V_N \, r_\mathrm{c}^{3/2}/\kappa$, and $\kappa$ being the parameter that enforces the correct dimensions.

Contracting~\eqref{eq:F_implicit} with $\bra P$, where we take $\ket P$ to be non-degenerate with $\ket N$, we find
\begin{equation}
    \braket{P|{F^V_N}} = \frac{\bra P V \ket N}{\epsilon_P - \epsilon_N}.
\end{equation}
Using this, we can rewrite the sum over states in~\eqref{eq:LoveAB} as:
\begin{eqnarray} \label{eq:F_sum}
    \sum_{\ket{P} \in \mathcal{S}_\perp} \frac{\bra P V \ket N \bra M U \ket P }{\epsilon_P - \epsilon_N} &=& \sum_{\ket{P} \in \mathcal{S}_\perp}  \bra M U \ket P \braket{P|{F^V_N}}  \, \nonumber \\
    &=& \bra M U \ket {F^V_N} - \sum_{\ket{\chi_P} \in \mathcal{S}_{\mathrm{d}}} \bra M U \ket{\chi_P} \braket{\chi_P|{F^V_N}} \,,
\end{eqnarray}
where, in the last row, we used the decomposition of the identity operator $\hat{\mathbb{I}} = \sum_P \ket P \bra P$ (including the continuum states). Thus, knowing $F^V_N(\bm{x})$, one can evaluate the sum in the perturbative expansion~\eqref{eq:uv_tidal}. \vskip 4pt

Let us now specialise to $\ket{N}=\ket{M}=\ket{a}=\ket{n00}$ and $V = \, \mathsf{r}^\ell Y_{\ell m}$, $U=V^\dag$~\footnote{Let us emphasise that $V$ is a non-Hermitian operator, in contrast to the physical potential $\frac{1}{2}\mu H_{00}$ that perturbs the states.}. From the selection rules, it follows that $\braket{\hat{V}}_N = 0$. Using the ansatz $\hat{F}^{\ell m}_a = \hat{\psi}_{a} Y_{\ell m} f^{(\ell)}_{n}(\mathsf{r})$ in~\eqref{eq:F_scaled}, we find
\begin{equation}
    (f^{(\ell)}_{n})'' + \left(\frac{2}{\mathsf{r}} + 2\frac{\hat{\mathcal{R}}_{n0}'}{\hat{\mathcal{R}}_{n0}}\right)(f^{(\ell)}_{n})' - \frac{\ell(\ell+1)}{\mathsf{r}^2} f^{(\ell)}_{n}  + 2 \mathsf{r}^\ell = 0 \,,
\end{equation}
with the prime denoting the radial derivative. Focusing further on the ground state, $n=1$, and the quadrupolar perturbation, $\ell = 2$, the solution to the above equation is
\begin{equation}
    f^{(2)}_1 = \frac{1}{6}  \mathsf{r}^2(2\mathsf{r}+3) + \frac{c_1 e^{2 \mathsf{r}} (\mathsf{r}-2)}{\mathsf{r}^3}+\frac{c_2 \left(-2 \mathsf{r}^3-6 \mathsf{r}^2-9 \mathsf{r}-6\right)}{4
   \mathsf{r}^3} \,.
\end{equation}
Requiring the solution to have finite norm sets $c_1 = c_2 = 0$, giving, via~\eqref{eq:F_sum}, an $m$-independent result 
\begin{equation}
    \sum_{\ket{P} \in \mathcal{S}_\perp} \frac{\bra{a}  r^2 Y_{2m}^\ast \ket P \bra P r^2 Y_{2m}  \ket{a}}{\epsilon_P - \epsilon_a} = \bra a r^2 Y_{2m}^\ast \ket {F^V_N} =  \frac{75}{8\pi}  \frac{1}{\mu \alpha^2} (\mu \alpha)^{-4}   \,, \nonumber
\end{equation}
leading to~\eqref{eq:love_100}.

\textit{Strong-field Love numbers.} The Love numbers~\eqref{eq:love_strong} and~\eqref{eq:love_strong_pm} of the good-basis states $\ket{\pm}$~\eqref{eq:eigen_val_vec} (with $\Pi=0$) follow by rotating the doublet indices of those in the hydrogenic basis, $\big(\lambda^{\braket{M|N}}\big)_\mathcal{B}$, where the subscript $\mathcal{B}$ indicates that only the Bohr and continuum states enter the sum (we suppress the tidal indices),
\begin{eqnarray}
\lambda^{\braket{\pm|\pm}} &=& \cos^2\!\varrho\,\big(\lambda^{\braket{21\pm1|21\pm1}}\big)_\mathcal{B}+\sin^2\!\varrho\,\big(\lambda^{\braket{21\mp1|21\mp1}}\big)_\mathcal{B} \nonumber\\ && \pm\sin\varrho\cos\varrho\left[\big(\lambda^{\braket{211|21{-}1}}\big)_\mathcal{B}+\big(\lambda^{\braket{21{-}1|211}}\big)_\mathcal{B}\right]\,,\nonumber\\
\lambda^{\braket{\pm|\mp}} &=& \cos^2\!\varrho\,\big(\lambda^{\braket{21\pm1|21\mp1}}\big)_\mathcal{B}-\sin^2\!\varrho\,\big(\lambda^{\braket{21\mp1|21\pm1}}\big)_\mathcal{B} \nonumber\\ && +\sin\varrho\cos\varrho\left[\big(\lambda^{\braket{21{-}1|21{-}1}}\big)_\mathcal{B}-\big(\lambda^{\braket{211|211}}\big)_\mathcal{B}\right]\,,
\end{eqnarray}
which in the strong-field limit, $\varrho=\pi/4$, give~\eqref{eq:love_strong} and~\eqref{eq:love_strong_pm}. The last term of~\eqref{eq:love_strong}, coupling the $(2,0)$ and $(2,\pm2)$ components, comes from the interference between $\ket{211}$ and $\ket{21{-}1}$ within $\ket{\pm}$: as these states are not axisymmetric, an $m=0$ tide induces an $m=\pm2$ quadrupole, and vice versa. \vskip 4pt

The hydrogenic-basis Love numbers follow from the DL method as above, with two modifications. First, as DPT is applied at the Bohr level, $\mathcal{S}_\mathrm{d}$ is the whole $n=2$ shell: we project it out of the source in~\eqref{eq:F_implicit} and fix the homogeneous solution by requiring $\ket{F^V_N}$ to be orthogonal to $\mathcal{S}_\mathrm{d}$, so that the last term in~\eqref{eq:F_sum} vanishes. Second, for $\ket{N}=\ket{21{\pm}1}$ the selection rules allow $\ell_P\in\{1,3\}$, so the ansatz for $\hat{F}^V_N$ carries both angular components, each with its own radial function. As the two states of the doublet are degenerate at the Bohr level, both terms of $\mathcal{K}^{MN}_{(\ell m)(\ell'm')}$ have the same denominator also for $M\neq N$, and via~\eqref{eq:LoveAB} we find
\begin{eqnarray}
N_\mathrm{c}\big(\lambda^{\braket{211|211}}_{(2,m),(2,m')}\big)_\mathcal{B} = N_\mathrm{c}\big(\lambda^{\braket{21{-}1|21{-}1}}_{(2,m),(2,m')}\big)_\mathcal{B} &=& \frac{15}{\pi}\frac{M_\mathrm{c}}{M}\frac{M^5}{\alpha^{10}} \nonumber\\ && \times\Big[68\left(\delta_{m,2}\delta_{m',2}+\delta_{m,-2}\delta_{m',-2}\right)+40\,\delta_{m,0}\delta_{m',0}\Big]\,, \nonumber\\
N_\mathrm{c}\big(\lambda^{\braket{211|21{-}1}}_{(2,m),(2,m')}\big)_\mathcal{B} &=& \frac{15}{\pi}\frac{M_\mathrm{c}}{M}\frac{M^5}{\alpha^{10}}\,14\sqrt{6}\left(\delta_{m,2}\delta_{m',0}+\delta_{m,0}\delta_{m',-2}\right)\,,
\end{eqnarray}
with $\big(\lambda^{\braket{21{-}1|211}}\big)_\mathcal{B}$ obtained via $(m,m')\to(-m,-m')$.
 \vskip 4pt

\section{Dissipative processes and (beyond) the adiabatic approximation} \label{sec:dis}

Scalar clouds are, by construction, non-Hermitian systems that can be generated (or dissipated) via BH SR~\cite{Arvanitaki:2009fg,East:2018glu} and other environmental processes~\cite{Hui:2019aqm,Hui:2022sri,Budker:2023sex}. In the former case, on which we mostly focus in this work, the growth of the cloud is at the expense of the BH mass and spin, implying a temporal evolution of the BH parameters $\{\alpha,\tilde{a}\}$ and consequently of the cloud spectrum [via~\eqref{eq:nreft_h0},~\eqref{eq:hf_spectrum}]. Other decay channels may also be relevant in tracking the cloud dynamics\footref{ft:gw}. We will include such effects by assigning a decay width to the individual states in~\eqref{eq:aa_var},
\begin{eqnarray} \label{eq:c_dis}
\dot{C}_N \Big|_\mathrm{dis} = -  \Gamma_N C_N \,.
\end{eqnarray}
In more detail, we assume that the GA can be treated as a sequence\footnote{Note that, in general, completeness is not guaranteed for a non-Hermitian system; see the discussion of this point in the context of scalar clouds in~\cite{Witte:2024drg} and references therein.} of states $\ket{\mathrm{Kerr}(T)} \otimes \sum_N \chi_N \left(\bm{x}; \mathscr{P}(T) \right)$, where $\mathscr{P} \equiv \{\alpha,\tilde{a},N_\mathrm{c} \}$, that vary on a ``long timescale'' $T \gg \Delta \epsilon_{NM}^{-1}$, where $\Delta\epsilon_{NM}$ represents a typical energy splitting in the system (adiabatic theorem~\cite{Weinberg_2015}). Specifically, for $\Gamma_N= \Gamma^\mathrm{SR}_N$, numerical relativity simulations confirm the validity of the adiabatic assumption for isolated clouds formed via SR~\cite{East:2018glu}, with $T^{-1} \simeq \Gamma^\mathrm{SR}_N$ and with the adiabatic co-evolution of the background BH parameters~\cite{Brito:2014wla,East:2018glu,Hui:2022sri}. At leading order, the SR decay timescale is given by Detweiler's approximation
\begin{eqnarray}  \label{eq:detw}
\Gamma^\mathrm{SR}_N \simeq 2\tilde{r}_+ C_{n \ell}\, g_{N} \, \alpha^{4\ell + 5}(\mu + \epsilon_N - m \Omega_H) \,,
\end{eqnarray}
with\footnote{Therefore, in the convention that we use, SR-growing states satisfy $\Gamma^\mathrm{SR}_N < 0$.} $g_{N}\equiv\prod_{k=1}^{\ell}\left[k^2(1-\tilde{a}^2)+(\tilde{a}m-2 \tilde{r}_+\alpha)^2\right]$, $C_{n\ell}\equiv\frac{2^{4\ell+1}(n+\ell)!}{n^{2\ell+4}(n-\ell-1)!}\left(\frac{\ell!}{(2\ell)!(2\ell+1)!}\right)^2$, $\tilde r_+= 1+\sqrt{1-\tilde a^2}$ and $M \Omega_H = \tilde a/(2\tilde r_+)$ (with $\epsilon_N$ including also the relativistic corrections)~\cite{Detweiler:1980uk,Dolan:2007mj,Arvanitaki:2010sy,Baumann:2019eav}. One can verify that parametrically $\max\{|\Gamma_N|,|\Gamma_M| \} \ll \Delta \epsilon_{NM}$ for all types of (non-)relativistic splittings~\eqref{eq:nreft_h0},~\eqref{eq:hf_spectrum}, as long as $\alpha < 1$. The adiabatic assumption is at its weakest for the fine splitting involving spherical states with $\ell_M=0$, where it may be only marginally satisfied already at intermediate values of $\alpha$. \vskip 4pt

Including the presence of the external perturbations (i.e.\ the orbital companion), the full spectrum will have an additional dependence on the orbital elements $\mathscr{P}'$. This dependence will propagate into the perturbative correction of the (instantaneous) eigenfrequencies and the wavefunctions $\ket{\chi_N \left(\mathscr{P},\mathscr{P}' \right)}$. On top of this, we will track the occupancy dynamics in full in Sec.~\ref{sec:pheno}, i.e.\ beyond the adiabatic approximation. \vskip 4pt

\bibliography{references.bib}
\bibliographystyle{apsrev4-1}

\end{document}